\newif\iftwocolumn\twocolumnfalse   % Plain one-column format, with appendix but no bios.

\iftwocolumn
    \documentclass[9.5pt,journal,finalsubmission,cspaper,compsoc]{IEEEtran}
\else
    \pdfoutput=1
    \documentclass{article}
    \date{3 March 2014} 
\fi

\usepackage{chngpage}
\usepackage{array}
\newcolumntype{L}[1]{>{\raggedright\let\newline\\\arraybackslash\hspace{0pt}}m{#1}}
\newcolumntype{C}[1]{>{\centering\let\newline\\\arraybackslash\hspace{0pt}}m{#1}}
\newcolumntype{R}[1]{>{\raggedleft\let\newline\\\arraybackslash\hspace{0pt}}m{#1}}

\usepackage{booktabs}
\usepackage{enumerate}
\usepackage{hyperref}
\usepackage{url}

\usepackage{listings}
\usepackage{multirow}
\usepackage{xspace}
\usepackage[table]{xcolor}
\usepackage{triotex} %% For namelabels

\usepackage{tikz}
\usetikzlibrary{shapes,arrows,positioning,fit,backgrounds}

\usepackage{pifont} % For check and mark symbols
\newcommand{\cmark}{\ding{51}}%
\newcommand{\xmark}{\ding{55}}%

\usepackage{graphicx}
\usepackage{subcaption}

\iftwocolumn
    \usepackage[nocompress]{cite}
\else
    \usepackage{appendix}
\fi

\iftwocolumn
	\newcommand{\tablefont}{}
\else
	\newcommand{\tablefont}{\footnotesize}
\fi

\makeatletter

\let\c@table\c@figure
\makeatother

\usepackage{boxedminipage}
\newenvironment{result}%
{\medskip
\noindent
\let\emph=\textbf
\begin{boxedminipage}{\columnwidth}\begin{center}\em}%
{\end{center}\end{boxedminipage}%
}

\iftwocolumn
    \newcommand{\figref}[1]{Fig.~\ref{#1}}  % Figure reference
\else
    \newcommand{\figref}[1]{Figure~\ref{#1}} % Figure reference
\fi
\newcommand{\autotest}{Au\-to\-Test\xspace}							% AutoTest tool
\newcommand{\afx}{Au\-to\-Fix\xspace}
\newcommand{\autofix}{\afx}  						% AutoFix tool
\newcommand{\website}{\url{http://se.inf.ethz.ch/research/autofix/}}

\newcommand{\code}[1]{\mbox{\lstinline|#1|}}           % Inline Eiffel code
\newcommand{\fakeF}[1]{\textit{#1}}  % Feature-like, for usage in comments

\renewcommand{\paragraph}[1]{\textbf{#1}\xspace}

\newcommand{\projectID}[1]{\texttt{#1}}
\newcommand{\efbase}{\projectID{Base}\xspace}
\newcommand{\txtlib}{\projectID{TxtLib}\xspace}
\newcommand{\cardgm}{\projectID{Cards}\xspace}
\newcommand{\elearn}{\projectID{ELearn}\xspace}

\newcommand{\expr}{\mathbb{E}}
\newcommand{\pred}{\mathbb{P}}
\newcommand{\subexp}{\mathsf{sub}}
\newcommand{\edist}[2]{\mathsf{eprox}({#1},{#2})}
\newcommand{\edep}[2]{\mathsf{edep}({#1},{#2})}
\newcommand{\cdist}[2]{\mathsf{cdist}({#1},{#2})}
\newcommand{\cdep}[2]{\mathsf{cdep}({#1},{#2})}
\newcommand{\locs}[1]{\mathsf{loc}({#1})}
\newcommand{\valueof}[3]{[\![{#1}]\!]_{#3}^{#2}}
\newcommand{\comp}{\mathsf{snap}}

\newcommand{\dscore}[1]{\mathsf{dyn}{#1}}

\newcommand{\npass}[1]{\#\mathsf{p}{#1}}
\newcommand{\nfail}[1]{\#\mathsf{f}{#1}}
\newcommand{\rank}[1]{\mathsf{susp}{#1}}
\newcommand{\modif}[1]{\mathsf{set}{#1}}
\newcommand{\replac}[1]{\mathsf{replace}{#1}}

\newcommand{\targs}[1]{\mathsf{targ}{#1}}

\newcommand{\ff}{\ensuremath{\mathfrak{f}}\xspace}
\newcommand{\perfspec}{\ensuremath{\mathfrak{S}}}

\newcommand{\runningemphasis}[2]{\emph\bgroup#1\egroup\xspace#2}

\newcommand{\tomainref}[1]{(Section~\ref{#1})}

\begin{document}

\iftwocolumn
    \title{Automated Fixing of Programs with Contracts}

    \author{Yu~Pei,~Carlo A. Furia,~Martin Nordio,~Yi Wei,~Bertrand~Meyer,~Andreas~Zeller%
    \thanks{Yu Pei, Carlo A. Furia, Martin Nordio, and Bertrand Meyer are with the Chair of Software Engineering, Department of Computer Science, ETH Z\"urich, Switzerland.}%
    \thanks{Yi Wei is with Microsoft Research Cambridge, UK.}%
    \thanks{Andreas Zeller is with the Software Engineering Chair, Saarland University, Germany.}}

    \markboth{IEEE Transactions on Software Engineering,~Vol.~1, No.~1,~January~2015}{Pei et al.: Automated Fixing of Programs with Contracts}

    \pubid{}
\else
    \clubpenalty=10000
    \widowpenalty = 10000
    \title{\textsc{Automated Fixing of \\ Programs with Contracts}}

    \newcommand{\ETH}{${}^*$\xspace}
    \newcommand{\MSR}{${}^\dag$\xspace}
    \newcommand{\UDS}{${}^\ddagger$\xspace}
    \newcommand{\smallemail}[1]{ {\scriptsize $\langle$#1$\rangle$}}

    \author{
    	Yu Pei\ETH $\cdot$ Carlo A. Furia\ETH
    		$\cdot$ Martin Nordio\ETH $\cdot$ Yi Wei\MSR \\[1mm]
    	Bertrand~Meyer\ETH $\cdot$ Andreas~Zeller\UDS\\[2mm]
      	{\ETH{}Chair of Software Engineering, ETH Zurich}\\
        {Z\"urich, Switzerland} \\
        {\{firstname.lastname\}@inf.ethz.ch}\\[2mm]
     		{\MSR{}Microsoft Research Cambridge} \\
       	{Cambridge, United Kingdom} \\
       	{yiwe@microsoft.com} \\[2mm]
     		{\UDS{}Software Engineering Chair, Saarland University} \\
       	{Saarbr\"ucken, Germany} \\
       	{zeller@cs.uni-saarland.de}
    }
\fi

% Commented out for arXiv \graphicspath{{analysis/R/subject-analysis/}{./analysis/R/success-rate-analysis/}{./analysis/R/time-cost-analysis/}}

\maketitle

\begin{abstract}
This paper describes \autofix, an automatic debugging technique that can fix faults in general-purpose software.
To provide high-quality fix suggestions and to enable automation of the whole debugging process, \autofix relies on the presence of simple specification elements in the form of contracts (such as pre- and postconditions). Using contracts enhances the precision of dynamic analysis techniques for fault detection and localization, and for validating fixes.
The only required user input to the \autofix supporting tool is then a faulty program annotated with contracts; the tool produces a collection of validated fixes for the fault ranked according to an estimate of their suitability.

In an extensive experimental evaluation, we applied \autofix to over 200 faults in four code bases of different maturity and quality (of implementation and of contracts).
\autofix successfully fixed 42\% of the faults, producing, in the majority of cases, corrections of quality comparable to those competent programmers would write; the used computational resources were modest, with an average time per fix below 20 minutes on commodity hardware.
These figures compare favorably to the state of the art in automated program fixing, and demonstrate that the \autofix approach is successfully applicable to reduce the debugging burden in real-world scenarios.
\end{abstract}

\iftwocolumn
\else
\fi

%%%%%%%%%%%%%%%%%%%%%%%%%%%%%%%%%%%%%%%%%%%%%%%%%%%%%%%%%%%%%%%%%%%%%%
\section{Introduction}
%%%%%%%%%%%%%%%%%%%%%%%%%%%%%%%%%%%%%%%%%%%%%%%%%%%%%%%%%%%%%%%%%%%%%%
\iftwocolumn\PARstart{T}{he} \else The\fi
programmer's ever recommencing fight against errors involves two
tasks: finding faults; and correcting them. Both are in dire need of
at least partial automation.

Techniques to \emph{detect} errors automatically are becoming increasingly
available and slowly making their way into industrial
practice~\cite{BesseyBCCFHHKME10,GodefroidLM12,Penix12}.  In contrast, automating the
whole debugging process---in particular, the synthesis of suitable
fixes---is still a challenging problem, and only recently have usable
techniques (reviewed in Section~\ref{sec:related}) started to appear.

\autofix, described in this paper, is a technique and supporting tool
that can generate corrections for faults of general-purpose software\footnote{As opposed to the domain-specific programs targeted by related repair techniques, which we review in Section~\ref{sec:doma-spec-models}.}
completely automatically.  \autofix targets programs annotated with
\emph{contracts}---simple specification elements in the form of
preconditions, postconditions, and class invariants.  Contracts
provide a specification of correct behavior that can be used not only
to detect faults automatically~\cite{meyer:programs:2009} but also to
suggest corrections.  The current implementation of \autofix is
integrated in the open-source Eiffel Verification
Environment~\cite{:eve:????}---the research branch of the EiffelStudio
IDE---and works on programs written in Eiffel; its concepts and
techniques are, however, applicable to any programming language
supporting some form of annotations (such as JML for Java or the .NET
CodeContracts libraries).

\autofix combines various program analysis techniques---such as
dynamic invariant inference, simple static analysis, and fault
localization---and produces a collection of suggested fixes, ranked
according to a heuristic measurement of relevance.  The dynamic
analysis for each fault is driven by a set of test cases that exercise
the routine (method) where the fault occurs.  While the \autofix
techniques are independent of how these test cases have been obtained,
all our experiments so far have relied on the \autotest random-testing
framework to generate the test cases, using the contracts as oracles.
This makes for a completely automatic debugging process that goes from
detecting a fault to suggesting a patch for it. The only user input is
a program annotated with the same contracts that programmers using a
contract-equipped language normally
write~\cite{polikarpova:comparative:2009, howSpecChange}.

In previous work, we presented the basic algorithms behind \autofix
and demonstrated them on some preliminary
examples~\cite{wei:automated:2010,PWFNM11-ASE11}.  The present paper
discusses the latest \autofix implementation, which combines and
integrates the previous approaches to improve the flexibility and
generality of the overall fixing technique.  The paper also includes,
in Section~\ref{sec:experiment}, an extensive experimental evaluation
that applied \autofix to over 200 faults in four code bases, including
both open-source software developed by professionals and student
projects of various quality.  \autofix successfully fixed 86 (or 42\%)
of the faults; inspection shows that 51 of these fixes are genuine
corrections of quality comparable to those competent programmers would
write. The other 35 fixes are not as satisfactory---because they may
change the intended program behavior---but are still useful patches
that pass all available regression tests; hence, they avoid program
failure and can be used as suggestions for further debugging.
\autofix required only limited computational resources to produce the
fixes, with an average time per fix below 20 minutes on commodity
hardware (about half of the time is used to generate the test cases
that expose the fault).  These results provide strong evidence that
\autofix is a promising technique that can correct many faults found
in real programs completely automatically, often with high reliability
and modest computational resources.

In the rest of the paper, Section~\ref{sec:overview} gives an overview
of \autofix from a user's perspective, presenting a fault fixed
automatically; the fault is included in the evaluation
(Section~\ref{sec:experiment}) and is used as running example.
Section~\ref{sec:prel-contr-tests} introduces some concepts and
notation repeatedly used in the rest of the paper, such as the
semantics of contracts and the program expressions manipulated by
\autofix.  Section~\ref{sec:fixgeneration} presents the \autofix
algorithm in detail through its successive stages: program state
abstraction, fault localization, synthesis of fix actions, generation
of candidate fixes, validation of candidates, and ranking heuristics.
Section~\ref{sec:experiment} discusses the experimental evaluation,
including a detailed statistical analysis of numerous important
measures.  Section~\ref{sec:related} presents related work and
compares it with our contribution.  Finally, Section~\ref{sec:future}
includes a summary and concluding remarks.

%%%%%%%%%%%%%%%%%%%%%%%%%%%%%%%%%%%%%%%%%%%%%%%%%%%%%%%%%%%%%%%%%%%%%%
\section{\afx in action}
\label{sec:overview}
%%%%%%%%%%%%%%%%%%%%%%%%%%%%%%%%%%%%%%%%%%%%%%%%%%%%%%%%%%%%%%%%%%%%%%

We begin with a concise demonstration of how \afx, as seen from a
user's perspective, fixes faults completely automatically.

\subsection{Moving items in sorted sets}
Class \code{TWO_WAY_SORTED_SET} is the standard Eiffel implementation of sets using a doubly-linked list. Figure~\ref{lst:moveItem} outlines features (members) of the class, some annotated with their pre- (\code{require}) and postconditions (\code{ensure}).\footnote{All annotations were provided by developers as part of the library implementation.}
As pictured in Figure~\ref{fig:set-overview}, the integer attribute \code{index} is an internal cursor useful to navigate the content of the set: the set elements occupy positions 1 to \code{count} (another integer attribute, storing the total number of elements in the set), whereas the indexes 0 and \code{count + 1} correspond to the positions \code{before} the first element and \code{after} the last.
\code{before} and \code{after} are also Boolean argumentless queries (member functions) that return \code{True} when the cursor is in the corresponding boundary positions.

\begin{figure}[!h]
  \centering
  \begin{tikzpicture}[
  item/.style={rectangle, minimum size=8mm,very thick,rounded corners=2mm,draw=green!50!black!50,font=\scriptsize, top color=white, bottom color=green!50!black!20},
  node distance=6mm,
  pin distance=7mm,
  every pin edge/.style={<-, shorten <=4mm, ultra thick,green!40!black!60},
  ]
  \lstset{basicstyle=\footnotesize}

  \node (i0) [item,dotted,label=above:\code{0}] {};
  \node (i1) [item,right=of i0,label=\code{1}] {$a$};
  \node (i2) [item,right=of i1,label=above:\code{2},pin=above:\code{index}] {$b$};
  \node (i3) [item,right=of i2,label=\code{3}] {$c$};
  \node (icm) [item,right=of i3,label=above:\code{4},label={[label distance=9pt]above:\code{count}}] {$d$};
  \node (iend) [item,right=of icm,dotted,label=above:\code{5},label={[label distance=8.5pt]above:\code{count + 1}}] {};
  \begin{scope}[-latex,green!40!black!70,thick]
    \foreach \na / \nb in {i0/i1,i1/i2,i2/i3,i3/icm,icm/iend}
    {
      \path (node cs:name=\na,angle=20) edge (node cs:name=\nb,angle=160);
      \path (node cs:name=\nb,angle=-160) edge (node cs:name=\na,angle=-20);
    }
  \end{scope}
\end{tikzpicture}
\caption{A doubly-linked list implementing
  \code{TWO_WAY_SORTED_SET}. The cursor \code{index} is on position
  \code{2}. The elements are stored in positions 1 to~4, whereas
  positions 0 (\code{before}) and 5 (\code{after}) mark the list's
  boundaries. \code{count} denotes the number of stored
  elements (i.e., four).}
\label{fig:set-overview}
\end{figure}
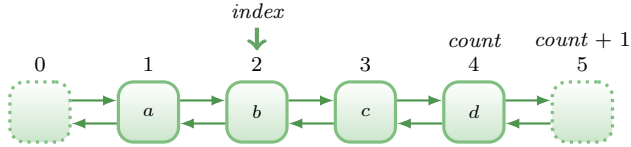

\begin{figure}[!p]
\centering
\begin{lstlisting}
  index: INTEGER  -- Position of internal cursor.

  count: INTEGER  -- Number of elements in the set.

  before: BOOLEAN -- Is (*\fakeF{index}*) = 0 ?   (*\label{ln:before}*)
    do Result := (index = 0) end

  after: BOOLEAN  -- Is (*\fakeF{index}*) = (*\fakeF{count}*) + 1 ?

  off: BOOLEAN    -- Is cursor (*\fakeF{before}*) or (*\fakeF{after}*) ?

  item: G         -- Item at current cursor position.
    require not off

  forth           -- Move cursor forward by one.
    require not after
    ensure index = old index + 1      (*\label{ln:forth-post}*)

  has (v: G): BOOLEAN  -- Does the set contain (*\fakeF{v}*) ?
    ensure Result implies count /= $\,$0     (*\label{ln:has-post}*)

  go_i_th (i: INTEGER) -- Move cursor to position (*\fakeF{i}*).
    require 0 <= i <= count + 1   (*\label{ln:goith-pre}*)

  put_left (v: G) -- Insert (*\fakeF{v}*) to the left of cursor.
    require not before            (*\label{ln:putleft-pre}*)

  move_item (v$\,$: G) -- Move (*\fakeF{v}*) to the left of cursor.
    require
       v /= Void                       (*\label{ln:pre:notVoid}*)
       has (v)                         (*\label{ln:pre:has}*)
    local idx: INTEGER ; found: BOOLEAN
    do
        idx := index                         (*\label{ln:begin}*)
        from start until found or after loop    (*\label{ln:loop-begin}*)
            found := (v = item)
            if not found then forth end
        end                                     (*\label{ln:loop-end}*)
        check found and not after end
        remove                                  (*\label{ln:remove}*)
        go_i_th (idx)                           (*\label{ln:goith}*)
        put_left (v)                            (*\label{ln:putleft}*)
    end
\end{lstlisting}
\caption{Some features of class \mbox{\code{TWO_WAY_SORTED_SET}}.}
\label{lst:moveItem}
\end{figure}

Figure~\ref{lst:moveItem} also shows the complete implementation of
routine \code{move_item}, which moves an element \code{v} (passed as
argument) from its current (unique) position in the set to the
immediate left of the internal cursor \mbox{\code{index}.}  For
example, if the list contains $\langle a, b, c, d \rangle$ and
\code{index} is 2 upon invocation (as in
Figure~\ref{fig:set-overview}), \mbox{\code{move_item (d)}} changes
the list to $\langle a, d, b, c \rangle$.  \code{move_item}'s
precondition requires that the actual argument \code{v} be a valid
reference (not \code{Void}, that is not \emph{null}) to an element already stored in the set
(\code{has(v)}).  After saving the cursor position as the local
variable \code{idx}, the loop in
lines~\ref{ln:loop-begin}--\ref{ln:loop-end} performs a linear search
for the element \code{v} using the internal cursor: when the loop
terminates, \mbox{\code{index}} denotes \mbox{\code{v}'s} position in
the set.  The three routine calls on
lines~\ref{ln:remove}--\ref{ln:putleft} complete the work:
\code{remove} takes \code{v} out of the set; \code{go_i_th} restores
\code{index} to its original value saved in \code{idx};
\code{put_left} puts \code{v} back in the set to the left of the
position \code{index}.

\subsection{An error in \code{move_item}} \label{sec:oneerror}
Running \autotest on class \code{TWO_WAY_SORTED_SET} for only a few minutes exposes, completely automatically, an error in the implementation of \mbox{\code{move_item}.}

The error is due to the property that calling \code{remove} decrements the \mbox{\code{count}} of elements in the set by one.
\autotest produces a test that calls \code{move_item} when \mbox{\code{index}} equals \mbox{\code{count + 1}}; after \mbox{\code{v}} is removed, this value is not a valid position because it exceeds the new value of \code{count} by two, while a valid cursor ranges between 0 and \code{count + 1}.
The test violates \mbox{\code{go_i_th}'s} precondition (line~\ref{ln:goith-pre}), which enforces the consistency constraint on \mbox{\code{index},} when \code{move_item} calls it on line~\ref{ln:goith}.

This fault is quite subtle, and the failing test represents only a special case of a more general faulty behavior that occurs whenever \code{v} appears in the set in a position to the left of the initial value of \code{index}:
even if \code{index <= count} initially, \code{put_left} will insert \code{v} in the wrong position as a result of \code{remove} decrementing \code{count}---which indirectly shifts the index of every element after \code{index} to the left by one.
For example, if \code{index} is \code{3} initially, calling \code{move_item (d)} on $\langle a, d, b, c\rangle$ changes the set to $\langle a, b, d, c\rangle$, but the correct behavior is leaving it unchanged.
Such additional inputs leading to erroneous behavior go undetected by \autotest because the developers of \code{TWO_WAY_SORTED_SET} provided an incomplete postcondition;
the class lacks a query to characterize the fault condition in general terms.\footnote{Recent work~\cite{PFM10-VSTTE10,EB2-website,PFPWM-ICSE13} has led to new versions of the libraries with strong (often complete) contracts, capturing all relevant postcondition properties.}

\subsection{Automatic correction of the error in \code{move_item}} \label{sec:correction-example}
\afx collects the test cases generated by \autotest that exercise routine \code{move_item}.
Based on them, and on other information gathered by dynamic and static analysis, it produces, after running only a few minutes on commodity hardware without any user input, up to 10 suggestions of fixes for the error discussed.
The suggestions include only \emph{valid} fixes: fixes that pass all available tests targeting \code{move_item}.
Among them, we find the ``\emph{proper}'' fix in Figure~\ref{lst:moveItemCorrections-one}, which completely corrects the error in a way that makes us confident enough to deploy it in the program.
\begin{figure}[!t]
\centering
\begin{lstlisting}
     if idx > index then     (*\label{ln:fix2-begin}*)
       idx := idx - 1
     end                     (*\label{ln:fix2-end}*)
\end{lstlisting}
\caption{Correction of the error in \code{move_item} automatically generated by \afx.}
\label{lst:moveItemCorrections-one}
\end{figure}
The correction consists of inserting the lines~\ref{ln:fix2-begin}--\ref{ln:fix2-end} in Figure~\ref{lst:moveItemCorrections-one} before the call to \code{go_i_th} on line~\ref{ln:goith} in Figure~\ref{lst:moveItem}.
The condition \code{idx > index} holds precisely when \code{v} was initially in a position to the left of \code{index}$\,$; in this case, we must decrement \code{idx} by one to accommodate the decreased value of \code{count} after the call to \code{remove}.
This fix completely corrects the error beyond the specific case reported by \autotest, even though \code{move_item} has no postcondition that formalizes its intended behavior.

\section{Preliminaries: contracts, tests, and predicates} \label{sec:prel-contr-tests}
To identify faults, distinguish between correct and faulty input, and abstract the state of objects at runtime, \afx relies on basic concepts which will now be summarized.

%----------------------------------------------------------------------------
\subsection{Contracts and correctness}
\label{sec:contracts}
%----------------------------------------------------------------------------

\afx works on Eiffel classes equipped with \emph{contracts}~\cite{meyer:object-oriented:2000}.
Contracts define the specification of a class and consist of \emph{assertions}: preconditions (\lstinline|require|), postconditions (\lstinline|ensure|), intermediate assertions (\lstinline|check|), and class invariants (translated for simplicity of presentation into additional pre- and postconditions in the examples of this paper).
Each assertion consists of one or more \emph{clauses}, implicitly conjoined and usually displayed on different lines; for example, \mbox{\code{move_item}'s} precondition has two clauses: \code{v /= Void} on line~\ref{ln:pre:notVoid} and \code{has(v)} on line~\ref{ln:pre:has}.

Contracts provide a criterion to determine the correctness of a routine:
every execution of a routine starting in a state satisfying the precondition (and the class invariant) must terminate in a state satisfying the postcondition (and the class invariant);
every intermediate assertion must hold in any execution that reaches it;
every call to another routine must occur in a state satisfying the callee's precondition.
Whenever one of these conditions is violated, we have a \emph{fault},\footnote{Since contracts provide a specification of correct behavior, contract violations are actual faults and not mere \emph{failures}.} uniquely identified by a location in the routine where the violation occurred and by the specific contract clause that is violated.
For example, the fault discussed in Section~\ref{sec:overview} occurs on line~\ref{ln:putleft} in routine \code{move_item} and violates the single precondition clause of \code{put_left}.

%----------------------------------------------------------------------------
\subsection{Tests and correctness}
\label{sec:test-cases-as-input}
%----------------------------------------------------------------------------
In this work, a test case $t$ is a sequence of object creations and routine invocations on the objects; if $r$ is the last routine called in $t$, we say that $t$ is a \emph{test case for} $r$.
A test case is \emph{passing} if it terminates without violating any contract and \emph{failing} otherwise.\footnote{Since execution cannot continue after a failure, a test case can only fail in the last call.}

Every session of automated program fixing takes as input a set $T$ of test cases, partitioned into sets $P$ (passing) and $F$ (failing).
Each session targets a single specific fault---identified by some failing location $f$ in some routine $r$ and by a violated contract clause $c$.
When we want to make the targeted fault explicit, we write $T_r$, $P_r$, and $F_r^{f, c}$.
For example, $F^{\ref{ln:putleft}, \text{\code{not before}}}_{\text{\code{move_item}}}$ denotes a set of test cases all violating \mbox{\code{put_left}'s} precondition at line~\ref{ln:putleft} in \mbox{\code{move_item}}.

The fixing algorithm described in Section~\ref{sec:fixgeneration} is independent of whether the test cases $T$ are generated automatically or written manually.
The experiments discussed in Section~\ref{sec:experiment} all use the random testing framework \autotest~\cite{meyer:programs:2009} developed in our previous work.
Relying on \autotest makes the whole process, from fault detection to fixing, completely automatic; our experiments show that even short \autotest sessions are sufficient to produce suitable test cases that \afx can use for generating good-quality fixes.

\subsection{Expressions and predicates} \label{sec:predicates}
\afx understands the causes of faults and builds fixes by constructing and analyzing a number of \emph{abstractions} of the program states.
Such abstractions are based on Boolean \emph{predicates} that \afx collects from three basic sources:
\begin{itemize}
\item argumentless Boolean queries;
\item expressions appearing in the program text or in contracts;
\item Boolean combinations of basic predicates (previous two items).
\end{itemize}

\subsubsection{Argumentless Boolean queries} \label{sec:abqs}

Classes are usually equipped with a set of argumentless Boolean-valued  functions (called \emph{Boolean queries} from now on), defining key properties of the object state:
a list is empty or not, the cursor is on boundary positions or before the first element ($\!$\code{off} and \code{before} in Figure~\ref{lst:moveItem}), a checking account is overdrawn or not.
For a routine $r$, $Q_r$ denotes the set of all calls to public Boolean queries on objects visible in $r$'s body or contracts.

Boolean queries characterize fundamental object properties.
Hence, they are good candidates to provide useful characterizations of object states: being argumentless, they describe the object state \emph{absolutely}, as opposed to in relation with some given arguments; they usually do not have preconditions, and hence are always defined; they are widely used in object-oriented design, which suggests that they model important properties of classes.
Some of our previous work~\cite{liu:using:2007,dallmeier:generating:2009} showed the effectiveness of Boolean queries as a guide to partitioning the state space for testing and other applications.

\subsubsection{Program expressions}

In addition to programmer-written Boolean queries, it is useful to build additional predicates by combining expressions extracted from the program text of failing routines and from failing contract clauses.
For a routine $r$ and a contract clause $c$, the set $E_{r,c}$ denotes all \emph{expressions} (of any type) that appear in $r$'s body or in $c$.
We normally compute the set $E_{r,c}$ for a clause $c$ that fails in some execution of $r$; for illustrative purposes, however, consider the simple case of the routine \code{before} and the contract clause \code{index > 1} in Figure~\ref{lst:moveItem}: $E_{\text{\code{before, index > 1}}}$ consists of the expressions \code{Result, index, index = 0, index > 1, 0, 1}.

Then, with the goal of collecting additional expressions that are applicable in the context of a routine $r$ for describing program state, the set $\expr_{r,c}$ extends $E_{r,c}$ by \emph{unfolding}~\cite{polikarpova:comparative:2009}: $\expr_{r,c}$ includes all elements in $E_{r,c}$ and, for every $e \in E_{r,c}$ of reference type $t$ and for every argumentless query $q$ applicable to objects of type $t$, $\expr_{r,c}$ also includes the expression \mbox{\code{e.q}} (a call of $q$ on target $e$).
In the example, $\expr_{{\text{\code{before}}}, {\text{\code{index > 1}}}} = E_{{\text{\code{before}}}, {\text{\code{index > 1}}}}$ because all the expressions in $E_{{\text{\code{before}}}, {\text{\code{index > 1}}}}$ are of primitive type (integer or Boolean), but this will no longer be the case for assertions involving references.

Finally, we combine the expressions in $\expr_{r,c}$ to form Boolean \emph{predicates}; the resulting set is denoted $B_{r,c}$.
The set $B_{r,c}$ contains all predicates built according to the following rules:
\begin{list}{}{}
\item[\textbf{Boolean expressions}:] $b$, for every Boolean $b \in \expr_{r,c}$ of Boolean type (including, in particular, the Boolean queries $Q_r$ defined in Section~\ref{sec:abqs});
\item[\textbf{Voidness checks}:] \mbox{\code{e = Void},} for every $e \in \expr_{r,c}$ of reference type;
\item[\textbf{Integer comparisons}:] $e \sim e'$, for every $e \in \expr_{r,c}$ of integer type, every \iftwocolumn\else\linebreak\fi $e' \in \expr_{r,c} \setminus \{e\} \cup \{0\}$ also of integer type,\footnote{The constant 0 is always included because it is likely to expose relevant cases.} and every comparison operator $\sim$ in $\{=, <, \iftwocolumn\linebreak\fi \leq\}$;
\item[\textbf{Complements}:] \code{not p}, for every $p \in B_{r,c}$.
\end{list}
In the example, $B_{{\text{\code{before}}}, {\text{\code{index > 1}}}}$ contains \code{Result} and \code{not Result}, since \code{Result} has Boolean type; the comparisons \code{index < 0}, \code{index <= 0}, \code{index = 0}, \code{index /= 0}, \code{index >= 0}, and \code{index > 0}; and the same comparisons between \code{index} and the constant 1.

\def\wprogram{13mm}
\iftwocolumn
\def\wtesting{14mm}
\def\wcomponents{18mm}
\def\wactions{10mm}
\def\wcandidates{25mm}
\def\wvalid{22mm}
\else
\def\wtesting{18mm}
\def\wcomponents{22mm}
\def\wactions{13mm}
\def\wcandidates{34mm}
\def\wvalid{29mm}
\fi

% Figure is placed here because double-column floats will only appear after the page where they are defined.
\begin{figure}[!tbp]
\centering
\begin{tikzpicture}[
  stage/.style={rectangle, minimum width=10mm,draw=black,align=center,font=\scriptsize, inner xsep=4pt},
  node distance=7mm and 5mm
  ]

  \node (program) [stage] {\textbf{Eiffel program} \\
\\
\begin{lstlisting}[numbers=none,basicstyle=\tiny,aboveskip=0pt,linewidth=\wprogram,xleftmargin=0pt]
class SET
  ...
end
\end{lstlisting}};

  \node (testing) [stage,below=of program] {\textbf{Test cases} \\
    \\
\begin{lstlisting}[numbers=none,basicstyle=\tiny,aboveskip=0pt,linewidth=\wtesting,xleftmargin=0pt]
forth ; move $\ \text{\normalsize \cmark}$
back ; move $\ \ \text{\normalsize \xmark}$
\end{lstlisting}
};

  \node (components) [stage,below=of testing] {\textbf{Suspicious}\\\textbf{snapshots} \\
    \\
\begin{lstlisting}[numbers=none,basicstyle=\tiny,aboveskip=0pt,linewidth=\wcomponents,xleftmargin=0pt] 1. $\text{line 42:}$ before
2. $\text{line 41:}$ idx > index
\end{lstlisting}
};

  \node (actions) [stage,right=of components] {\textbf{Fix actions} \\
    \\
\begin{lstlisting}[numbers=none,basicstyle=\tiny,aboveskip=0pt,linewidth=\wactions,xleftmargin=0pt]
1. forth
2. idx := 1
\end{lstlisting}
};

  \node (candidates) [stage,right=of actions] {\textbf{Candidate fixes} \\
    \\
\begin{lstlisting}[numbers=none,basicstyle=\tiny,aboveskip=0pt,linewidth=\wcandidates,xleftmargin=0pt]
1. $\text{line 42:}$
 > if before then forth
2. $\text{line 41:}$
 > if idx > index then idx := 1
\end{lstlisting}
};

  \node (valid) [stage,below=of candidates] {\textbf{Valid fixes} \\
    \\
\begin{lstlisting}[numbers=none,basicstyle=\tiny,aboveskip=0pt,linewidth=\wvalid,xleftmargin=0pt]
$\text{line 42:}$
 > if before then forth $\text{\normalsize \cmark}$
\end{lstlisting}
};

  \begin{pgfonlayer}{background}
    \node [label={[label distance=-5mm]above:\textbf{\normalsize\autofix}},fit={(components)(actions)(candidates)},inner xsep=2mm,inner ysep=5mm,draw,dotted,rounded corners] {};
  \end{pgfonlayer}

  \begin{scope}[->,thick]
  \path (program) edge (testing);
  \path (testing) edge (components);
  \path (components) edge (actions);
  \path (actions) edge (candidates);
  \path (candidates) edge (valid);
\end{scope}
\end{tikzpicture}
\caption{How \autofix works.  Given an Eiffel program with
  contracts~(Section~\ref{sec:contracts}), we generate passing and
  failing test cases that target a faulty routine
  (Section~\ref{sec:test-cases-as-input}).  By comparing the program
  state during passing and failing runs, \autofix identifies
  \emph{suspicious snapshots} (Sections~\ref{sec:state-assess}--\ref{sec:fault-analysis}) that
  denote likely locations and causes of failures.  For each suspicious
  snapshot, \autofix generates \emph{fix actions}
  (Section~\ref{sec:synthesis-of-fix-actions}) that can change the
  program state of the snapshot.  Injecting fix actions into the
  original program determines a collection of \emph{candidate fixes}
  (Section~\ref{sec:generating-candidate-fixes}).  The candidates that
  pass the regression test suite are \emph{valid}
  (Section~\ref{sec:validating-fixes}) and output to the user.  }
\label{fig:overview}
\end{figure}
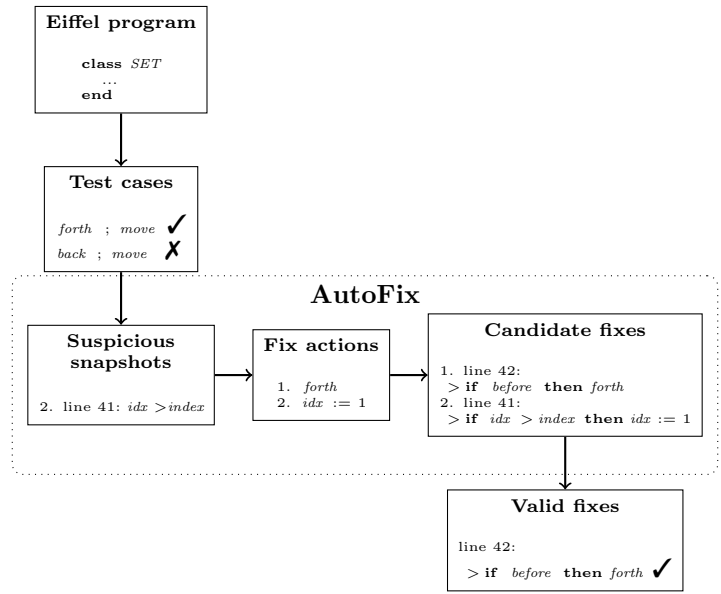

\subsubsection{Combinations of basic predicates}
One final source of predicates comes from the observation that the values of Boolean expressions describing object states are often correlated.
For example, \code{off} always returns \code{True} on an empty set (Figure~\ref{lst:moveItem}); thus, the implication \code{count = 0 implies off} describes a correlation between two predicates that partially characterizes the semantics of routine \code{off}.

Considering all possible implications between predicates is impractical and leads to a huge number of often irrelevant predicates.
Instead, we define the set $\pred_{r, c}$ as the superset of $B_{r,c}$ that also includes:
\begin{itemize}
\item All \emph{implications} appearing in $c$, in \emph{contracts} of $r$, or in contracts of any routine appearing in $B_{r,c}$;
\item For every implication \code{a implies b} collected from contracts, its \emph{mutations} \code{not a implies b}, \code{a implies not b}, \code{b implies a} obtained by ne\-gat\-ing the antecedent $a$, the consequent $b$, or both.
\end{itemize}
These implications are often helpful in capturing the object state in faulty runs.

The collection of implications and their mutations may contain \emph{redundancies} in the form of implications that are co-implied (they are always both true or both false).
Redundancies increase the size of the predicate set without providing additional information.
To prune redundancies, we use the automated theorem prover Z3~\cite{de:moura:z3::2008}: we iteratively remove redundant implications until we reach a fixpoint.
In the remainder, we assume $\pred_{r,c}$ has pruned out redundant implications using this procedure.

%%%%%%%%%%%%%%%%%%%%%%%%%%%%%%%%%%%%%%%%%%%%%%%%%%%%%%%%%%%%%%
\section{How \afx works}
\label{sec:fixgeneration}
%%%%%%%%%%%%%%%%%%%%%%%%%%%%%%%%%%%%%%%%%%%%%%%%%%%%%%%%%%%%%%

\figref{fig:overview} summarizes the steps of \afx processing, from failure to fix. The following subsections give the details.

\afx starts with a set of test cases, some passing and some failing, that expose a specific fault.
The fault being fixed is characterized by a program location $f$ and by a violated contract clause $c$ (Section~\ref{sec:test-cases-as-input}); the presentation in this section leaves $f$ and $c$ implicit whenever clear from the context.
The notion of \emph{snapshot} (described in Section~\ref{sec:state-assess}) is the fundamental abstraction for characterizing and understanding the behavior of the program in the passing or failing test cases; \afx uses snapshots to model correct and incorrect behavior.
Fixing a fault requires finding a suitable location where to modify the program to remove the source of the error.
Since each snapshot refers to a specific program location, \emph{fault localization} (described in Section~\ref{sec:fault-analysis}) boils down to ranking snapshots according to a combination of static and dynamic analyses that search for the origins of faults.

Once \afx has decided where to modify the program, it builds a code snippet that changes the program behavior at the chosen location.
\afx synthesizes such \emph{fix actions}, described in Section~\ref{sec:synthesis-of-fix-actions}, by combining the information in snapshots with heuristics and behavioral abstractions that amend common sources of programming errors.

\afx injects fix actions at program locations according to simple conditional schema; the result is a collection of \emph{candidate fixes} (Section~\ref{sec:generating-candidate-fixes}).
The following \emph{validation} phase (Section~\ref{sec:validating-fixes}) determines which candidate fixes pass all available test cases and can thus be retained.

In general, \afx builds several valid fixes for the same fault; the valid fixes are \emph{ranked} according to heuristic measures of ``quality'' (Section~\ref{sec:ranking}), so that the best fixes are likely to emerge in top positions.

The latest implementation of \afx combines two approaches developed in previous work: model-based techniques~\cite{wei:automated:2010} and code-based techniques~\cite{PWFNM11-ASE11}.

%---------------------------------------------------------------------------
\subsection{Program state abstraction: snapshots}
\label{sec:state-assess}
%---------------------------------------------------------------------------
The first phase of the fixing algorithm constructs abstractions of the passing and failing runs that assess the program behavior in different conditions.
These abstractions rely on the notion of \emph{snapshot}\footnote{In previous work~\cite{PWFNM11-ASE11}, we used the term ``component'' instead of ``snapshot''.}: a triple
\[
\left\langle \ell, p, v \right\rangle ,
\]
consisting of a program location $\ell$, a Boolean predicate $p$, and a Boolean value~$v$.
A snapshot abstracts one or more program executions that reach location $\ell$ with $p$ evaluating to $v$.
For example, $\langle \ref{ln:pre:has}, \text{\code{v = Void}}, \text{\code{False}}\rangle$ describes that the predicate \code{v = Void} evalutes to \code{False} in an execution reaching line 31.

Consider a routine $r$ failing at some location $f$ by violating a contract clause~$c$.
Given a set $T_r$ of test cases for this fault, partitioned into passing $P_r$ and failing $F_r^{f, c}$ as described in Section~\ref{sec:test-cases-as-input}, \afx constructs a set $\comp(T_r)$ of snapshots.
The snapshots come from two sources: invariant analysis (described in Section~\ref{sec:comp-inv-analysis}) and enumeration (Section~\ref{sec:comp-enumeration}).

We introduce some notation to define snapshots.
A test case $t \in T_r$ describes a sequence $\locs{t} = \ell_1, \ell_2, \ldots$ of executed program locations.
For an expression $e$ and a location $\ell \in \locs{t}$, $\valueof{e}{\ell}{t}$ is the value of $e$ at $\ell$ in $t$, if $e$ can be evaluated at $\ell$ (otherwise, $\valueof{e}{\ell}{t}$ is undefined).

\subsubsection{Invariant analysis} \label{sec:comp-inv-analysis}
An \emph{invariant} at a program location $\ell$ with respect to a set
of test cases is a collection of predicates that all hold at
$\ell$ in every run of the tests.\footnote{The class invariants mentioned in Section~\ref{sec:contracts} are a special case.} \autofix uses
Daikon~\cite{ernst:dynamically:1999} to infer invariants that
characterize the \emph{passing} and \emph{failing} runs; their
difference determine some snapshots that
highlight possible failure causes.\footnote{Using Daikon is an implementation choice made to take advantage of its useful collection of invariant templates, which includes Boolean combinations beyond those described in Section~\ref{sec:predicates}.}

For each location $\ell$ reached by \emph{some} tests in $T_r$, we compute the \emph{passing invariant} $\pi_\ell$ as the collection of predicates that hold in all passing tests $P_r \subset T_r$; and the \emph{failing invariant} $\phi_\ell$ as the collection of predicates that hold in all failing tests in $F_r^{f, c} \subseteq T_r$.
\autofix uses only invariants built out of publicly visible predicates in $\pred_{r,c}$.
The predicates in $\Pi = \{ p \mid p \in \phi_\ell \text{ and } \neg p \in \pi_\ell\}$ characterize potential causes of errors, as $\Pi$ contains predicates that hold in failing runs but not in passing runs.\footnote{Since the set of predicates used by \autofix is closed under complement (Section~\ref{sec:predicates}), $\Pi$ is equivalently computed as the negations of the predicates in $\{ p \mid p \in \pi_\ell \text{ and } \neg p \in \phi_\ell\}$.}
Correspondingly, the set $\comp(T_r)$ includes all components
\[
\left\langle \ell, \bigwedge_{p \in \overline{\Pi}} p, \text{\code{True}} \right\rangle,
\]
for every non-empty subset $\overline{\Pi}$ of $\Pi$ that profiles potential error causes.

The rationale for considering differences of sets of predicates is similar to the ideas behind the predicate elimination strategies in ``cooperative bug isolation'' techniques~\cite{Liblit:2004:CBI}.
The dynamic analysis described in Section~\ref{sec:dynamic-analysis} would assign the highest dynamic score to snapshots whose predicates correspond to the deterministic bug predictors in cooperative bug isolation.

\subsubsection{Enumeration} \label{sec:comp-enumeration}
For each test $t \in T_r$, each predicate $p \in \pred_{r, c}$, and each location $\ell \in \locs{t}$ reached in $t$'s execution where the value of $p$ is defined, the set $\comp(T_r)$ of snapshots includes
\[
\langle \ell, p, \valueof{p}{\ell}{t} \rangle\,,
\]
where $p$ is evaluated at $\ell$ in $t$.

In the case of the fault of routine \code{move_item} (discussed in Section~\ref{sec:overview}), the snapshots include, among many others, $\langle \ref{ln:begin}, \text{\code{v = Void}}, \text{\code{False}} \rangle$ (every execution has \code{v /= Void} when it reaches line \ref{ln:begin}) and $\langle \ref{ln:goith}, \text{\code{idx > index}}, \text{\code{True}} \rangle$ (executions failing at line~\ref{ln:goith} have \code{idx > index}).

Only considering snapshots corresponding to actual test executions avoids a blow-up in the size of $\comp(T_r)$.
In our experiments (Section~\ref{sec:experiment}), the number of snapshots enumerated for each fault ranged from about a dozen to few hundreds; those achieving a high suspiciousness score (hence actually used to build fixes, as explained in Section~\ref{sec:combining-analysis}) typically targeted only one or two locations $\ell$ with different predicates $p$.

%----------------------------------------------------------------------------
\subsection{Fault localization}
\label{sec:fault-analysis}
%----------------------------------------------------------------------------
The goal of the fault localization phase is to determine which snapshots in $\comp(T_r)$ are reliable characterizations of the reasons for the fault under analysis.
Fault localization in \afx computes a number of heuristic measures for each snapshot, described in the following subsections; these include simple syntactic measures such as the distance between program statements (Section~\ref{sec:static-analysis}) and metrics based on the runtime behavior of the program in the passing and failing tests (Section~\ref{sec:dynamic-analysis}).

The various measures are combined in a \emph{ranking} of the snapshots (Section~\ref{sec:combining-analysis}) to estimate their ``\emph{suspiciousness}'': each triple $\langle \ell, p, v\rangle$ is assigned a score $\rank{\langle \ell, p, v\rangle}$ which assesses how suspicious the snapshot is.
A high ranking for a snapshot $\langle \ell, p, v\rangle$ indicates that the fault is likely to originate at location $\ell$ when predicate $p$ evaluates to $v$.
The following phases of the fixing algorithm only target snapshots achieving a high score in the ranking.

\subsubsection{Static analysis} \label{sec:static-analysis}
The static analysis performed by \autofix is based on simple measures of proximity and similarity: \emph{control dependence} measures the distance, in
terms of number of instructions, between two program locations;
\emph{expression dependence} measures the syntactic similarity between
two predicates.
Both measures are variants of standard notions used in compiler construction~\cite{Allen:1970:CFA:800028.808479,CC}.
\afx uses control dependence to estimate the proximity
of a location to where a contract violation is triggered; the algorithm then
differentiates further among expressions evaluated at nearby program
locations according to syntactic similarity between each expression
and the violated contract clause.  Static analysis provides
coarse-grained measures that are only useful when combined with the
more accurate dynamic analysis (Section~\ref{sec:dynamic-analysis}) as
described in Section~\ref{sec:combining-analysis}.

%----------------------------------------------------------------------------
\paragraph{Control dependence.} \label{sec:control-dependence}
%----------------------------------------------------------------------------
\afx uses control dependence to rank locations (in snapshots) according to proximity to the location of failure.
For two program locations $\ell_1, \ell_2$, write $\ell_1 \leadsto \ell_2$ if $\ell_1$ and $\ell_2$ belong to the same routine and there exists a directed path from $\ell_1$ to $\ell_2$ on the control-flow graph of the routine's body; otherwise, $\ell_1 \not\leadsto \ell_2$.
The \emph{control distance} $\cdist{\ell_1}{\ell_2}$ of two program locations is the length of the shortest directed path from $\ell_1$ to $\ell_2$ on the control-flow graph if $\ell_1 \leadsto \ell_2$, and $\infty$ if $\ell_1 \not\leadsto \ell_2$.
For example, $\cdist{\ref{ln:remove}}{\ref{ln:putleft}} = 2$ in Figure~\ref{lst:moveItem}.

    Correspondingly,  when $\ell \leadsto \jmath$, the \emph{control dependence} $\cdep{\ell}{\jmath}$ is the normalized score:
    \[
    \cdep{\ell}{\jmath} = 1 - \frac{\cdist{\ell}{\jmath}}{\max \{\cdist{\lambda}{\jmath} \mid \lambda \in r \text{ and } \lambda \leadsto \jmath\}}\,,
    \]
    where $\lambda$ ranges over all locations in routine $r$ (where $\ell$ and $\jmath$ also appear); otherwise, $\ell \not\leadsto \jmath$ and $\cdep{\ell}{\jmath} = 0$.

Ignoring whether a path in the control-flow graph is feasible when computing control-dependence scores does not affect the overall precision of \autofix's heuristics: Section~\ref{sec:combining-analysis} shows how static analysis scores are combined with a score obtained by dynamic analysis; when the latter is zero (the case for unfeasible paths, which no test can exercise), the overall score is also zero regardless of static analysis scores.

%----------------------------------------------------------------------------
\paragraph{Expression dependence.} \label{sec:dependence}
%----------------------------------------------------------------------------
\afx uses expression dependence to rank expressions (in snapshots) according to similarity to the \emph{contract clause} violated in a failure.
Expression dependence is meaningful for expressions evaluated in the same local environment (that is, with strong control dependence), where the same syntax is likely to refer to identical program elements.
Considering only syntactic similarity is sufficient because \autofix will be able to affect the value of any assignable expressions (see Section~\ref{sec:synthesis-of-fix-actions}).
For an expression $e$, define the set $\subexp(e)$ of its sub-ex\-pres\-sions as follows:
\begin{itemize}
\item $e \in \subexp(e)$;
\item if $e' \in \subexp(e)$ is a query call of the form $t.q\,(a_1, \ldots, a_m)$ for $m \geq 0$, then $t \in \subexp(e)$ and $a_i \in \subexp(e)$ for all $1 \leq i \leq m$.
\end{itemize}
This definition also accommodates infix operators (such as Boolean connectives and arithmetic operators), which are just syntactic sugar for query calls; for example $a$ and $b$ are both sub-expressions of $a + b$, a shorthand for \code{a.plus (b)}.
Unqualified query calls are treated as qualified call on the implicit target \mbox{\lstinline|Current|.}

The \emph{expression proximity} $\edist{e_1}{e_2}$ of two expressions $e_1, e_2$ measures how similar $e_1$ and $e_2$ are in terms of shared sub-expressions; namely, 
$
\edist{e_1}{e_2} = \iftwocolumn\linebreak\fi \left| \subexp(e_1) \cap \subexp(e_2) \right|.
$
For example, \iftwocolumn the expression proximity\fi $\edist{\text{\code{i <= count}}}{\text{\code{0 <= i <= count + 1}}}$ is $2$, corresponding to the shared sub-expressions \code{i} and \code{count}.
The larger the expression proximity between two expressions is, the more similar they are.

Correspondingly, the \emph{expression dependence} $\edep{p}{c}$ is the normalized score:
\[
\edep{p}{c}  \ =\ \frac{\edist{p}{c}}{\max \{\edist{\pi}{c} \mid {\pi \in \pred_{r,c}}\}}\,,
\]
measuring the amount of evidence that $p$ and $c$ are syntactically similar.
In routine \code{before} in Figure~\ref{lst:moveItem}, for example, $\edep{\text{\code{index}}}{\text{\code{index = 0}}}$ is $1/3$ because $\edist{\text{\code{index}}}{\text{\code{index = 0}}} = 1$ and \code{index = 0} itself has the maximum expression proximity to \code{index = 0}.

\subsubsection{Dynamic analysis} \label{sec:dynamic-analysis}

Our dynamic analysis borrows techniques from generic fault localization~\cite{wong:family:2010} to determine which locations are likely to host the cause of failure.
Each snapshot receives a \emph{dynamic score} $\dscore{\langle \ell, p, v \rangle}$, roughly measuring how often it appears in failing runs as opposed to passing runs.
A high dynamic score is empirical evidence that the snapshot characterizes the fault and suggests what has to be changed; we use static analysis (Section~\ref{sec:static-analysis}) to differentiate further among snapshots that receive similar dynamic scores.

%----------------------------------------------------------------------------
\paragraph{Principles for computing the dynamic score.}
%----------------------------------------------------------------------------
Consider a failure violating the contract clause $c$ at location $f$ in some routine $r$. %;
For a test case $t \in T_r$ and a snapshot $\langle \ell, p, v \rangle$ such that $\ell$ is a location in $r$'s body, write $\langle \ell, p, v \rangle \in t$ if $t$ reaches location $\ell$ at least once and $p$ evaluates to $v$ there:
\[
\langle \ell, p, v \rangle \in t \quad\text{ iff }\quad \exists \ell_i \in \locs{t}, \ell = \ell_i, \text{ and } v = \valueof{p}{\ell_i}{t} \,.
\]

For every test case $t \in T_r$ such that $\langle \ell, p, v \rangle \in t$, $\sigma(t)$ describes $t$'s contribution to the dynamic score of $\langle \ell, p, v \rangle$: a large $\sigma(t)$ should denote evidence that $\langle \ell, p, v \rangle$ is a likely ``source'' of error if $t$ is a failing test case, and evidence against it if $t$ is passing.
We choose a $\sigma$ that meets the following requirements:
\begin{enumerate}[(a)]
	\item \label{pr1} If there is at least one failing test case $t$ such that $\langle \ell, p, v \rangle \in t$, the overall score assigned to $\langle \ell, p, v \rangle$ must be positive: the evidence provided by failing test cases cannot be canceled out completely.
	\item \label{pr2} The magnitude of each failing (resp.~passing) test case's contribution $\sigma(t)$ to the dynamic score assigned to $\langle \ell, p, v \rangle$ decreases as more failing (resp.~passing) test cases for that snapshot are available: the evidence provided by the first few test cases is crucial, while repeated outcomes carry a lower weight.
	\item \label{pr3} The evidence provided by one failing test case alone is stron\-ger than the evidence provided by one passing test case.
\end{enumerate}
The first two principles correspond to ``Heuristic III'' of Wong et al.~\cite{wong:family:2010}, whose experiments yielded better fault localization accuracy than most alternative approaches.
According to these principles, snapshots appearing only in failing test cases are more likely to be fault causes.

\afx's dynamic analysis assigns scores starting from the same basic principles as Wong et al.'s, but with differences suggested by the ultimate goal of automatic fixing:
our dynamic score ranks snapshots rather than just program locations, and assigns weight to test cases differently.
Contracts help find the location responsible for a fault: in many cases, it is close to where the contract violation occurred;
on the other hand, automatic fixing requires gathering information not only about the location but also about the state ``responsible'' for the fault.
This observation led to the application of fault localization principles on snapshots in \afx.
It is also consistent with recent experimental evidence~\cite{QiMLW13} suggesting that the behavior of existing fault localization techniques on the standard benchmarks used to evaluate them is not always a good predictor of their performance in the context of automated program repair; hence the necessity of adapting to the specific needs of automated fixing.\footnote{The results of Wong et al.'s heuristics in Qi et al.'s experiments~\cite{QiMLW13} are not directly applicable to \autofix (which uses different algorithms and adapts Wong et al.'s heuristics to its specific needs); replication belongs to future work.}

%----------------------------------------------------------------------------
\paragraph{Dynamic score.}
\label{sec:score-from-dynamic}
%----------------------------------------------------------------------------
Assume an arbitrary order on the test cases and let $\sigma(t)$ be $\alpha^i$ for the $i$-th failing test case $t$ and $\beta \alpha^i$ for the $i$-th passing test case.
Selecting $0 < \alpha < 1$ decreases the contribution of each test case exponentially, which meets principle (\ref{pr2}); then, selecting $0 < \beta < 1$ fulfills principle (\ref{pr3}).

The evidence provided by each test case adds up:
\[
\dscore{\langle \ell, p, v \rangle} \!= \gamma + \sum\! \left\{ \sigma(u)\! \mid\! {u \in F_r^{f, c}} \right\}  - \sum \left\{ \sigma(v) \mid {v \in P_r}\right\},
\]
for some $\gamma \geq 0$; the chosen ordering is immaterial.
We compute the score with the closed form of geometric progressions:
\iftwocolumn
    \begin{align*}
    \npass{\langle \ell, p, v \rangle} &\! =\! \left| \left\{ t \in P_r \mid \langle \ell, p, v \rangle \in t \right\} \right|\,, \\
    \nfail{\langle \ell, p, v \rangle} &\! =\! \left| \left\{ t \in F_r^{f, c} \mid \langle \ell, p, v \rangle \in t \right\} \right|\,, \\
    \dscore{\langle \ell, p, v \rangle} &\! =\! \gamma + \frac{\alpha}{1 - \alpha} \left( 1 - \beta + \beta \alpha^{\npass{\langle \ell, p, v \rangle}} - \alpha^{\nfail{\langle \ell, p, v \rangle}} \right),
    \end{align*}
\else
    \begin{align*}
    \npass{\langle \ell, p, v \rangle} &\quad =\quad \left| \left\{ t \in P_r \mid \langle \ell, p, v \rangle \in t \right\} \right|\,, \\
    \nfail{\langle \ell, p, v \rangle} &\quad =\quad \left| \left\{ t \in F_r^{f, c} \mid \langle \ell, p, v \rangle \in t \right\} \right|\,, \\
    \dscore{\langle \ell, p, v \rangle} &\quad =\quad \gamma + \frac{\alpha}{1 - \alpha} \left( 1 - \beta + \beta \alpha^{\npass{\langle \ell, p, v \rangle}} - \alpha^{\nfail{\langle \ell, p, v \rangle}} \right),
    \end{align*}
\fi
where $\npass{\langle \ell, p, v \rangle}$ and $\nfail{\langle \ell, p, v \rangle}$ are the number of passing and failing test cases that determine the snapshot $\langle \ell, p, v \rangle$.
It is straightforward to prove that $\dscore{\langle \ell, p, v \rangle}$ is positive if $\nfail{\langle \ell, p, v \rangle} \geq 1$, for every nonnegative $\alpha, \beta, \gamma$ such that $0 < \alpha + \beta < 1$; hence the score meets principle (\ref{pr1}) as well.

Since the dynamic score $\dscore{}$ varies exponentially only with the number of passing and failing test cases, the overall success rate of the \autofix algorithm is affected mainly by the number of tests but not significantly by variations in the values of $\alpha$ and $\beta$.
A small empirical trial involving a sample of the faults used in the evaluation of Section~\ref{sec:experiment} confirmed this expectation of robustness; it also suggested selecting the values $\alpha = 1/3$, $\beta = 2/3$, and $\gamma = 1$ as defaults in the current implementation of \afx, which tend to produce slightly shorter running times on average (up to 10\% improvement).
With these values, one can check that $2/3 < \dscore{\langle \ell, p, v \rangle} < 3/2$, and $1 < \dscore{\langle \ell, p, v \rangle} < 3/2$ if at least one failing test exercises the snapshot.

\subsubsection{Overall score} \label{sec:combining-analysis}
\afx combines the various metrics into an overall score $\rank{\langle \ell, p, v\rangle}$.
The score puts together static and dynamic metrics with the idea that the latter give the primary source of evidence, whereas the less precise evidence provided by static analysis is useful to discriminate among snapshots with similar dynamic behavior.

Since the static measures are normalized ratios, and the dynamic score is also fractional, we may combine them by harmonic mean~\cite{chou:statistical:1975}:
\[
\rank{\langle \ell, p, v \rangle}  = \frac{3}{\edep{p}{c}^{-1} \!+\! \cdep{\ell}{f}^{-1} \!+\! \dscore{\langle \ell, p, v \rangle}^{-1}} \,.
\]
Our current choice of parameters for the dynamic score (Section~\ref{sec:score-from-dynamic}) makes it  dominant in determining the overall score $\rank{\langle \ell, p, v \rangle}$:
while expression and control dependence vary between $0$ and $1$, the dynamic score has minimum $1$ (for at least one failing test case and indefinitely many passing).
This range difference is consistent with the principle that dynamic analysis is the principal source of evidence.

For the fault of Figure~\ref{lst:moveItem}, the snapshot $\langle \ref{ln:goith}, \text{\code{idx > index}}, \text{\code{True}} \rangle$ receives a high overall score.
\afx targets snapshots such as this in the fix action phase.

%------------------------------------------------------------------------------
\subsection{Fix action synthesis}
\label{sec:synthesis-of-fix-actions}
%------------------------------------------------------------------------------
A snapshot $\langle \ell, p, v\rangle$ in $\comp(T_r)$ with a high score $\rank{\langle \ell, p, v\rangle}$ suggests that the ``cause'' of the fault under analysis is that expression $p$ takes value $v$ when the execution reaches $\ell$.
Correspondingly, \afx tries to build fixing \emph{actions} (snippets of instructions) that \emph{modify} the value of $p$ at $\ell$, so that the execution can hopefully continue without triggering the fault.
This view reduces fixing to a program synthesis problem: find an action \code{snip} that satisfies the specification:
\iftwocolumn
\begin{lstlisting}[numbers=none]
       require p = v do snip ensure p /= v end .
\end{lstlisting}
\else
\begin{lstlisting}[numbers=none]
                 require p = v do snip ensure p /= v end .
\end{lstlisting}
\fi

\afx uses two basic strategies for generating fixing actions: setting and replacement.
Setting (described in Section~\ref{sec:setting}) consists of modifying the value of variables or objects through assignments or routine calls.
Replacement (described in Section~\ref{sec:replacement}) consists of modifying the value of expressions directly where they are used in the program.
Three simple heuristics, with increasing specificity, help prevent the combinatorial explosion in the generation of fixing actions:
\begin{enumerate}
\item Since the majority of program fixes are short and simple~\cite{dallmeier:extraction:2007,Martinez2013}, we only generate fixing actions that consist of simple instructions;
\item We select the instructions in the actions according to context (the location that we are fixing) and common patterns, and based on behavioral models of the classes (Section~\ref{sec:behavioral});
\item For integer expressions, we also deploy constraint solving techniques to build suitable derived expressions (Section~\ref{sec:lin-constraints}).
\end{enumerate}

We now describe actions by setting and replacements, which are the basic mechanisms \afx uses to synthesize actions, as well as the usage of behavioral models and constraint solving.
To limit the number of candidates, \autofix uses no more than one basic action in each candidate fix.

\subsubsection{Actions by setting} \label{sec:setting}
One way to change the value of a predicate is to modify the value of its constituent expressions by assigning new values to them or by calling modifier routines on them.
For example, calling routine \code{forth} on the current object has the indirect effect of setting predicate \code{before} to \code{False}.

Not all expressions are directly modifiable by setting; an expression $e$ is \emph{modifiable} at a location $\ell$ if: $e$ is of reference type (hence we can use $e$ as target of routine calls); or $e$ is of integer type and the assignment \mbox{\code{e := 0}} can be executed at $\ell$; or $e$ is of Boolean type and the assignment \code{e := True} can be executed at~$\ell$.
For example, \code{index} is modifiable everywhere in routine \code{move_item} because it is an attribute of the enclosing class;
the argument \code{i} of routine \code{go_i_th}, instead, is not modifiable within its scope because arguments are read-only in Eiffel.

Since the Boolean predicates of snapshots may not be directly modifiable, we also consider sub-expressions of any type.
The definition of sub-expression (introduced in Section~\ref{sec:dependence}) induces a partial order $\preceq$: $e_1 \preceq e_2$ iff $e_1 \in \subexp(e_2)$ that is $e_1$ is a sub-expression of $e_2$; correspondingly, we define the \emph{largest} expressions in a set as those that are only sub-expressions of themselves.
For example, the largest expressions of integer type in $\subexp(\text{\code{idx < index or after}})$ are \code{idx} and \code{index}.

A snapshot $\langle \ell, p, v \rangle$ induces a set of target expressions that are modifiable in the context given by the snapshot.
For each type (Boolean, integer, and reference), the set $\targs{\langle \ell, p \rangle}$ of \emph{target expressions} includes the largest expressions of that type among $p$'s sub-expressions $\subexp(p)$ that are modifiable at $\ell$.
For example, $\targs{\langle \ref{ln:goith}, \text{\code{idx > Current.index}}\rangle}$ in Figure~\ref{lst:moveItem} includes the reference expression \code{Current}, the integer expressions \mbox{\code{Current.index}}  and \code{idx}, but no Boolean expressions (\code{idx > Current.index} is not modifiable because it is not a valid L-value of an assignment).

Finally, the algorithm constructs the set $\modif{\langle \ell, p \rangle}$ of \emph{settings} induced by a snapshot $\langle \ell, p, v \rangle$ according to the target types as follows; these include elementary assignments, as well as the available routine calls.

\paragraph{Boolean targets.}
For $e \in \targs{\langle \ell, p \rangle}$ of Boolean type, $\modif{\langle \ell, p \rangle}$ includes the assignments $e := d$ for $d$ equal to the constants \code{True} and \code{False} and to the complement expression \code{not e}.

\paragraph{Integer targets.}
For $e \in \targs{\langle \ell, p \rangle}$ of integer type, $\modif{\langle \ell, p \rangle}$ includes the assignments $e := d$ for $d$ equal to the constants $0$, $1$, and $-1$, the ``shifted'' expressions $e+1$ and $e-1$, and the expressions deriving from integer constraint solving (discussed in Section~\ref{sec:lin-constraints}).

\paragraph{Reference targets.}
For $e \in \targs{\langle \ell, p \rangle}$ of reference type, if $e.c\,(a_1, \ldots, a_n)$  is a call to a command (procedure) $c$ executable at $\ell$, include $e.c\,(a_1, \ldots, a_n)$ in $\modif{\langle \ell, p \rangle}$.
(Section~\ref{sec:behavioral} discusses how behavioral models help select executable calls at $\ell$ with chances of affecting the program state indicated by the snapshot.)

In the example of Section~\ref{sec:overview}, the fault's snapshot $\langle \ref{ln:goith}, \text{\code{idx > index}}, \iftwocolumn\else\linebreak\fi \text{\code{True}} \rangle$ determines the settings $\modif{\langle \ref{ln:goith}, \text{\code{idx > index}} \rangle}$  that include assignments of $0$, $1$, and~$-1$ to \code{idx} and \code{index}, and unit increments and decrements of the same variables.

\subsubsection{Actions by replacement} \label{sec:replacement}
In some cases, assigning new values to an expression is undesirable or infeasible.
For example, expression~$i$ in routine \code{go_i_th} of Figure~\ref{lst:moveItem} does not have any modifiable sub-expression.
In such situations, \emph{replacement} directly substitutes the usage of expressions in existing instructions.
Replacing the argument \code{idx} with \code{idx - 1} on line~\ref{ln:goith} modifies the effect of the call to \code{go_i_th} without directly changing any local or global variables.

Every location $\ell$ labels either a primitive instruction (an assignment or a routine call) or a Boolean condition (the branching condition of an \code{if} instruction or the exit condition of a \code{loop}).
Correspondingly, we define the set $\subexp(\ell)$ of sub-expressions of a \emph{location} $\ell$ as follows:
\begin{itemize}
\item if $\ell$ labels a Boolean condition $b$ then $\subexp(\ell) = \subexp(b)$;
\item if $\ell$ labels an assignment $v := e$ then $\subexp(\ell) = \subexp(e)$;
\item if $\ell$ labels a routine call $t.c\, (a_1, \ldots, a_n)$ then
 \[\subexp(\ell) = \bigcup \{\subexp(a_i) \mid 1 \leq i \leq n\,\}\,.\]
\end{itemize}

Then, a snapshot $\langle \ell, p, v \rangle$ determines a set $\replac{\langle\ell, p\rangle}$ of \emph{replacements}: instructions obtained by replacing one of the sub-expressions of the instruction at $\ell$ according to the same simple heuristics used for setting.
More precisely, we consider expressions $e$ among the largest ones of Boolean or integer type in $\subexp(p)$ and we modify their occurrences in the instruction at $\ell$.
Notice that if $\ell$ labels a conditional or loop, we replace $e$ only in the Boolean condition, not in the body of the compound instruction.

\paragraph{Boolean expressions.}
For $e$ of Boolean type, $\replac{\langle\ell, p\rangle}$ includes the instructions obtained by replacing each occurrence of $e$ in $\ell$ by the constants \code{True} and \code{False} and by the complement expression \code{not e}.

\paragraph{Integer expressions.}
For $e$ of integer type, $\replac{\langle\ell, p\rangle}$ includes the instructions obtained by replacing each occurrence of $e$ in $\ell$ by the constants $0$, $1$, and $-1$, by the ``shifted'' expressions $e+1$ and $e-1$, and by the expressions deriving from integer constraint solving (Section~\ref{sec:lin-constraints}).

Continuing the example of the fault of Section~\ref{sec:overview}, the snapshot $\langle \ref{ln:goith}, \iftwocolumn\else\linebreak\fi \text{\code{idx > index}}, \text{\code{True}} \rangle$ induces the replacement set $\replac{\langle \ref{ln:goith}, \text{\code{idx > index}}\rangle}$ including \code{go_i_th (idx - 1)}, \code{go_i_th (idx + 1)}, as well as \code{go_i_th (0)}, \code{go_i_th (1)}, and \code{go_i_th (-1)}.

\subsubsection{Behavioral models} \label{sec:behavioral}
Some of the fixing actions generated by \afx try to modify the program state by calling routines on the current or other objects.
This generation is not blind but targets operations applicable to the target objects that can modify the value of the predicate $p$ in the current snapshot $\langle \ell, p, v \rangle$.
To this end, we exploit the \emph{finite-state behavioral model} abstraction to quickly find out the most promising operations or operation sequences.

\begin{figure}[!htb]
\centering
\lstset{basicstyle=\footnotesize}
\begin{tikzpicture}[->,>=stealth',shorten >=1pt,node distance=7mm and 1.5cm,semithick,align=center]
 \tikzstyle{state}=[ellipse,draw,minimum width=85pt,inner sep=0pt]
\node[state] (from1) {\code{is_empty}\\\code{before}\\\code{not after}};
\node[state,below=of from1] (from2) {\code{not is_empty}\\\code{before}\\\code{not after}};
\node[state,right=of from1] (to1) {\code{is_empty}\\\code{not before}\\\code{not after}};
\node[state,below=of to1] (to2) {\code{not is_empty}\\\code{not before}\\\code{not after}};
\path (from1) edge node [above] {\code{forth}} (to1);
\path (from2) edge node [above] {\code{forth}} (to2);
\end{tikzpicture}
\caption{Behavioral model of routine \code{forth}.}
\label{fig:transition}
\end{figure}
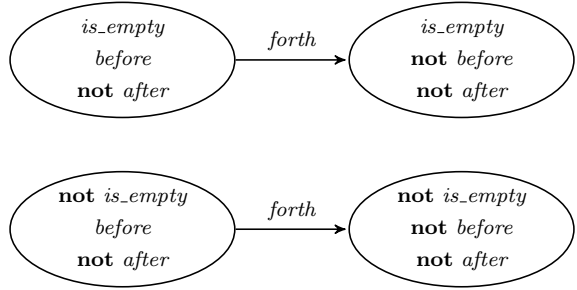

Using techniques we previously developed for Pachika~\cite{dallmeier:generating:2009}, \afx extracts a simple behavioral model from \emph{all} passing runs of the class under consideration.
The behavioral model represents a \emph{predicate abstraction} of the class behavior.
It is a finite-state automaton whose states are labeled with predicates that hold in that state, and transitions are labeled with routine names, connecting observed pre-state to observed post-states.

As an example, \figref{fig:transition} shows a partial behavioral model for the \code{forth} routine in Figure~\ref{lst:moveItem}.
This behavioral model shows, among other things, that \code{not before} always holds after calls to \code{forth} in any valid initial state.
By combining this information with the snapshot $\langle \ref{ln:putleft}, \text{\code{before}}, \text{\code{True}}\rangle$, we can surmise that invoking \code{forth} on line \ref{ln:putleft} mutates the current object state so that it avoids the possible failure cause \code{before = True}.

In general, the built behavioral abstraction is neither complete nor sound because it is based on a finite number of test runs. Nonetheless, it is often sufficiently precise
to reduce the generation of routine calls to those that are likely to affect the snapshot state in the few cases where enumerating all actions by setting (Section~\ref{sec:setting}) is impractical.

\subsubsection{Constraint solving} \label{sec:lin-constraints}
In contract-based development, numerous assertions take the form of Boolean combinations of linear inequalities over program variables and constants.
The precondition of \code{go_i_th} on line \ref{ln:goith-pre} in Figure~\ref{lst:moveItem} is an example of such \emph{linearly constrained assertions} (or \emph{linear assertions} for short).
Such precondition requires that the argument \code{i} denote a valid position inside the set.

When dealing with integer expressions extracted from linear assertions, we deploy specific techniques to generate fixing actions in addition to the basic heuristics discussed in the previous sections (such as trying out the ``special'' values $0$ and $1$).
The basic idea is to \emph{solve} linear assertions for extremal values compatible with the constraint.
Given a snapshot $\langle \ell,\lambda, v\rangle$ such that $\lambda$ is a linear assertion, and an integer expression $j$ appearing in $\lambda$, \afx uses Mathematica to solve $\lambda$  for maximal and minimal values of $j$ as a function of the other parameters (numeric or symbolic) in $\lambda$.
To increase the quality of the solution, we strengthen $\lambda$ with linear assertions from the class invariants that share identifiers with $\lambda$.
In the example of \code{go_i_th}, the class invariant \code{count >= 0} would be added to $\lambda$ when looking for extrema.
The solution consists, in this case, of the extremal values~\code{0} and \code{count + 1}, which are both used as replacements (Section~\ref{sec:replacement}) of variable \code{i}.

%----------------------------------------------------------------------------
\subsection{Candidate fix generation}
\label{sec:generating-candidate-fixes}
%----------------------------------------------------------------------------
Given a ``suspicious'' snapshot $\langle \ell, p, v \rangle$ in $\comp(T_r)$, the previous section showed how to generate fix actions that can mutate the value of $p$ at location $\ell$.
Injecting any such fix actions at location $\ell$ gives a modified program that is a \emph{candidate fix}: a program where the faulty behavior may have been corrected.
We inject fix actions in program in two phases.
First, we select a \emph{fix schema}---a template that abstracts common instruction patterns (Section~\ref{sec:schema}).
Then, we \emph{instantiate} the fix schema with the snapshot's predicate $p$ and some fixing action it induces (Section~\ref{sec:instantiation}).

Whereas the space of all possible fixes generated with this approach is potentially huge, \autofix only generates candidate fixes for the few most suspicious snapshots (15~most suspicious ones, in the current implementation).
In our experiments, each snapshot determines at most 50~candidate fixes (on average, no more than 30), which can be validated in reasonable time (see Section~\ref{sec:cost}).

\subsubsection{Fix schemas}
\label{sec:schema}
\afx uses a set of predefined templates called \emph{fix schemas}.
The four fix schemas currently supported are shown in Figure~\ref{lst:fixSkeleton};\footnote{Recent work~\cite{Martinez2013} has demonstrated that these simple schemas account for a large fraction of the manually-written fixes found in open-source projects.} they consist of conditional wrappers that apply the fix actions only in certain conditions (with the exception of schema~\emph{a} which is unconditional).
In the schemas, \code{fail} is a placeholder for a predicate, \code{snippet} is a fixing action, and \code{old_stmt} are the statements in the original program where the fix is injected.

\lstset{numbers=none}
\begin{figure}[!h]
\begin{tabular}{p{.45\columnwidth} p{.45\columnwidth}}
\begin{lstlisting}
   (a) snippet
       old_stmt


   (c) if not fail then
           old_stmt
       end
\end{lstlisting}
&
\begin{lstlisting}
   (b) if fail then
           snippet
       end
       old_stmt

   (d) if fail then
           snippet
       else
           old_stmt
       end
\end{lstlisting}
\end{tabular}
\caption{Fix schemas implemented in \afx.}
\label{lst:fixSkeleton}
\end{figure}

%----------------------------------------------------------------------------
\subsubsection{Schema instantiation}
\label{sec:instantiation}
%----------------------------------------------------------------------------
For a state snapshot $\langle \ell, p, v \rangle$, we instantiate the schemas in Figure~\ref{lst:fixSkeleton}  as follows:
\begin{list}{}{}
\item[\textit{fail}] becomes $p = v$, the snapshot's predicate and value.

\item[\textit{snippet}] becomes any fix action by setting ($\modif{\langle \ell, p \rangle}$ in Section~\ref{sec:setting}) or by replacement ($\replac{\langle\ell, p\rangle}$ in Section~\ref{sec:replacement}).

\item[\textit{old\_stmt}] is the instruction at location $\ell$ in the original program.
\end{list}
The instantiated schema \emph{replaces} the instruction at position $\ell$ in the program being fixed; the modified program is a \emph{candidate fix}.

For example, consider again the snapshot $\langle \ref{ln:goith}, \text{\code{idx > index}}, \text{\code{True}} \rangle$, which receives a high ``suspiciousness'' score for the fault described in Section~\ref{sec:overview} and which induces, among others, the fix action consisting of decrementing \code{idx}.
The corresponding instantiation of fix schema (b) in Figure~\ref{lst:fixSkeleton} is then: \code{fail} becomes \code{idx > index = True}, \code{snippet} becomes \code{idx := idx - 1}, and \code{old_stmt} is the instruction \code{go_i_th (idx)} on line~\ref{ln:goith-pre} in Figure~\ref{lst:moveItem}.
Injecting the instantiated schema (replacing line~\ref{ln:goith-pre}) yields the candidate fix in Figure~\ref{lst:moveItemCorrections-one}, already discussed in Section~\ref{sec:overview}.

%----------------------------------------------------------------------------
\subsection{Fix validation}
\label{sec:validating-fixes}
%----------------------------------------------------------------------------

The generation of candidate fixes, described in the previous Sections~\ref{sec:synthesis-of-fix-actions} and \ref{sec:generating-candidate-fixes}, involves several heuristics and is ``best effort'': there is no guarantee that the candidates actually correct the error (or even that they are executable programs).
Each candidate fix must pass a validation phase which determines whether its deployment removes the erroneous behavior under consideration.
The validation phase regressively runs each candidate fix through the full set $T_r$ of passing and failing test cases for the routine $r$ being fixed.
A fix is \emph{validated} (or \emph{valid}) if it passes all the previously failing test cases $F_r^{f, c}$ and it still passes the original passing test cases $P_r$.
\afx only reports valid fixes to users, ranked as described in Section~\ref{sec:ranking}.

The correctness of a program is defined relative to its specification;
in the case of automated program fixing, this implies that the validated fixes are only as good as the available tests or, if these are generated automatically, as the available contracts.
In other words, evidently incomplete or incorrect contracts may let inappropriate candidate fixes pass the validation phase.

To distinguish between fixes that merely pass the validation phase because they do not violate any of the available contracts and high-quality fixes that developers would confidently deploy, we introduce the notion of \emph{proper} fix.
Intuitively, a proper fix is one that removes a fault without introducing other faulty or unexpected behavior.
More rigorously, assume we have the complete behavioral specification $\perfspec_r$ of a routine $r$; following our related work~\cite{PFM10-VSTTE10,PFPWM-ICSE13}, $\perfspec_r$ is a pre-/postcondition pair that characterizes the effects of executing $r$ on every query (attribute or function) of its enclosing class.
A valid fix is \emph{proper} if it satisfies $\perfspec_r$; conversely, it is \emph{improper} if it is valid but not proper.

While we have demonstrated~\cite{PFPWM-ICSE13} that it is possible to formalize complete behavioral specifications in many interesting cases (in particular, for a large part of the EiffelBase library used in the experiments of Section~\ref{sec:experiment}), the line between proper and improper may be fuzzy under some circumstances when the notion of ``reasonable'' behavior is disputable or context-dependent.
Conversely, there are cases---such as when building a proper fix is very complex or exceedingly expensive---where a valid but improper fix is still better than no fix at all because it removes a concrete failure and lets the program continue its execution.

In spite of these difficulties of principle, the experiments in Section~\ref{sec:experiment} show that the simple contracts normally available in Eiffel programs are often good enough in many practical cases to enable \afx to suggest fixes that we can confidently classify as \emph{proper}, as they meet the expectations of real programmers familiar with the code base under analysis.

%----------------------------------------------------------------------------
\subsection{Fix ranking}
\label{sec:ranking}
%----------------------------------------------------------------------------
The \afx algorithm often finds \emph{several} valid fixes for a given fault.
While it is ultimately the programmer's responsibility to select which one to deploy, flooding them with many fixes defeats the purpose of automated debugging, because understanding what the various fixes actually do and deciding which one is the most appropriate is tantamount to the effort of designing a fix in the first place.

To facilitate the selection, \afx ranks the valid fixes according to the ``suspiciousness'' score $\rank{\langle \ell, p, v \rangle}$ of the snapshot $\langle \ell, p, v \rangle$ that determined each fix.\footnote{Since all fixing actions are comparatively simple, they do not affect the ranking of valid fixes, which is only based on suspiciousness of snapshots.}
Since multiple fixing actions may determine valid fixes for the same snapshot, ties in the ranking are possible.
The experiments in Section~\ref{sec:experiment} demonstrate that high-quality proper fixes often rank in the top 10 positions among the valid ones; hence \afx users only have to inspect the top fixes to decide with good confidence if any of them is deployable.

%=====================================
\section{Experimental evaluation}
\label{sec:experiment}
%=====================================

We performed an extensive experimental evaluation of the behavior and performance of \afx by applying it to over 200 faults found in various Eiffel programs.
The experiments characterize the reproducible \emph{average} behavior of \afx in a variety of conditions that are indicative of \emph{general} usage.
To ensure generalizable results, the evaluation follows stringent rules: the experimental protocol follows recommended guidelines~\cite{arcuri:practical:2011} to achieve \emph{statistically significant} results in the parts that involve randomization; the faults submitted to \afx come from four code bases of \emph{different quality and maturity}; the experiments characterize usage with \emph{limited computational resources}.

Two additional features distinguish this experimental evaluation from those of most related work (see Section~\ref{sec:related}).
First, the experiments try to capture the usage of \afx as a fully automatic tool where user interaction is limited to selecting a project, pushing a button, and waiting for the results.
The second feature of the evaluation is that it includes a detailed inspection of the quality of the automatically generated fixes, based on the distinction between \emph{valid} and \emph{proper} fixes introduced in Section~\ref{sec:validating-fixes}.

\subsection{Experimental questions and summary of findings}
Based on the high-level goals just presented, the experimental evaluation addresses the following questions:
\begin{description}
\item[\namelabel{rq:how-many}{Q1}\nref{rq:how-many}]
  \emph{How many} faults can \afx correct, and what are their characteristics?

\item[\namelabel{rq:quality}{Q2}\nref{rq:quality}]
  What is the \emph{quality} of the fixes produced by \afx?

\item[\namelabel{rq:cost}{Q3}\nref{rq:cost}]
  What is the \emph{cost} of fixing faults with \afx?

\item[\namelabel{rq:robustness}{Q4}\nref{rq:robustness}]
  How \emph{robust} is \afx's performance in an ``average'' run?
\end{description}

The \textbf{main findings} of the evaluation are as follows:
\begin{itemize}
\item \afx produced valid fixes for 86 (or 42\%) out of 204 randomly detected faults in various programs.

\item Of the $86$ valid fixes produced by \afx, $51$ (or $59\%$) are proper, that is of quality comparable to those produced by professional programmers.

\item \afx achieves its results with limited computational resources: \autofix ran no more than $15$ minutes per fault in $93.1\%$ of the experiments; its median running time in all our experiments was $3$ minutes, with a standard deviation of $6.3$ minutes.

\item \afx's behavior is, to a large extent, robust with respect to variations in the test cases produced by \autotest: 48 (or 56\%) of the faults that \autofix managed to fix at least once were fixed (with possibly different fixes) in over 95\% of the sessions.
If we ignore the empty sessions where \autotest did not manage to reproduce a fault, \autofix produced a valid fix 41\% of all non-empty sessions---when \autofix is successful, it is \emph{robustly} so.
\end{itemize}

%=========================================================================
\subsection{Experimental setup}
\label{sec:experimental-setup}
%=========================================================================

All the experiments ran on the computing facilities of the Swiss National Supercomputing Centre consisting of Transtec Lynx CALLEO High-Performance Servers 2840 with 12 physical cores and 48 GB of RAM.
Each experiment session used exclusively one physical core at 1.6 GHz and 4 GB of RAM, whose computing power is similar to that of a commodity personal computer.
Therefore, the experiments reflect the performance of \autofix in a standard programming environment.

We now describe the code bases and the faults targeted by the experiments (Section~\ref{sec:subjects}), then present the experimental protocol (Section~\ref{sec:experimental-protocol}).

%=========================================================================
\subsubsection{Experimental subjects}
\label{sec:subjects}
%=========================================================================

The experiments targeted a total of $204$ contract-violation faults
collected from four code bases of different quality and maturity. The following discussion analyzes whether such a setup provides a sufficiently varied collection of
subjects that exercise \autofix in different conditions.

\paragraph{Code bases.}  The experiments targeted four code bases:

\begin{itemize}

\item \efbase is a data structure library. It consists of the standard data structure classes from the EiffelBase and Gobo projects, distributed with the EiffelStudio IDE and developed by professional programmers over many years.
    
\item \txtlib is a library to manipulate text documents, developed at ETH Zurich by second-year bachelor's students with some programming experience.

\item \cardgm is an on-line card gaming system, developed as project for \textsc{dose}, a distributed software engineering course organized by ETH~\cite{NordioDOSE2011} for master's students. Since this project is a collaborative effort involving groups in different countries, the students who developed \cardgm had heterogeneous, but generally limited, skills and experience with Eiffel programming and using contracts; their development process had to face the challenges of team distribution.

\item \elearn is an application supporting e-learning, developed in another edition of \textsc{dose}.
\end{itemize}

 \begin{table}[!htbp]
   \caption{Size and other metrics of the code bases (the dot is the decimal mark; the comma is the thousands separator).}
   \label{tab:subject-programs-size}%
   \begin{adjustwidth}{-2.2cm}{-2.2cm}
   \centering
   \tablefont
     \begin{tabular}{lr r@{.}l rrrrr}
     \toprule
     \multicolumn{1}{c}{\textbf{Code base}} & \multicolumn{1}{c}{\textbf{\#C}} & \multicolumn{2}{c}{\textbf{\#kLOC}} & \multicolumn{1}{c}{\textbf{\#R}} & \multicolumn{1}{c}{\textbf{\#Q}}  & \multicolumn{1}{c}{\textbf{\#Pre}} & \multicolumn{1}{c}{\textbf{\#Post}} & \multicolumn{1}{c}{\textbf{\#Inv}} \\
     \midrule
     \efbase & 11 & 26&548 & 1,504 & 169 & 1,147 & 1,270 & 209 \\
     \txtlib & 10 & 12&126 & 780 & 48 & 97 & 134 & 11  \\
     \cardgm & 32 & 20&553 & 1,479 & 81 & 157 & 586 & 58  \\
     \elearn & 27 & 13&693 & 1,055 & 20 & 144 &  148 & 38  \\
     \toprule
     \textbf{Total} & \textbf{80} & \textbf{72}&\textbf{920} & \textbf{4,818} & \textbf{318} & \textbf{1,545} & \textbf{2,138} & \textbf{316} \\
     \bottomrule
     \end{tabular}%
   \end{adjustwidth}
 \end{table}%

Table~\ref{tab:subject-programs-size} gives an idea of the complexity of the programs selected for the experiments, in terms of number of classes (\#C), thousands of lines of code (\#kLOC), number of routines (\#R), Boolean queries (\#Q), and number of contract clauses in preconditions (\#Pre), postconditions (\#Post), and class invariants (\#Inv).

The data suggests that \efbase classes are significantly more complex than the classes in other code bases, but they also offer a better interface with more Boolean queries that can be used by \autofix (Section~\ref{sec:predicates}).
The availability of \emph{contracts} also varies significantly in the code bases, ranging from 0.76 precondition clauses per routine in \efbase down to only 0.11 precondition clauses per routine in \cardgm.
This diversity in the quality of interfaces and contracts ensures that the experiments are representative of \afx's behavior in different conditions; in particular, they demonstrate the performance also with software of low quality and with very few contracts, where fault localization can be imprecise and unacceptable behavior may be incorrectly classified as passing for lack of precise oracles (thus making it more difficult to satisfactorily fix the bugs that are exposed by other contracts).

\paragraph{Faults targeted by the experiments.}
To select a collection of faults for our fixing experiments, we performed a preliminarily run of \autotest~\cite{meyer:programs:2009} on the code bases and recorded information about all faults found that consisted of contract violations.
These include violations of preconditions, postconditions, class invariants, and intermediate assertions (\code{check} instructions), but also violations of \emph{implicit} contracts, such as dereferencing a void  pointer and accessing an array element using an index that is out of bounds, and application-level memory and I/O errors such as a program terminating without closing an open file and buffer overruns.
In contrast, we ignored lower-level errors such as disk failures or out-of-memory allocations, since these are only handled by the language runtime. %, and hence there are little chances of fixing them by changing the source code.
Running \autotest for two hours on each class in the code bases provided a total of $204$ \emph{unique} contract-violation faults (identified as discussed in Section~\ref{sec:test-cases-as-input}).
Table~\ref{tab:subject-faults} counts these unique faults for each code base (\#Faults), and also shows the breakdown into void-dereferencing faults (\#Void), precondition violations (\#Pre), postcondition violations (\#Post), class invariant violations (\#Inv), and check violations (\#Check), as well as the number of faults per kLOC ($\frac{\mathrm{\#F}}{\mathrm{kLOC}}$).
The figures in the last column give a rough estimate of the quality of the code bases, confirming the expectation that software developed by professional programmers adheres to higher quality standards.

 \begin{table}[!htbp]
   \tablefont
   \caption{Faults used in the fixing experiments.}
   \label{tab:subject-faults}%
   \centering
\iftwocolumn
	\setlength{\tabcolsep}{3pt}
     \begin{tabular}{lrrrrrr R{0.4cm}@{.}L{0.3cm}}
%     \begin{tabular}{L{1.4cm}@{}R{1.1cm}@{}R{1cm}@{}R{0.8cm}@{}R{1cm}@{}R{0.8cm}@{}R{1.2cm}@{}R{0.6cm}@{.}L{0.3cm}}
     \toprule
     \textbf{Code base} & \textbf{\#Faults} & \textbf{\#Void} & \textbf{\#Pre} & \textbf{\#Post} & \textbf{\#Inv} & \textbf{\#Check} & \multicolumn{2}{c}{$\frac{\textbf{\#F}}{\textbf{kLOC}}$} \\
     \midrule
\else
     \begin{tabular}{lrrrrrr r@{.}l}
     \toprule
     \multicolumn{1}{c}{\textbf{Code base}} & \multicolumn{1}{c}{\textbf{\#Faults}} & \multicolumn{1}{c}{\textbf{\#Void}} & \multicolumn{1}{c}{\textbf{\#Pre}} & \multicolumn{1}{c}{\textbf{\#Post}} & \multicolumn{1}{c}{\textbf{\#Inv}} & \multicolumn{1}{c}{\textbf{\#Check}} & \multicolumn{2}{c}{$\frac{\textbf{\#F}}{\textbf{kLOC}}$} \\
     \midrule
\fi
     \efbase & 60 & 0 & 23 & 32 & 0 & 5 &  2&3\\
     \txtlib & 31 & 12 & 14 & 1 & 0 & 4 & 2&6 \\
     \cardgm & 63 & 24 & 21 & 8 & 10 & 0 & 3&1 \\
     \elearn & 50 & 16 & 23 & 8 & 3 & 0 & 3&7 \\
     \toprule
     \textbf{Total} & \textbf{204} & \textbf{52} & \textbf{81} & \textbf{49} & \textbf{13} & \textbf{9} & \textbf{2}&\textbf{8} \\
     \bottomrule
     \end{tabular}%
 \end{table}%

The use of \autotest for selecting faults has two principal consequences for this study:
\begin{itemize}
\item On the negative side, using \autotest reduces the types of programs we can include in the experiments, as the random testing algorithm implemented in \autotest has limited effectiveness with functionalities related to graphical user interfaces, networking, or persistence.

\item On the positive side, using \autotest guards against bias in the selection of faults in the testable classes, and makes the experiments representative of the primary intended usage of \autofix: a completely automatic tool that can handle the whole debugging process autonomously.
\end{itemize}
To ensure that the faults found by \autotest are ``real'', we asked, in related work~\cite{PFPWM-ICSE13}, some of the maintainers of \efbase to inspect 10 faults, randomly selected among the 60 faults in \efbase used in our experiments; their analysis confirmed all of them as real bugs requiring to be fixed.
Since Eiffel developers write both programs and their contracts, it is generally safe to assume that a contract violation exposes a genuine fault, since a discrepancy between implementation and specification must be reconciled somehow; this assumption was confirmed in all our previous work with \autotest.

%===============================================================
\subsubsection{Experimental protocol}
\label{sec:experimental-protocol}
%===============================================================
The ultimate goal of the experiments is to determine the typical behavior of \autofix in general usage conditions under constrained computational resources and a completely automatic process.
Correspondingly, the experimental protocol involves a large number of repetitions, to ensure that the average results are statistically significant representatives of a typical run, and combines \autotest and \autofix sessions, to minimize the dependency of the quality of fixes produced by \autofix on the choice of test cases, and to avoid requiring users to provide test cases.

For each unique fault \ff identified as in Section~\ref{sec:subjects}, we ran 30 \autotest sessions of 60 minutes each, with the faulty routine as primary target.
Each session produces a sequence of test cases generated at different times.
Given a fault \ff in a routine $r$, we call \emph{$m$-minute series on \ff} any prefix of a testing sequence generated by \autotest on $r$.
A series may include both passing and failing test cases.
In our analysis we considered series of $m =$ 5, 10, 15, 20, 30, 40, 50, and 60 minutes.
The process determined 30 $m$-minute series (one per session) for every $m$ and for every fault \ff; each such series consists of a set $T = P \cup F$ of passing $P$ and failing $F$ test cases.

Since the \autofix algorithms are deterministic, an $m$-minute series on some fault \ff uniquely determines an \autofix session using the tests in $T$ to fix the fault~\ff.
The remainder of the discussion talks of \emph{$m$-minute fixing session on \ff} to denote the unique \autofix session run using some given $m$-minute series on \ff.
In all, we recorded the fixes produced by 270 ($= 9 \times 30$) fixing sessions of various lengths on each fault; in each session, we analyzed at most 10 fixes---those ranked in the top 10 positions---and discarded the others (if any).

%==========================================================
\subsection{Experimental results}
\label{sec:experimental-results}
%==========================================================

\graphicspath{{analysis/R/subject-analysis/}{./analysis/R/success-rate-analysis/}{./analysis/R/time-cost-analysis/}}

The experimental data were analyzed through statistical techniques.
Section~\ref{sec:feasible-faults} discusses how many valid fixes \autofix produced in the experiments, and Section~\ref{sec:fix-quality} how many of these were proper fixes.
Section~\ref{sec:cost} presents the average \autofix running times.
Section~\ref{sec:success-rate} analyzes the performance of \autofix over multiple sessions to assess its average behavior and its robustness.

%==========================================================
\subsubsection{How many faults \autofix can fix}
\label{sec:feasible-faults}
%==========================================================

It is important to know for how many faults \autofix managed to construct \emph{valid} fixes in \emph{some} of the repeated experiments. The related questions of whether these results are sensitive to the testing time or depend on chance are discussed in the following sections.

 \begin{table}[hbp]
   \tablefont
   \caption{Number of faults fixed by \autofix (\emph{valid} fixes).}
   \label{tab:feasible-faults}%
   \centering
  \setlength{\tabcolsep}{2pt}
     \begin{tabular}{lr@{\hspace{3pt}}r|r@{\hspace{3pt}}rr@{\hspace{3pt}}rr@{\hspace{3pt}}rr@{\hspace{3pt}}rr@{\hspace{3pt}}r}
     \toprule
     \textbf{Code base}  & \multicolumn{2}{c|}{\textbf{\#Fixed}} & \multicolumn{2}{c}{\textbf{\#Void}} & \multicolumn{2}{c}{\textbf{\#Pre}} & \multicolumn{2}{c}{\textbf{\#Post}} & \multicolumn{2}{c}{\textbf{\#Inv}} & \multicolumn{2}{c}{\textbf{\#Check}} \\
     \midrule
     \efbase  & {26} & {(43\%)} & -- & (--) & 18 & (78\%) & 7 & (22\%) & -- & (--) & 1 & (20\%) \\
    \txtlib & {14} & {(45\%)} & 5 & (42\%) & 5 & (36\%) & 0 & (0\%) & -- & (--) & 4 & (100\%) \\
     \cardgm & {31} & {(49\%)} & 14 & (58\%) & 13 & (62\%) & 4 & (50\%) & 0 & (0\%) & -- & (--) \\
     \elearn & {15} & {(30\%)} & 4 & (25\%) & 9 & (39\%) & 2 & (25\%) & 0 & (0\%) & -- & (--) \\
     \toprule
    \textbf{Total} & \textbf{86} & \textbf{(42\%)} & \textbf{23} & \textbf{(44\%)} & \textbf{45} & \textbf{(56\%)} & \textbf{13} & \textbf{(27\%)} & \textbf{0} & \textbf{(0\%)} & \textbf{5} & \textbf{(56\%)}  \\
     \bottomrule
     \end{tabular}%
 \end{table}%

\paragraph{When \autofix succeeds.}
The second column of Table~\ref{tab:feasible-faults} lists the total number of unique faults for which \autofix was able to build a \emph{valid} fix and \emph{rank} it among the top 10 during \emph{at least one} of the 55080 (270 sessions for each of the 204 unique faults) fixing sessions, and the percentage relative to the total number of unique faults in each code base.
The other columns give the breakdown into the same categories of fault as in Table~\ref{tab:subject-faults}.
Overall, \autofix succeeded in fixing 86 (or 42\%) of the faults.
Section~\ref{sec:success-rate} discusses related measures of \emph{success rate}, that is the percentage of sessions that produced a valid fix.

The fixing process is in general non-monotonic; that is, there are faults $\ff$ on which there exists some successful $m$-minute fixing session but no successful \mbox{$n$-minute} fixing sessions for some $n > m$.
The reason is the randomness of \autotest:
a short \autotest run may produce better, if fewer, tests for fixing than a longer run, which would have more chances of generating spurious or redundant passing tests.
Non-monotonic behavior is, however, very infrequent: we observed it only for two faults (one in \efbase and one in \cardgm) which were overly sensitive to the kinds of test cases generated.
In both cases, the faults were fixed in all sessions but those corresponding to a single intermediate testing time (respectively, 15 and 20 minutes).
This corroborates the idea that non-monotonicity is an ephemeral effect of randomness of test-case generation, and suggests that it is not a significant issue in practice.

\paragraph{When \autofix fails.}
To understand the limitations of our technique, we manually analyzed all the faults for which \autofix always failed, and identified four scenarios that prevent success.
Table~\ref{tab:faults-wo-fix} lists the number of faults not fixed (column \#NotFixed) and the breakdown into the scenarios described next.

 \begin{table}[bhtp]
   \tablefont
   \caption{Types of faults that \autofix could not fix.}
   \label{tab:faults-wo-fix}%
   \centering
  	 \setlength{\tabcolsep}{3pt}
     \begin{tabular}{lr|rrrrrr}
     \toprule
     \multicolumn{1}{c}{\textbf{Code base}} & \multicolumn{1}{c|}{\textbf{\#NotFixed}} & \multicolumn{1}{c}{\textbf{\#NoFail}} & \multicolumn{1}{c}{\textbf{\#Complex}} & \multicolumn{1}{c}{\textbf{\#Contract}} & \multicolumn{1}{c}{\textbf{\#Design}} \\
     \midrule
     \efbase & {34} & 3 & 8 & 10 & 13 \\
     \txtlib & {17} & 1 & 5 & 10 & 1 \\
     \cardgm & {32} & 6 & 4 & 16 & 6 \\
     \elearn & {35} & 0 & 13 & 14 & 8 \\
     \toprule
     \textbf{Total} & \textbf{118} & \textbf{10} & \textbf{30} & \textbf{50} & \textbf{28} \\
     \bottomrule
     \end{tabular}%
 \end{table}%

 \runningemphasis{Faults hard to reproduce.}{A small portion of the faults
      identified during the preliminary 2-hour sessions
      (Section~\ref{sec:subjects}) could not be reproduced during the
      shorter \autotest sessions used to provide input to \autofix
      (Section~\ref{sec:experimental-protocol}). Without failing test
      cases\footnote{As a side remark, \autofix managed to fix 19 faults
        for which \autotest could generate \emph{only failing} tests; 7 of
        those fixes are even proper.} the \autofix algorithms cannot
      possibly be expected to work. Column \#NoFail in
      Table~\ref{tab:faults-wo-fix} lists the faults that we could not
      reproduce, and hence could not fix, in the
      experiments.\footnote{Even if \autotest were given enough time to
        generate failing tests, \autofix would still not succeed on these
        faults due to complex patch required (4 faults) or incorrect
        contracts (6 faults).}}

    \runningemphasis{Complex patches required.}{
      	While a significant fraction of fixes are simple~\cite{dallmeier:extraction:2007}, some faults
        require complex changes to the implementation (for example, adding
        a loop or handling special cases differently).
        Such patches are
        currently out of the scope of \autofix; column \#Complex of
        Table~\ref{tab:faults-wo-fix} lists the faults that would require
        complex patches.}

      \runningemphasis{Incorrect or incomplete contracts.}{\autofix assumes
        contracts are correct and tries to fix implementations based on
        them. In practice, however, contracts contain errors too; in such
        cases, \autofix may be unable to satisfy an incorrect
        specification with changes to the code.  A related problem occurs
        when contracts are missing some constraints---for example about
        the invocation order of routines---that are documented informally
        in the comments; faults generated by violating such
        informally-stated requisites are spurious, and \autofix's attempts
        thus become vain.  Column \#Contract of
        Table~\ref{tab:faults-wo-fix} lists the faults involving incorrect
        or incomplete contracts that \autofix cannot fix.  (In recent
        work~\cite{SpeciFix}, we developed a fixing technique that
        suggests changes to incorrect or inconsistent contracts to remove
        faults.)}

      \runningemphasis{Design flaws.}{The design of a piece of software may include
      inconsistencies and dependencies between components; as a
      consequence fixing some faults may require changing elements of the
      design---something currently beyond what \autofix can do. The design
      flaws that \autofix cannot correct often involve inheritance; for
      example, a class \code{LINKED_SET} in \efbase inherits from
      \code{LINKED_LIST} but does not uniformly changes its contracts to
      reflect the fact that a set does not have duplicates while a list
      may. Fixing errors such as this requires a substantial makeover of
      the inheritance hierarchy, of the interfaces, or both. Column
      \#Design of Table~\ref{tab:faults-wo-fix} lists the faults due to
      design flaws that \autofix cannot fix.}

\paragraph{Which fix schemas are used.}
Not all four schemas available to \afx (Section~\ref{sec:schema}) are as successful at generating valid fixes.
Table~\ref{tab:roll-of-schema-valid} shows the number of faults successfully fixed using each of the schemas \emph{a}, \emph{b}, \emph{c}, and \emph{d} in Figure~\ref{lst:fixSkeleton}.
For reference, column \#F shows the total number of faults in each code base; since two valid fixes for the same fault may use different schemas, the total number of faults fixed with any schema is larger than the numbers in column \#F.
Schemas \emph{b} and \emph{d} are the most successful ones, producing valid fixes for 79\% and 75\% of the 86 fixable faults; together, they can fix \emph{all} the 86 faults.
This means that the most effectively deployable fixing strategies are: ``execute a repair action when a suspicious state holds'' (schema~\emph{b}); and ``execute an alternative action when a suspicious state holds, and proceed normally otherwise'' (schema~\emph{d}).

\begin{table}[htbp]
	\tablefont
  \caption{Number of faults fixed using each of the fix schemas in Figure~\ref{lst:fixSkeleton}.}
  \label{tab:roll-of-schema-valid}%
    \centering
  	 \setlength{\tabcolsep}{3pt}
    \begin{tabular}{lrrrrr}
    \toprule
    \textbf{Code base} & \multicolumn{1}{c}{\textbf{\#F}} & \multicolumn{1}{c}{\textbf{Schema (a)}} & \multicolumn{1}{c}{\textbf{Schema (b)}} & \multicolumn{1}{c}{\textbf{Schema (c)}} & \multicolumn{1}{c}{\textbf{Schema (d)}} \\
    \midrule
    \efbase  & 26   & 9        & 18       & 18       & 23 \\
    \txtlib  & 14   & 0        & 12       & 0        & 6 \\
    \cardgm  & 31   & 0        & 27       & 6        & 25 \\
    \elearn  & 15   & 0        & 11       & 4        & 11 \\
    \midrule
    \textbf{Total} & \textbf{86} & \textbf{9} & \textbf{68} & \textbf{28} & \textbf{65} \\
    \bottomrule
    \end{tabular}%
\end{table}%

\begin{result}
In our experiments, \autofix produced \\ valid fixes for 86 (42\%) of 204 faults.
\end{result}

%==========================================================
\subsubsection{Quality of fixes}
\label{sec:fix-quality}
%==========================================================
What is the quality of the valid fixes produced by \autofix in our experiments?
We manually inspected the valid fixes and determined how many of them can be considered \emph{proper}, that is genuine corrections that remove the root of the error (see Section~\ref{sec:validating-fixes}).

Since what constitutes correct behavior might be controversial in some corner cases, we tried to leverage as much information as possible to determine the likely intent of developers, using comments, inspecting client code, and consulting external documentation when available.
In other words, we tried to classify a valid fix as proper only if it really meets the expectations of real programmers familiar with the code base under analysis.
Whenever the notion of proper was still undetermined, we tried to be conservative as much as possible.
While we cannot guarantee that the classification is indisputable, we are confident it is overall very reasonable and sets high standards of quality.

 \begin{table}[htbp]
   \tablefont
   \caption{Number of faults fixed by \autofix (\emph{proper} fixes).}
   \label{tab:achievable-faults}%
   \centering
  	 \setlength{\tabcolsep}{2pt}
\begin{tabular}{lr@{\hspace{3pt}}r|r@{\hspace{3pt}}rr@{\hspace{3pt}}rr@{\hspace{3pt}}rr@{\hspace{3pt}}rr@{\hspace{3pt}}r}
     \toprule
     \textbf{Code base} & \multicolumn{2}{c|}{\textbf{\#Fixed}} & \multicolumn{2}{c}{\textbf{\#Void}} & \multicolumn{2}{c}{\textbf{\#Pre}} &
     	\multicolumn{2}{c}{\textbf{\#Post}} & \multicolumn{2}{c}{\textbf{\#Inv}} & \multicolumn{2}{c}{\textbf{\#Check}} \\
     \midrule
     \efbase & \textbf{12} & \textbf{(20\%)} & -- & (--) & 12 & (52\%) & 0 & (0\%) & -- & (--) & 0 & (0\%) \\
     \txtlib & \textbf{9} & \textbf{(29\%)} & 4 & (33\%) & 2 & (14\%) &  0 & (0\%) & -- & (--) & 3 & (75\%) \\
     \cardgm & \textbf{18} & \textbf{(29\%)} & 10 & (42\%) & 8 & (38\%) & 0 & (0\%) & 0 & (0\%) & -- & (--) \\
     \elearn & \textbf{12} & \textbf{(24\%)} & 3 & (19\%) & 7 & (30\%) & 2 & (25\%) & 0 & (0\%) & -- & (--) \\
     \toprule
     \textbf{Total} & \textbf{51} & \textbf{(25\%)} & \textbf{17} & \textbf{(33\%)} & \textbf{29} & \textbf{(36\%)} & \textbf{2} & \textbf{(4\%)} & \textbf{0} & \textbf{(0\%)} & \textbf{3} & \textbf{(33\%)} \\
     \bottomrule
     \end{tabular}%
 \end{table}%

The second column of Table~\ref{tab:achievable-faults} lists the total number of unique faults for which \autofix was able to build a \emph{proper} fix and \emph{rank} it among the top 10 during \emph{at least one} of the fixing sessions, and the percentage relative to the total number of faults in code base.
The other columns give the breakdown into the same categories of fault as in Tables~\ref{tab:subject-faults} and \ref{tab:feasible-faults}.
Overall, \autofix produces proper fixes in the majority (59\% of 86 faults) of cases where it succeeds, corresponding to 25\% of all unique faults considered in the experiments; these figures suggest that the quality of fixes produced by \autofix is often high.

The quality bar for proper fixes is set quite high: many valid but non-proper fixes could still be usefully deployed, as they provide effective work-arounds that can at least avoid system crashes and allow executions to continue.
Indeed, this kind of ``first-aid'' patches is the primary target of related approaches described in Section~\ref{sec:dynamic-patching}.

We did not analyze the ranking of proper fixes within the top 10 valid fixes reported by \afx.
The ranking criteria (Section~\ref{sec:ranking}) are currently not precise enough to guarantee that proper fixes consistently rank higher than improper ones.
Even if the schemas used by \afx lead to textually simple fixes, analyzing up to 10 fixes may introduce a significant overhead; nonetheless, especially for programmers familiar with the code bases\footnote{During the data collection phase for this paper, it took the first author 3 to 6 minutes to understand and assess each valid fix for a given fault.}, the time spent analyzing fixes is still likely to trade off favorably against the effort that would be required by a manual debugging process that starts from a single failing test case.
Future work will empirically investigate the human effort required to evaluate and deploy fixes produced by \autofix.

\paragraph{Which fix schemas are used.}
The effectiveness of the various fix schemas becomes less evenly distributed when we look at proper fixes.
Table~\ref{tab:roll-of-schema-proper} shows the number of faults with a proper fix using each of the schemas \emph{a}, \emph{b}, \emph{c}, and \emph{d} in Figure~\ref{lst:fixSkeleton}; it is the counterpart of Table~\ref{tab:roll-of-schema-valid} for proper fixes.
schema~\emph{a} is used in no proper fix, whereas schema~\emph{b} is successful with 78\% of the 51 faults for which \afx generates a proper fix; schemas \emph{b} and \emph{d} together can fix 44 out of those 51 faults.
These figures demonstrate that unconditional fixes (schema~\emph{a}) were not useful for the faults in our experiments.
Related empirical research on manually-written fixes~\cite{Pan2009} suggests, however, that there is a significant fraction of faults whose natural corrections consist of unconditionally adding an instruction; this indicates that schema~\emph{a} may still turn out to be applicable to code bases other than those used in our experiments (or that \autofix's fault localization based on Boolean conditions in snapshots naturally leads to conditional fixes).

\begin{table}[htbp]
	\tablefont
  \caption{Number of faults with proper fixes using each of the fix schemas in Figure~\ref{lst:fixSkeleton}.}
  \label{tab:roll-of-schema-proper}%
    \centering
  	 \setlength{\tabcolsep}{3pt}
\begin{tabular}{l rrrrr}
    \toprule
    \textbf{Code base} & \multicolumn{1}{c}{\textbf{\#F}} & \multicolumn{1}{c}{\textbf{Schema (a)}} & \multicolumn{1}{c}{\textbf{Schema (b)}} & \multicolumn{1}{c}{\textbf{Schema~(c)}} & \multicolumn{1}{c}{\textbf{Schema~(d)}} \\
    \midrule
    \efbase  & 12 & 0  & 7  & 5  & 7 \\
    \txtlib  &  9 & 0  & 8  & 0  & 0 \\
    \cardgm  & 18 & 0  & 18 & 0  & 3 \\
    \elearn  & 12 & 0  & 7  & 4  & 3 \\
    \midrule
    \textbf{Total} & \textbf{51} & \textbf{0} & \textbf{40} & \textbf{9} & \textbf{13} \\
    \bottomrule
    \end{tabular}%
\end{table}%

\begin{result}
In our experiments, \autofix produced \emph{proper} fixes \\(of quality comparable to programmer-written fixes) \\for 51 (25\%) of 204 faults.
\end{result}

\subsubsection{Time cost of fixing}
\label{sec:cost}

Two sets of measures quantify the cost of \afx in terms of running time.
The first one is the average running time for \afx alone;
the second one is the average total running time per fix produced, including both testing and fixing.

\begin{figure}[!htp]
	\centering
   	\includegraphics*[width=0.87\linewidth]{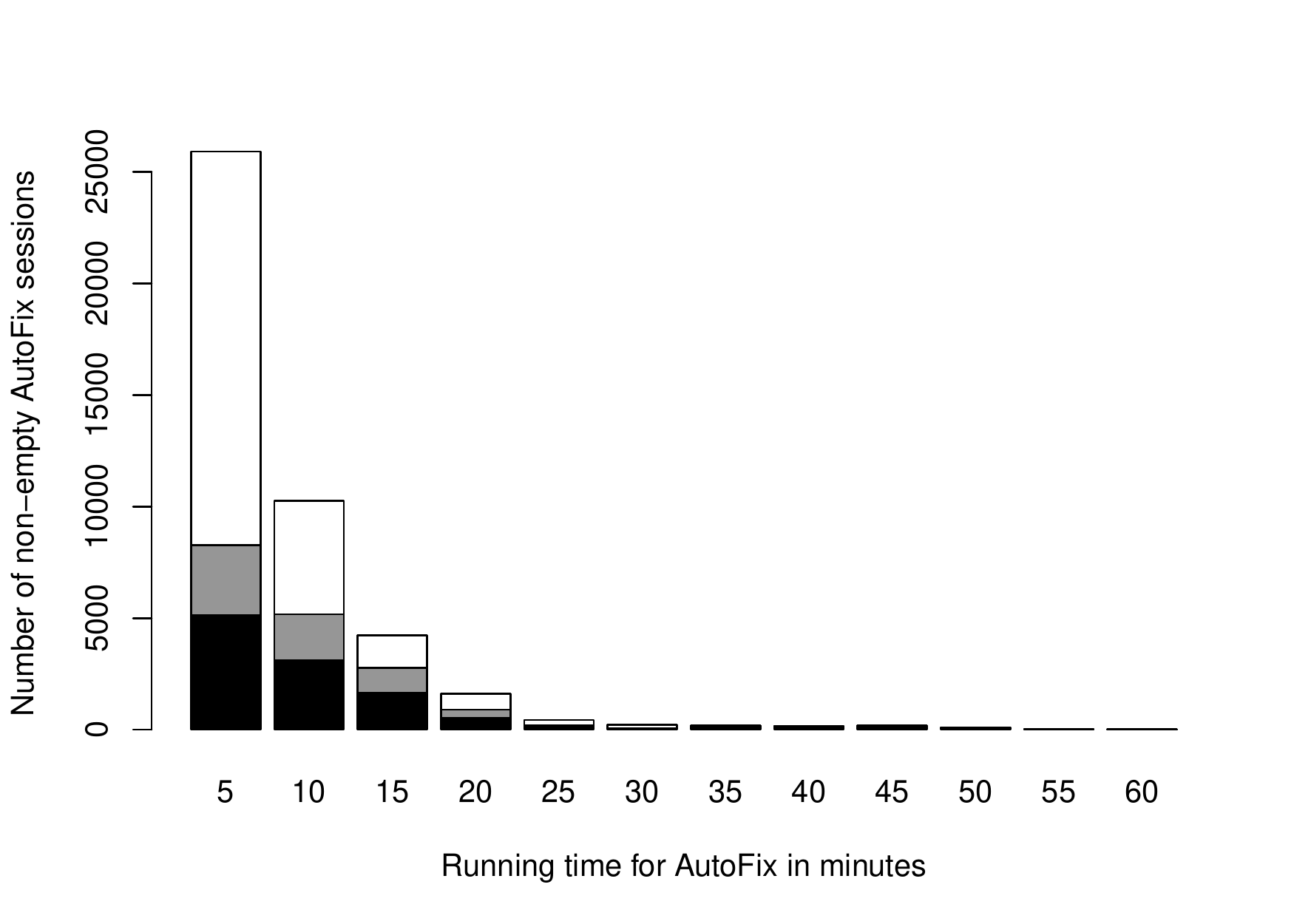}
 	\caption{Distribution of running times for \autofix, independent of the length of the preliminary \autotest sessions (black bars: sessions with proper fixes; gray bars: sessions with valid fixes; white bars: all sessions).}
	\label{fig:cost-distribution-by-fault}
\end{figure}

\paragraph{Fixing time per fault.}
Figure~\ref{fig:cost-distribution-by-fault} shows the distribution of running times for \autofix (independent of the length of the preliminary \autotest sessions) in all the experiments.\footnote{\autofix ran with a timeout of 60 minutes, which was reached only for two faults.}
A bar at position $x$ whose black component reaches height $y_B$, gray component reaches height $y_G \geq y_B$, and white component reaches height $y_W \geq y_G$ denotes that $y_W$ fixing sessions terminated in a time between $x-5$ and $x$ minutes; $y_G$ of them produced a valid fix; and $y_B$ of them produced a proper fix.
The pictured data does not include the 11670 ``empty'' sessions where \autotest failed to supply any failing test cases, which terminated immediately without producing any fix.
The distribution is visibly skewed towards shorter running times, which demonstrates that \autofix requires limited amounts of time in general.

Table~\ref{tab:time-cost} presents the same data about non-empty fixing sessions in a different form: for each amount of \autofix running time (first column), it displays the number and percentage of sessions that terminated in that amount of time (\#Sessions), the number and percentage of those that produced a valid fix (\#Valid), and the number and percentage of those that produced a proper fix (\#Proper).
\begin{table}[htbp]
  \tablefont
  \caption{Distribution of running times for \autofix.}
  \label{tab:time-cost}%
  \centering
  \setlength{\tabcolsep}{4pt}
    \begin{tabular}{p{10pt}rrr@{\hspace{3pt}}rr@{\hspace{3pt}}rr@{\hspace{3pt}}r}
    \toprule
    \multicolumn{3}{c}{\textbf{min. Fixing}} & \multicolumn{2}{c}{\textbf{\#Sessions}} & \multicolumn{2}{c}{\textbf{\#Valid}} & \multicolumn{2}{c}{\textbf{\#Proper}}   \\
    \midrule
   & $5$ &   & 25905 & (59.7\%) & 8275  & (31.9\%) & 5130  & (19.8\%) \\
   & $10$ &  & 36164 & (83.4\%) & 13449 & (37.2\%) & 8246  & (22.8\%) \\
   & $15$ &  & 40388 & (93.1\%) & 16220 & (40.2\%) & 9892  & (24.5\%) \\
   & $20$ &  & 42003 & (96.9\%) & 17114 & (40.7\%) & 10432 & (24.8\%) \\
   & $25$ &  & 42436 & (97.9\%) & 17295 & (40.8\%) & 10543 & (24.8\%) \\
   & $30$ &  & 42650 & (98.4\%) & 17371 & (40.7\%) & 10607 & (24.9\%) \\
   & $40$ &  & 43025 & (99.2\%) & 17670 & (41.1\%) & 10799 & (25.1\%) \\
   & $50$ &  & 43318 & (99.9\%) & 17918 & (41.4\%) & 11013 & (25.4\%) \\
   \midrule
   & $60$ &  & 43365 & (100.0\%) & 17954 & (41.4\%) & 11046 & (25.5\%) \\
    \bottomrule
    \end{tabular}%
\end{table}%
Table~\ref{tab:average-time-stats} shows the minimum, maximum, mean, median, standard deviation, and skewness of the running times (in minutes) across: all fixing sessions, all non-empty sessions, all sessions that produced a valid fix, and all sessions that produced a proper fix.
\begin{table}[htbp]
  \tablefont
  \caption{\autofix running time statistics (times are in minutes).}
  \label{tab:average-time-stats}%
  \centering
  \setlength{\tabcolsep}{3pt}
    \begin{tabular}{lrrrrrr}
    \toprule
    & \multicolumn{1}{c}{\textbf{min}} & \multicolumn{1}{c}{\textbf{max}} & \multicolumn{1}{c}{\textbf{mean}} & \multicolumn{1}{c}{\textbf{median}} & \multicolumn{1}{c}{\textbf{stddev}} & \multicolumn{1}{c}{\textbf{skew}}   \\
    \midrule
    \textbf{All} & 0.0 & 60 & 4.8 & 3.0 & 6.3 & 3.2 \\
    \textbf{Non-empty} & 0.0 & 60 & 6.1 & 4.0 & 6.5 & 3.2 \\
    \textbf{Valid} & 0.5 & 60 & 7.8 & 5.5 & 7.6 & 2.8 \\
    \textbf{Proper} & 0.5 & 60 & 8.1 & 5.4 & 8.3 & 2.9 \\
    \bottomrule
    \end{tabular}%
\end{table}%

\paragraph{Total time per fix.}
The \emph{total} running time of a fixing session also depends on the time spent generating input test cases; the session will then produce a variable number of valid fixes ranging between zero and ten (remember that we ignore fixes not ranked within the top 10).
To have a finer-grained measure of the running time based on these factors, we define the \emph{unit fixing time} of a combined session that runs \autotest for $t_1$ and \autofix for $t_2$ and produces $v > 0$ valid fixes as $(t_1 + t_2)/v$.
Figure~\ref{fig:testandfix-unittime-distribution-valid} shows the distribution of unit fixing times in the experiments: a bar at position $x$ reaching height $y$ denotes that $y$ sessions produced at least one valid fix each, spending an average of $x$ minutes of testing and fixing on each.
The distribution is strongly skewed towards short fixing times, showing that the vast majority of valid fixes is produced in 15 minutes or less.
Table~\ref{tab:testandfix-unittime-stats} shows the statistics of unit fixing times for all sessions producing valid fixes, and for all sessions producing proper fixes.
Figure~\ref{fig:testandfix-unittime-distribution-proper} shows the same distribution of unit fixing times as Figure~\ref{fig:testandfix-unittime-distribution-valid} but for proper fixes.
This distribution is also skewed towards shorted fixing times, but much less so than the one in Figure~\ref{fig:testandfix-unittime-distribution-valid}: while the majority of valid fixes can be produced in 35 minutes or less, proper fixes require more time on average, and there is a substantial fraction of proper fixes requiring longer times up to about 70 minutes.

\iftwocolumn
	\begin{figure*}[!tp]
		\centering
		\begin{subfigure}[b]{0.45\textwidth}
			\includegraphics*[width=0.95\linewidth]{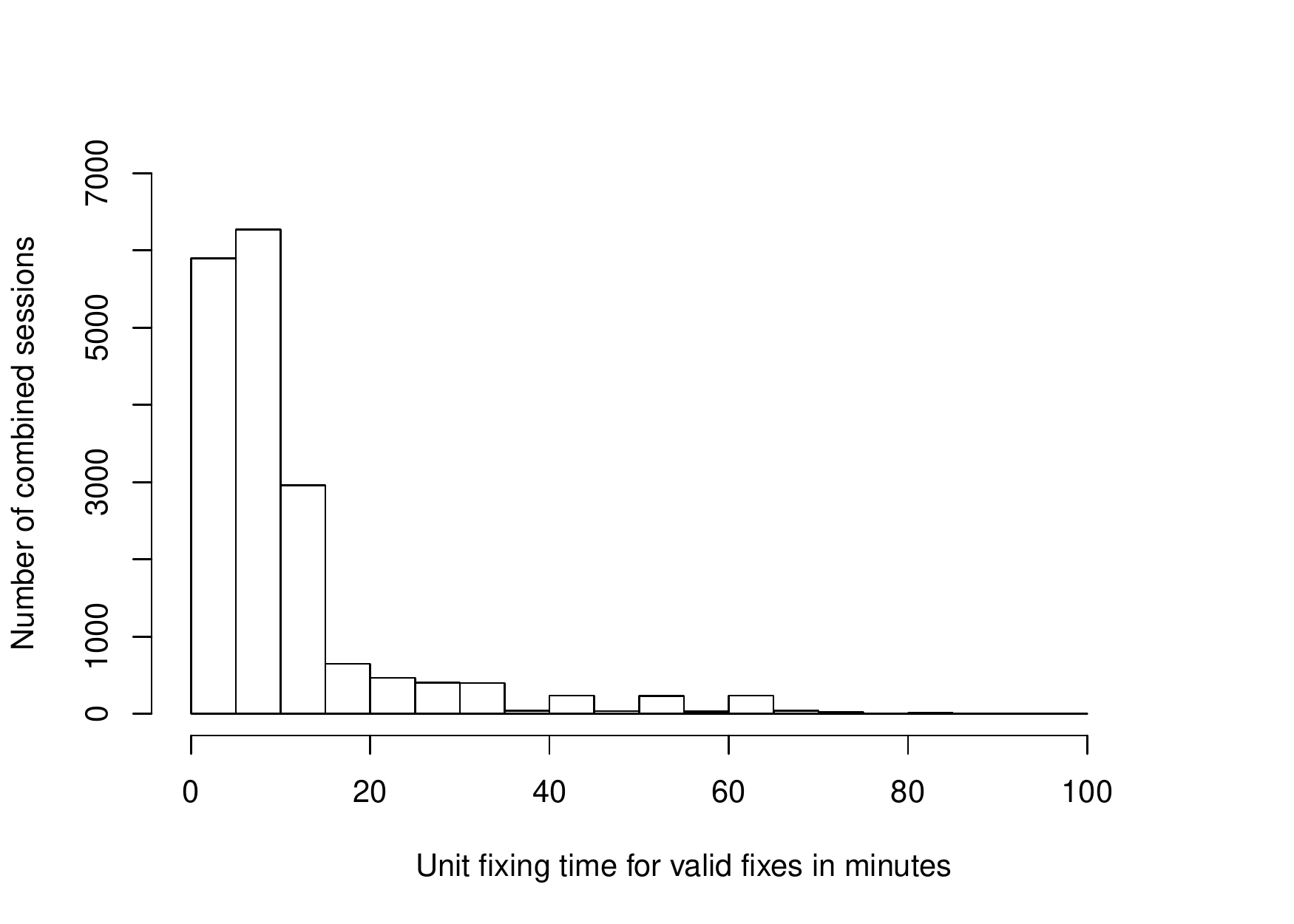}
		 	\caption{Distribution of unit fixing times for \emph{valid} fixes.}
			\label{fig:testandfix-unittime-distribution-valid}
	    \end{subfigure}
		\begin{subfigure}[b]{0.45\textwidth}
			\includegraphics*[width=0.95\linewidth]{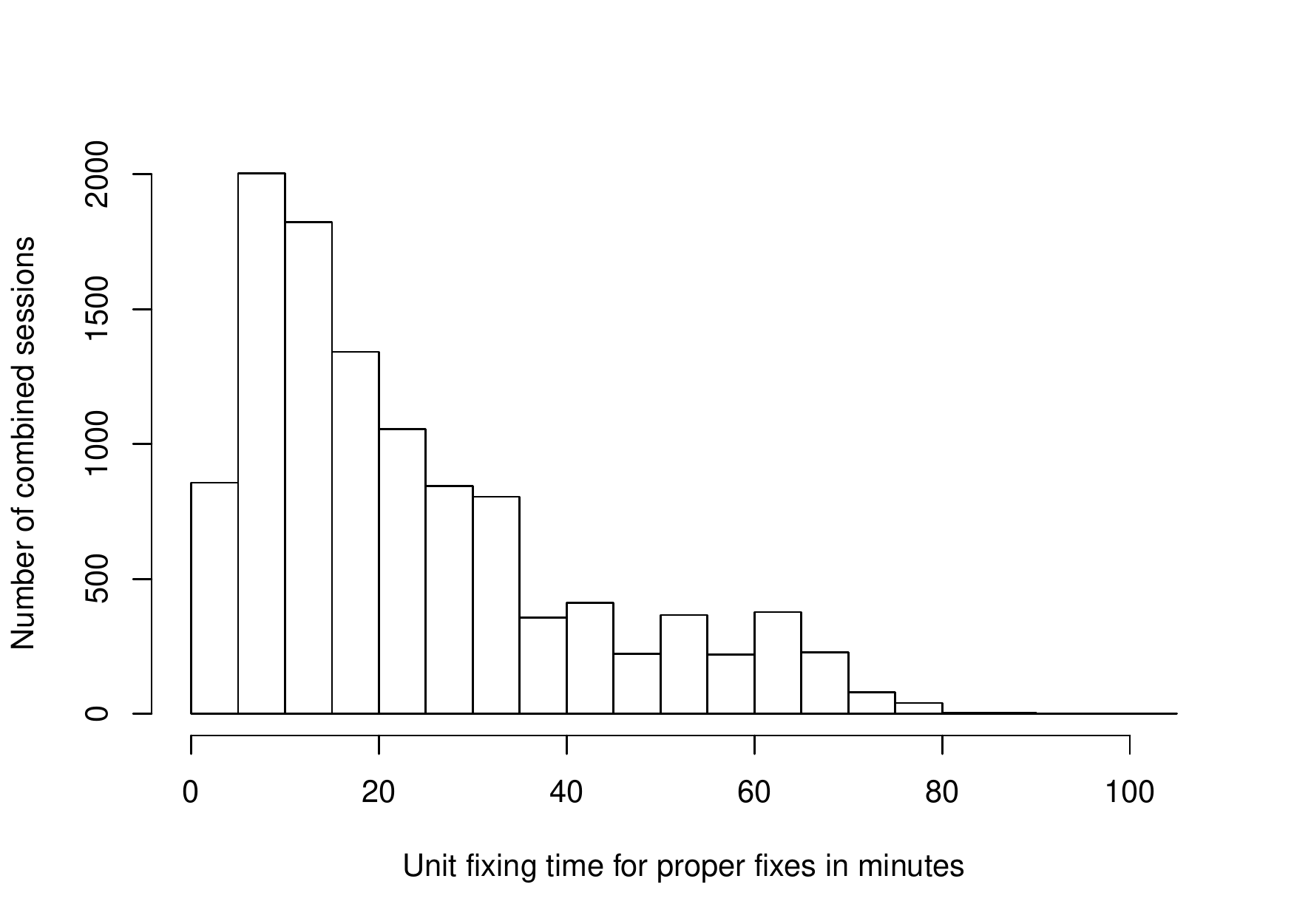}
	 		\caption{Distribution of unit fixing times for \emph{proper} fixes.}
			\label{fig:testandfix-unittime-distribution-proper}
		\end{subfigure}
	 	\caption{Distribution of unit fixing times, including the time spent in the preliminary \autotest sessions.}
		\label{fig:testandfix-unittime-distribution}
	\end{figure*}
\else
	\begin{figure}[bhtp]
		\centering
		\includegraphics*[width=0.8\linewidth]{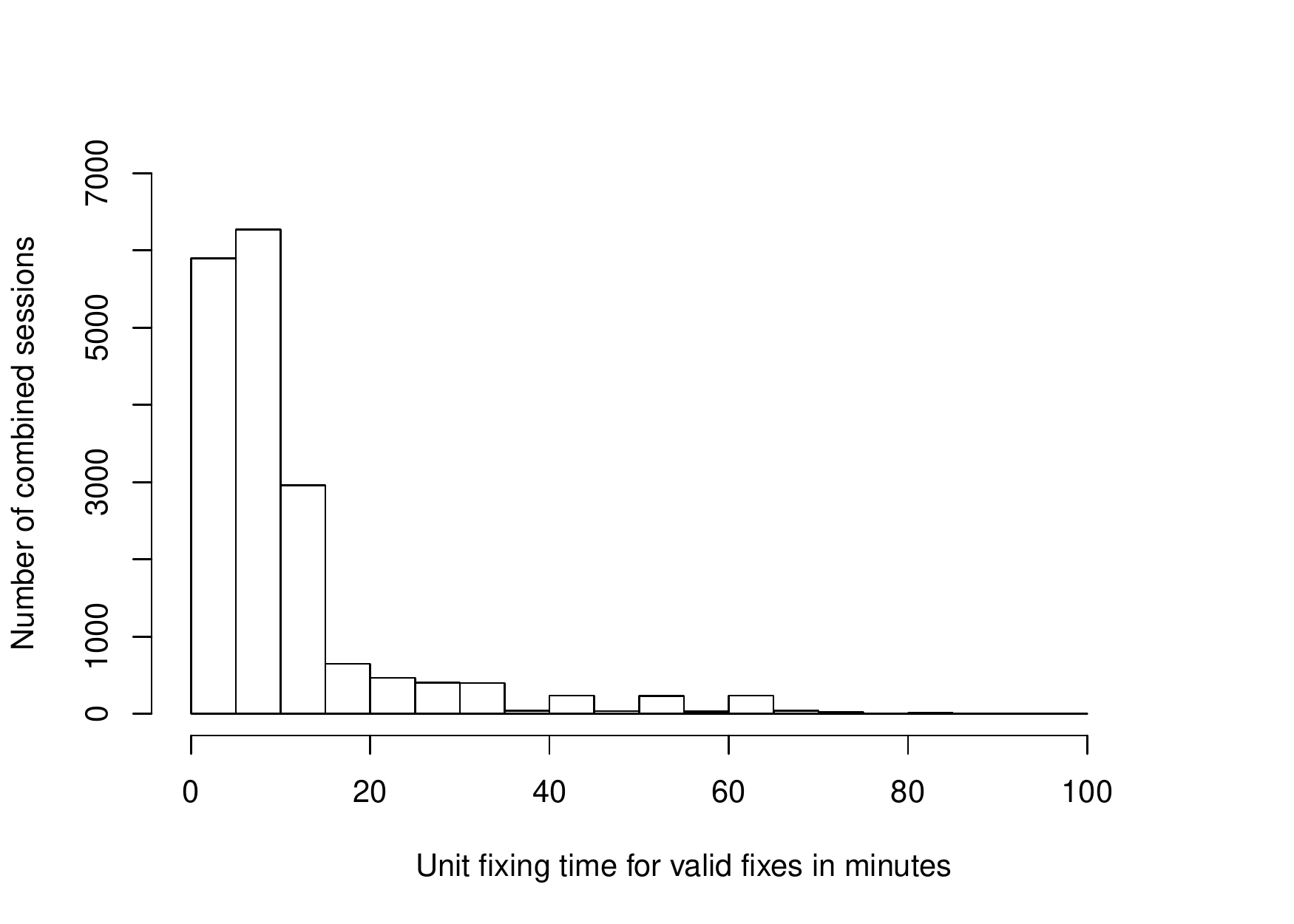}
	 	\caption{Distribution of unit fixing times for \emph{valid} fixes (which including the time spent in the preliminary \autotest sessions).}
		\label{fig:testandfix-unittime-distribution-valid}
	\end{figure}%
	
	\begin{figure}[bhtp]
		\centering
		\includegraphics*[width=0.8\linewidth]{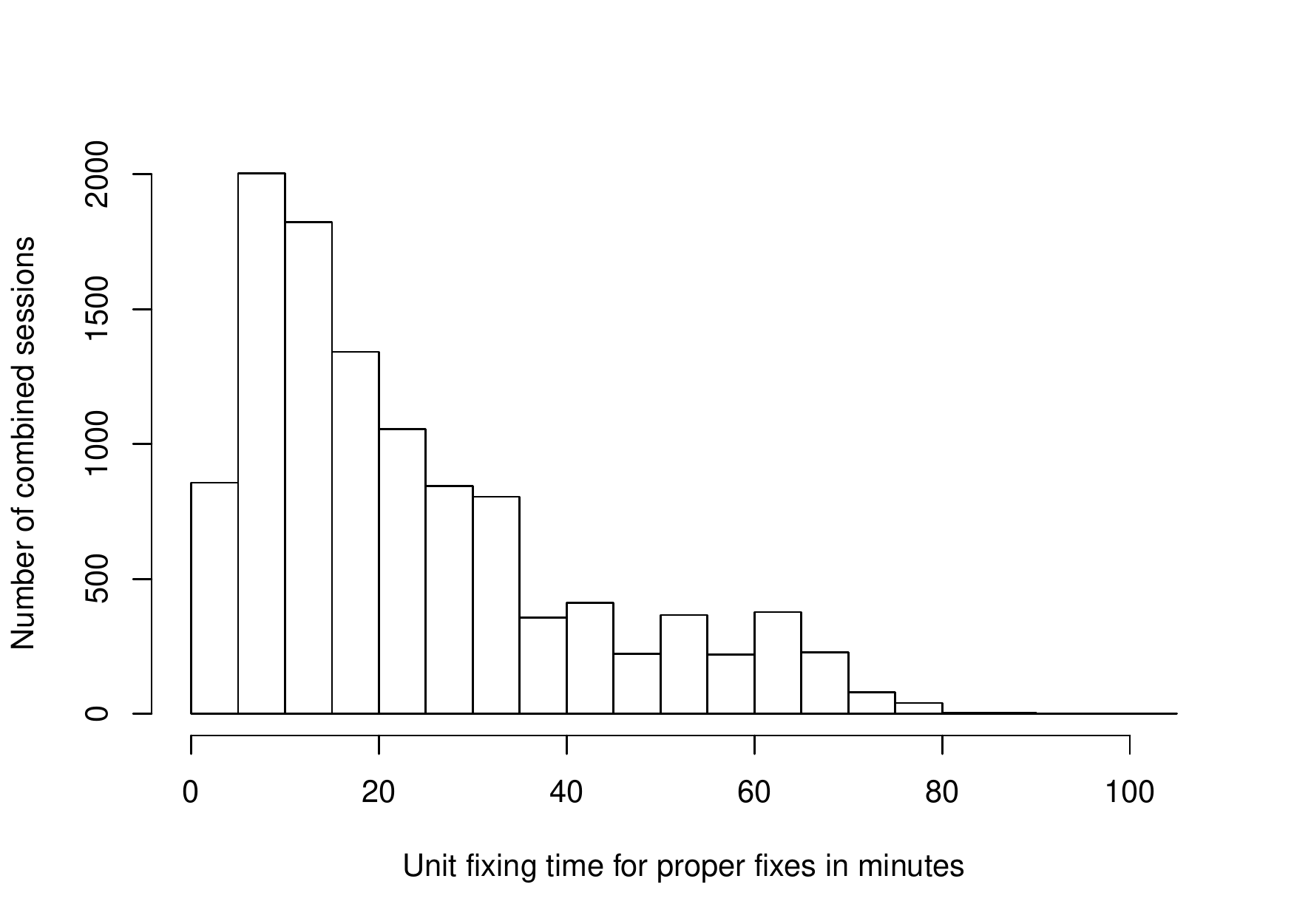}
	 	\caption{Distribution of unit fixing times for \emph{proper} fixes (including the time spent in the preliminary \autotest sessions).}
		\label{fig:testandfix-unittime-distribution-proper}
	\end{figure}%
\fi

\begin{table}[htbp]
  \tablefont
  \caption{Unit fixing times statistics (times are in minutes and include the time spent in the preliminary \autotest sessions).}
  \label{tab:testandfix-unittime-stats}%
  \centering
  \setlength{\tabcolsep}{3pt}
    \begin{tabular}{lrrrrrr}
    \toprule
    & \multicolumn{1}{c}{\textbf{min}} & \multicolumn{1}{c}{\textbf{max}} & \multicolumn{1}{c}{\textbf{mean}} & \multicolumn{1}{c}{\textbf{median}} & \multicolumn{1}{c}{\textbf{stddev}} & \multicolumn{1}{c}{\textbf{skew}}   \\
    \midrule
    \textbf{Valid} & 0.7 & 98.6 & 10.8 & 6.9 & 12.1 & 2.9 \\
    \textbf{Proper} & 1.0 & 101.1 & 23.5 & 17.9 & 17.9 & 1.1 \\
    \bottomrule
    \end{tabular}%
\end{table}%

The unit fixing time is undefined for sessions producing no fixes, but we can still account for the time spent by fruitless fixing sessions by defining the \emph{average unit fixing time} of a group of sessions as the total time spent testing and fixing divided by the total number of valid fixes produced (assuming we get at least one valid fix).
Table~\ref{tab:average-unit-time-cost} shows, for each choice of testing time, the average unit fixing time for valid fixes (second column) and for proper fixes (third column); the last line reports the average unit fixing time over all sessions: 19.9~minutes for valid fixes and 74.2 minutes for proper fixes.

\begin{table}[htbp]
  \tablefont
  \caption{Average unit fixing times for different testing times (times are in minutes).}
  \label{tab:average-unit-time-cost}%
  \centering
  \setlength{\tabcolsep}{4pt}
    \begin{tabular}{rrr}
    \toprule
    \multicolumn{1}{c}{\textbf{min. Testing}} & \multicolumn{1}{c}{\textbf{min. Valid}} & \multicolumn{1}{c}{\textbf{min. Proper}}   \\
    \midrule
    $5$    & 6.0 & 22.0  \\
    $10$   & 8.9 & 32.5  \\
    $15$   & 11.9 & 43.7  \\
    $20$   & 14.6 & 54.0  \\
    $25$   & 17.7 & 65.3  \\
    $30$   & 20.4 & 76.7  \\
    $40$   & 26.1 & 97.3  \\
    $50$   & 31.9 & 121.6  \\
    $60$   & 37.3 & 143.5  \\
    \toprule
    \multicolumn{1}{c}{\textbf{All}} & \textbf{19.9} & \textbf{74.2} \\
    \bottomrule
    \end{tabular}%
\end{table}%

Looking at the big picture, the fixing times are prevalently of moderate magnitude, suggesting that \autofix (and its usage in combination with \autotest) can make an efficient usage of computational time and quickly produce useful results in most cases.
The experimental results also suggest practical guidelines to use \autofix and \autotest: as a rule of thumb, running \autotest for five to ten minutes has a fair chance of producing test cases for \autofix to correct an ``average'' fault.

\begin{result}
In our experiments, \autofix took on average less than\\ 20 minutes per valid fix, including the time required \\ to generate suitable tests with \autotest.
\end{result}

%==========================================================
\subsubsection{Robustness}
\label{sec:success-rate}
%==========================================================

The last part of the evaluation analyzes the robustness and repeatability of \autofix sessions.
The \autofix algorithm is purely deterministic, given as input an annotated program and a set of passing and failing test cases exposing a fault in the program.
In our experiments, however, all the tests come from \autotest, which operates a randomized algorithm, so that different runs of \autotest may produce test suites of different quality.
We want to assess the robustness of \autofix with respect to different choices of input test suites, that is how \autofix's output depends on the test cases supplied.
Assessing robustness is important to demonstrate that our evaluation is indicative of \emph{average} usage, and its results do not hinge on having used a particularly fortunate selection of tests.

Our experiments consisted of many repeated runs of \autotest, each followed by \autofix runs using the generated test as input.
To assess robustness we fix the testing time, and we measure the percentage of \autofix runs, on each of the repeated testing sessions terminating within the allotted testing time, that produced a valid fix.
A high percentage shows that \autofix was successful in most of the repeated testing runs, and hence largely independent of the specific performance of \autotest; to put it differently, a random testing session followed by a fixing sessions has a high chance of producing a valid fix.

\begin{figure*}[!tp]
	\centering
	\begin{subfigure}[b]{\iftwocolumn0.35\else0.45\fi\textwidth}
	    \includegraphics[width=\textwidth]{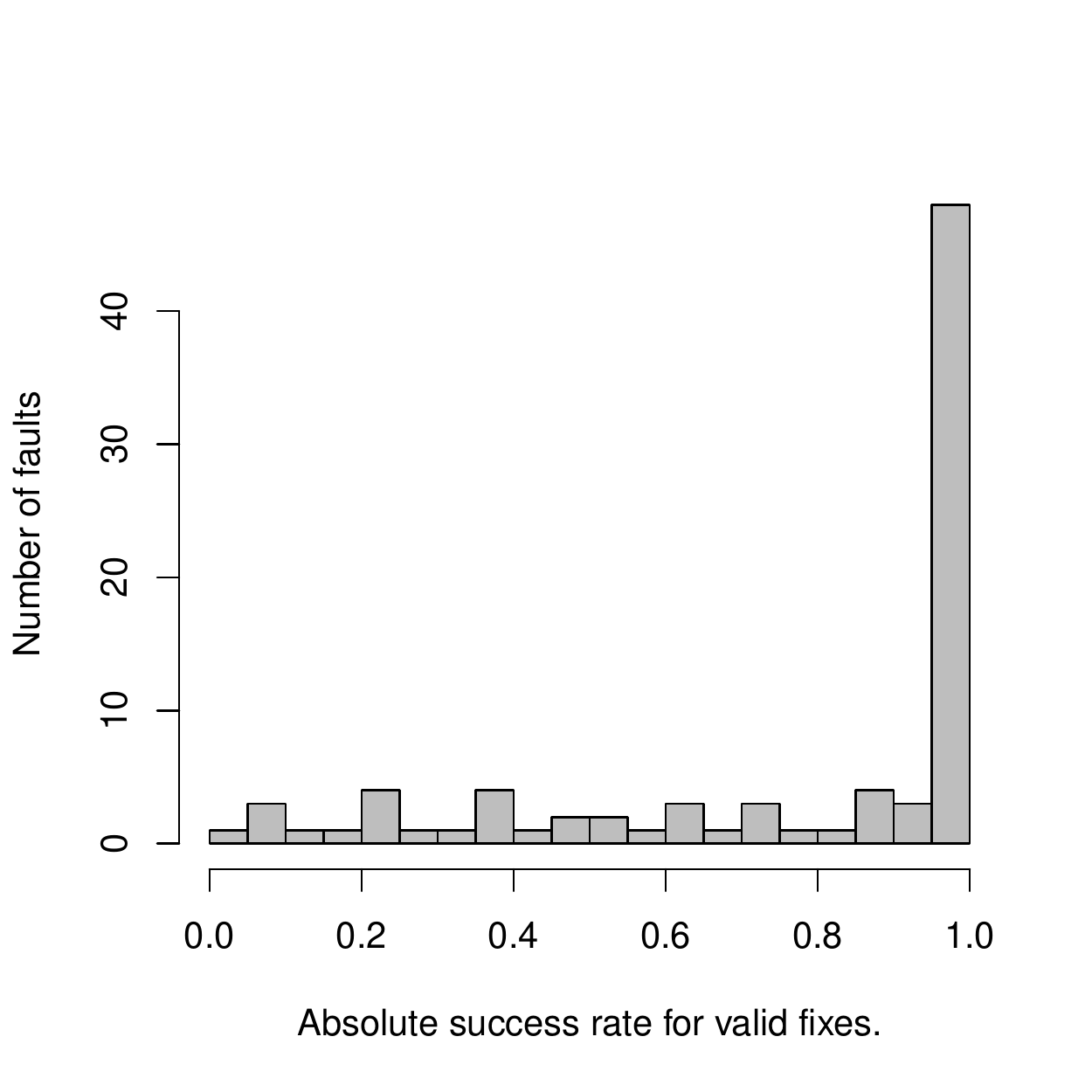}
       \caption{Absolute success rate.}
    	\label{fig:dist-abs-valid}
    \end{subfigure}
	\begin{subfigure}[b]{\iftwocolumn0.35\else0.45\fi\textwidth}
	    \includegraphics[width=1\textwidth]{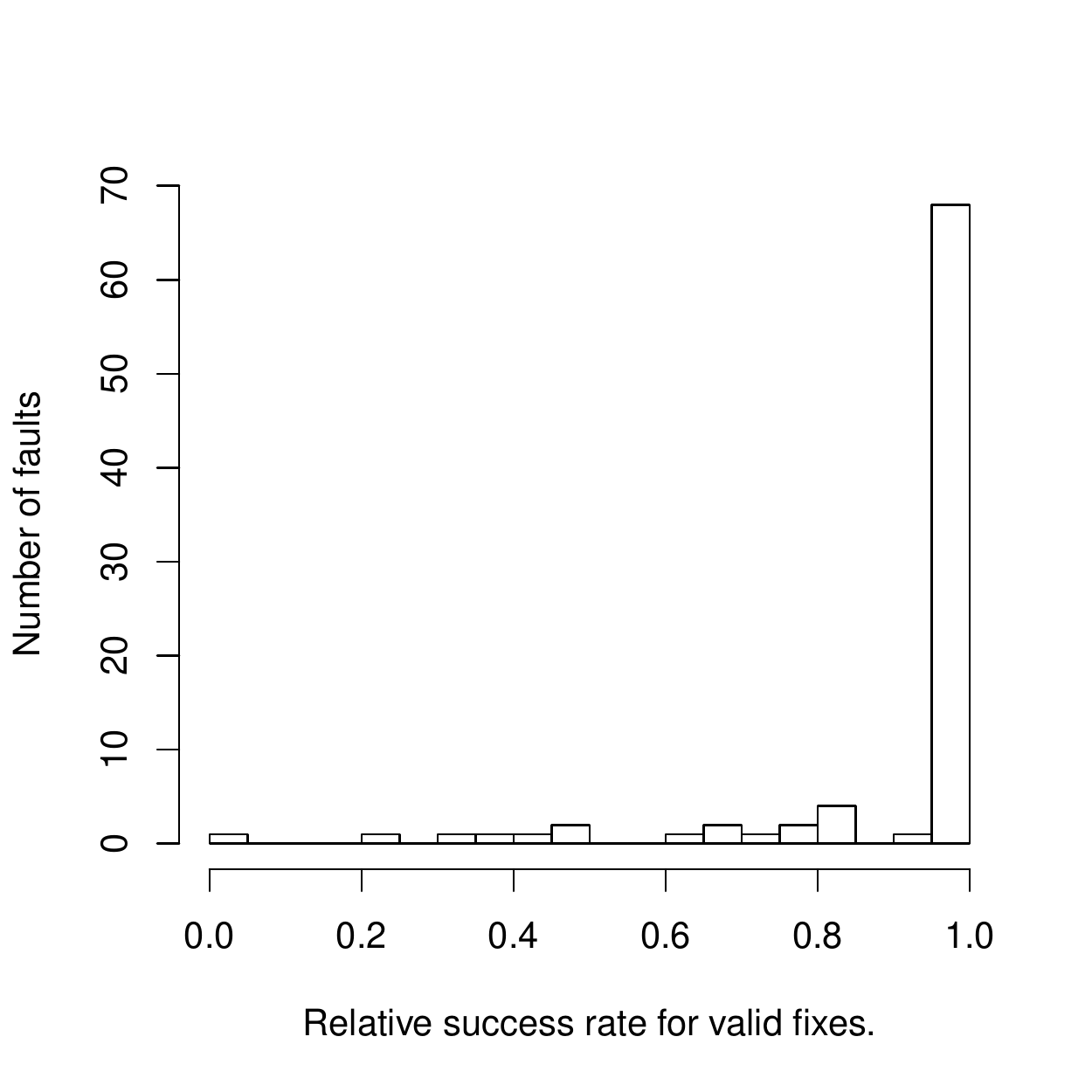}
       \caption{Relative success rate.}
    	\label{fig:dist-rel-valid}
	\end{subfigure}
 	\caption{Distribution of success rates for \emph{valid} fixes.}
	\label{fig:success-rate}
\end{figure*}

\begin{figure*}[!tp]
	\centering
	\begin{subfigure}[b]{\iftwocolumn0.35\else0.45\fi\textwidth}
	    \includegraphics[width=\textwidth]{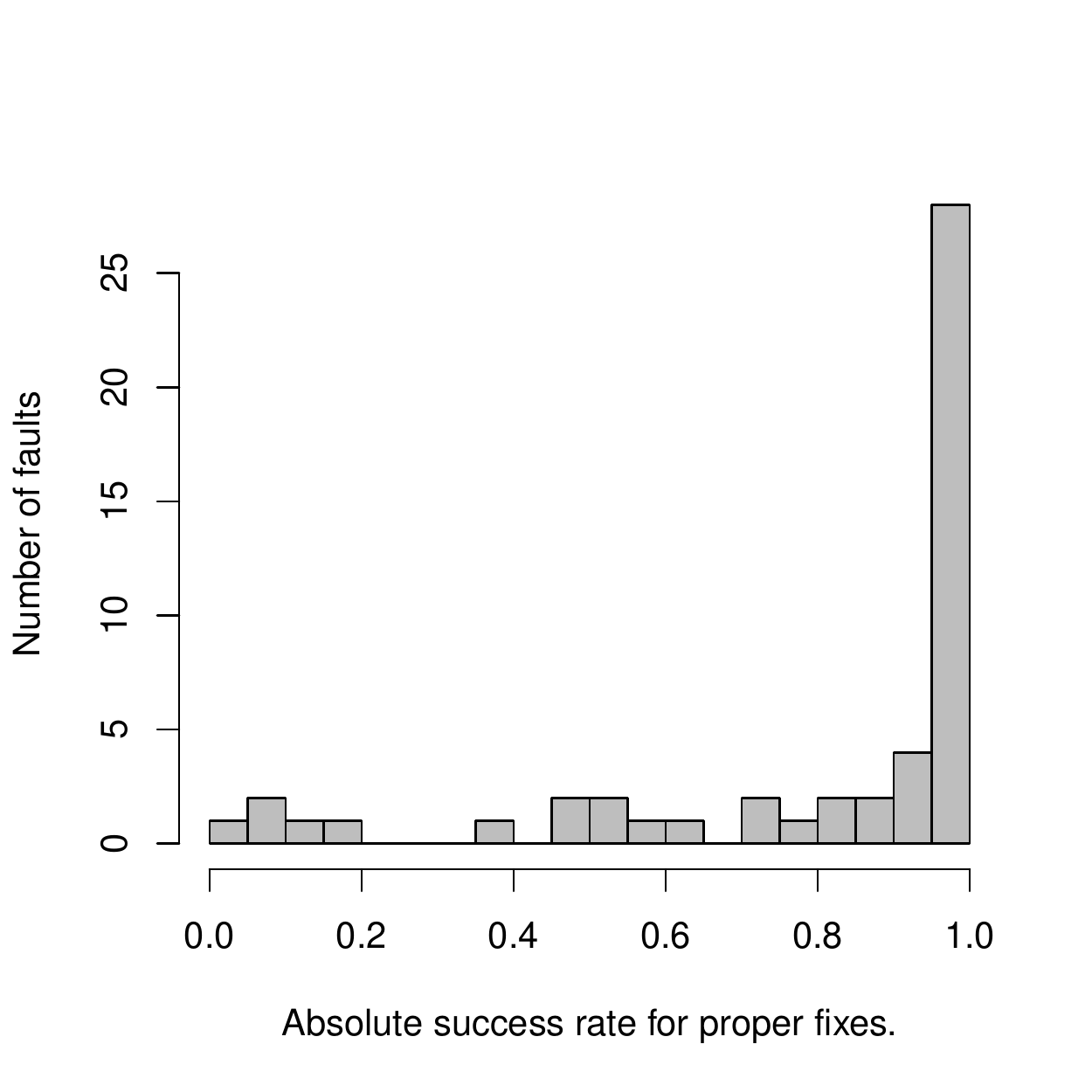}
       \caption{Absolute success rate.}
    	\label{fig:dist-abs-proper}
    \end{subfigure}
	\begin{subfigure}[b]{\iftwocolumn0.35\else0.45\fi\textwidth}
	    \includegraphics[width=\textwidth]{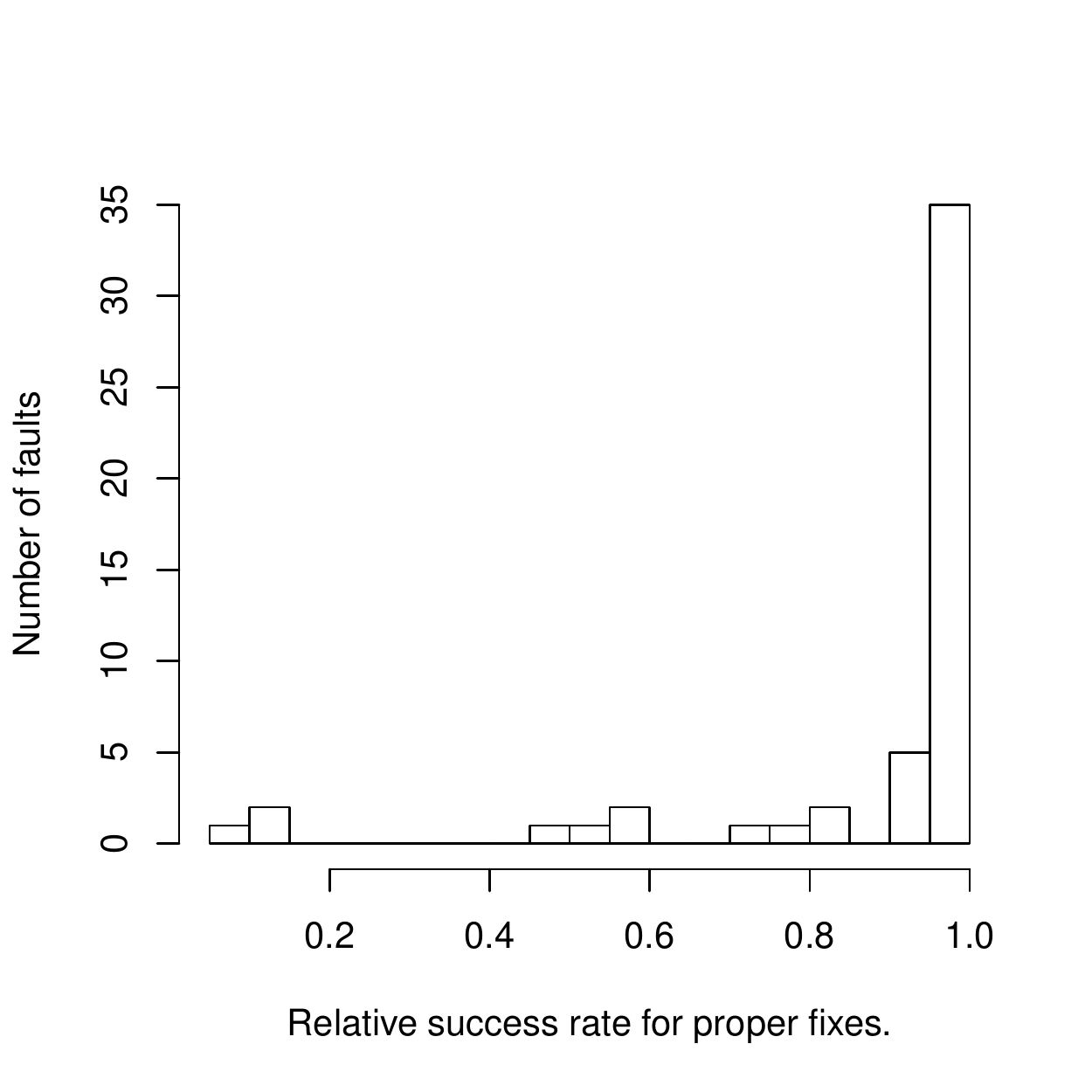}
       \caption{Relative success rate.}
    	\label{fig:dist-rel-proper}
	\end{subfigure}
 	\caption{Distribution of success rates for \emph{proper} fixes.}
	\label{fig:success-rate-proper}
\end{figure*}%

Formally, to measure the robustness with respect to choice of test cases, we introduce the notion of \emph{success rate}: given a fault \ff and a testing time $m$, the \emph{$m$-minute absolute success rate on \ff} is defined as the percentage of $m$-minute fixing sessions on \ff that produce at least one valid fix; the \emph{relative success rate} is defined similarly but the percentage is relative only to non-empty fixing sessions (where \autotest produced at least one failing test case).
Figure~\ref{fig:success-rate} shows the distribution of the absolute (Figure~\ref{fig:dist-abs-valid}) and relative (Figure~\ref{fig:dist-rel-valid}) success rates for all ``fixable'' faults---for which \autofix produced a valid fix at least once in our experiments---for any testing time $m$.
The graphs demonstrate that \autofix has repeatable behavior with a strong majority of faults, largely insensitive to the specific input test cases.
The relative success rates, in particular, exclude the empty \autotest sessions (which are concentrated on some ``hard to reproduce faults'' as discussed in Section~\ref{sec:feasible-faults}) and thus characterize the robustness of \autofix's behavior on the ``approachable'' faults.
(The fact that a classification into ``approachable'' and ``hard'' faults for \autofix naturally emerges further indicates that the kinds of faults used in this evaluation are varied.)

To have a quantitative look at the same data, Table~\ref{tab:success-rate} displays, for each testing time $m$, the number of faults that were fixed successfully---producing a valid fix---in at least $X\%$ of the $m$-minute fixing sessions, for percentages $X = 50, 80, 90, 95$.\footnote{All else being equal, the number of fixed faults is larger when considering \emph{relative} success rates: a relative success rate of $X\% = r/n$ corresponds to $r$ successful fixing sessions out of $n$ non-empty sessions; an absolute success rate of $X\% = a/(n+e)$ for the same testing time corresponds to $a$ successful fixing sessions out of $n$ non-empty sessions and $e$ empty sessions; since $r/n = a/(n+e)$ and $e \geq 0$, it must be $r \geq a$; hence the number of unique faults is also larger in general for the relative rate.}
Each table entry also shows, in parentheses, the percentage of the fixed faults, relative to the 86 fixable faults that \autofix fixed at least once; the data is shown for both the relative and the absolute success rate.
For example, \autofix was successful at least 95\% of the times with 56\% of all fixable faults; or even with 79\% of all fixable faults provided with at least one failing test case.
The last line displays the statistics over all testing sessions of any length.
The aggregated data over all fixing sessions for all faults is the following: 32\% of all sessions and 41\% of all non-empty sessions produced a valid fix.
These success rates suggest a high repeatability of fixing.

Figure~\ref{fig:success-rate-proper} and Table~\ref{tab:success-rate-proper}  display similar data about successful sessions that produced at least one \emph{proper} fix, with percentages relative to all faults for which \autofix produced a proper fix at least once in our experiments.
The aggregated data over all fixing sessions for all faults is the following: 20\% of all sessions and 25\% of all non-empty sessions produced a proper fix; these percentages are quite close to the 25\% of all faults for which \autofix produces at least once a proper fix (Table~\ref{tab:achievable-faults}).
The data for proper fixes is overall quite similar to the one for valid fixes. The absolute figures are a bit smaller, given that the requirement of proper fixes is more demanding, but still support the hypothesis that \autofix's behavior is often robust and largely independent of the quality of provided test cases.

\begin{table*}[!t]
  \caption{Repeatability of \autofix on faults that produced some \emph{valid} fixes.}
  \label{tab:success-rate}%
  \centering
  \tablefont
  \setlength{\tabcolsep}{1.6pt}
	\iftwocolumn\else\begin{adjustwidth}{-8mm}{-7mm}\fi
    \begin{tabular}{r| *{3}{rrrr|} rrrr}
    \toprule
    \multicolumn{1}{c}{\textbf{Success rate:}}
    & \multicolumn{4}{c}{\textbf{50\%}} & \multicolumn{4}{c}{\textbf{80\%}} & \multicolumn{4}{c}{\textbf{90\%}} & \multicolumn{4}{c}{\textbf{95\%}} \\
\midrule
    \multicolumn{1}{c|}{\textbf{min. Testing}} & \multicolumn{2}{c}{\textit{relative}} & \multicolumn{2}{c|}{\textit{absolute}} & \multicolumn{2}{c}{\textit{relative}} & \multicolumn{2}{c|}{\textit{absolute}} & \multicolumn{2}{c}{\textit{relative}} & \multicolumn{2}{c|}{\textit{absolute}} & \multicolumn{2}{c}{\textit{relative}} & \multicolumn{2}{c}{\textit{absolute}}\\
    \midrule
    5     & 83 &(97\%)  & 58 &(67\%)  & 80 &(93\%)  & 49 &(57\%)  & 78 &(91\%)  & 46 &(53\%)  & 75 &(87\%)  & 40 &(47\%) \\
    10    & 83    &(97\%)  & 62    &(72\%)  & 77    &(90\%)  & 56   &(65\%)  & 75    &(87\%)  & 51  &(59\%)  & 69 &(80\%)  & 45  &(52\%) \\
    15    & 81   &(94\%)  & 65   &(76\%)  & 76   &(88\%)  & 58 &(67\%)  & 71 &(83\%)  & 52  &(60\%)  & 68  &(79\%)  & 48 &(56\%) \\
    20    & 82 &(95\%)  & 68  &(79\%)  & 76  &(88\%)  & 58  &(67\%)  & 70  &(81\%)  & 54 &(63\%) & 67  &(78\%)  & 51 &(59\%) \\
    25    & 80  &(93\%)  & 68  &(79\%)  & 72  &(84\%)  & 58  &(67\%)  & 70  &(81\%)  & 56  &(65\%) & 65    &(76\%)  & 51 &(59\%) \\
    30    & 81   &(94\%)  & 69   &(80\%)  & 74  &(86\%)  & 59  &(69\%)  & 70 &(81\%)  & 56 &(65\%)  & 68 &(79\%)  & 53 &(62\%) \\
    40    & 79  &(92\%)  & 69  &(80\%)  & 71 &(83\%)  & 61  &(71\%)  & 68 &(79\%)  & 58  &(67\%)  & 65  &(76\%)  & 55  &(64\%) \\
    50    & 79  &(92\%)  & 70 &(81\%)  & 73 &(85\%)  & 62 &(72\%)  & 69 &(80\%)  & 59 &(69\%)  & 63   &(73\%)  & 53 &(62\%) \\
    60    & 78 &(91\%)  & 71  &(83\%)  & 73  &(85\%)  & 61  &(71\%)  & 68  &(79\%)  & 59 &(69\%)  & 67   &(78\%)  & 57 &(66\%) \\
    \midrule
    \multicolumn{1}{c}{\textbf{All}} & \textbf{79}&\textbf{(92\%)} & \textbf{67}&\textbf{(78\%)} & \textbf{73}&\textbf{(85\%)} & \textbf{56}&\textbf{(65\%)} & \textbf{69}&\textbf{(80\%)} & \textbf{51}&\textbf{(59\%)} & \textbf{68}&\textbf{(79\%)} & \textbf{48}&\textbf{(56\%)} \\
    \bottomrule
    \end{tabular}%
	\iftwocolumn\else\end{adjustwidth}\fi
\end{table*}%

\begin{table*}[!t]
  \caption{Repeatability of \autofix on faults that produced some \emph{proper} fixes.}
  \label{tab:success-rate-proper}%
  \centering
  \tablefont
  \setlength{\tabcolsep}{1.6pt}
	\iftwocolumn\else\begin{adjustwidth}{-8mm}{-7mm}\fi
    \begin{tabular}{r| *{3}{rrrr|} rrrr}
    \toprule
    \multicolumn{1}{c}{\textbf{Success rate:}}
    & \multicolumn{4}{c}{\textbf{50\%}} & \multicolumn{4}{c}{\textbf{80\%}} & \multicolumn{4}{c}{\textbf{90\%}} & \multicolumn{4}{c}{\textbf{95\%}} \\
    \midrule
    \multicolumn{1}{c|}{\textbf{min. Testing}} & \multicolumn{2}{c}{\textit{relative}} & \multicolumn{2}{c|}{\textit{absolute}} & \multicolumn{2}{c}{\textit{relative}} & \multicolumn{2}{c|}{\textit{absolute}} & \multicolumn{2}{c}{\textit{relative}} & \multicolumn{2}{c|}{\textit{absolute}} & \multicolumn{2}{c}{\textit{relative}} & \multicolumn{2}{c}{\textit{absolute}}\\
    \midrule
    5     & 45 &(88\%) & 35 &(69\%) & 42 &(82\%) & 31 &(61\%) & 41 &(80\%) & 41 &(80\%) & 39 &(76\%) & 24 &(47\%) \\
    10    & 47 &(92\%) & 41 &(80\%) & 43 &(84\%) & 35 &(69\%) & 42 &(82\%) & 42 &(82\%) & 36 &(71\%) & 27 &(53\%) \\
    15    & 47 &(92\%) & 41 &(80\%) & 43 &(84\%) & 37 &(73\%) & 39 &(76\%) & 39 &(76\%) & 36 &(71\%) & 29 &(57\%) \\
    20    & 47 &(92\%) & 43 &(84\%) & 43 &(84\%) & 37 &(73\%) & 40 &(78\%) & 40 &(78\%) & 35 &(69\%) & 27 &(53\%) \\
    25    & 48 &(94\%) & 44 &(86\%) & 42 &(82\%) & 37 &(73\%) & 39 &(76\%) & 39 &(76\%) & 34 &(67\%) & 28 &(55\%) \\
    30    & 46 &(90\%) & 43 &(84\%) & 42 &(82\%) & 37 &(73\%) & 41 &(80\%) & 41 &(80\%) & 39 &(76\%) & 32 &(63\%) \\
    40    & 47 &(92\%) & 45 &(88\%) & 41 &(80\%) & 39 &(76\%) & 39 &(76\%) & 39 &(76\%) & 34 &(67\%) & 32 &(63\%) \\
    50    & 47 &(92\%) & 45 &(88\%) & 42 &(82\%) & 39 &(76\%) & 39 &(76\%) & 39 &(76\%) & 33 &(65\%) & 31 &(61\%) \\
    60    & 47 &(92\%) & 45 &(88\%) & 41 &(80\%) & 39 &(76\%) & 40 &(78\%) & 40 &(78\%) & 34 &(67\%) & 31 &(61\%) \\
    \midrule
    \multicolumn{1}{c}{\textbf{All}} & \textbf{47}&\textbf{(92\%)} & \textbf{43}&\textbf{(84\%)} & \textbf{42}&\textbf{(82\%)} & \textbf{36}&\textbf{(71\%)} & \textbf{40}&\textbf{(78\%)} & \textbf{40}&\textbf{(78\%)} & \textbf{35}&\textbf{(69\%)} & \textbf{28}&\textbf{(55\%)} \\
    \bottomrule
    \end{tabular}%
	\iftwocolumn\else\end{adjustwidth}\fi
\end{table*}%

\begin{result}
In our experiments, \autofix produced valid fixes \\ in 41\% of the sessions with valid input tests.
\end{result}

\subsection{Limitations and threats to validity} \label{sec:threats-validity}

\paragraph{Limitations.}
\autofix relies on a few assumptions, which may restrict its practical applicability.

\runningemphasis{Contracts}{or a similar form of annotation must be available in
      the source code. The simple contracts that programmers
      write~\cite{howSpecChange} are sufficient for \autofix; and having
      to write contracts can be traded off against not having to write
      test cases. Requiring contracts does not limit the applicability of
      our technique to Eiffel, given the increasing availability of
      support for contracts in mainstream programming languages.  However,
      the software projects that use contracts in their development is
      still a small minority~\cite{howSpecChange}, which restricts broader
      applicability of \autofix on the software that is currently
      available without additional annotation effort.

      Whether writing contracts is a practice that can become part of
      mainstream software development is a long-standing question. Our
      previous experience is certainly encouraging, in that using
      contracts does not require highly-trained programmers, and involves
      efforts that can be traded off against other costs (e.g.,
      maintenance~\cite{tichy-assertions}) and are comparable to those
      required by other more accepted practices. For example, EiffelBase's
      contracts-to-code ratio is around 0.2~\cite{PFPWM-ICSE13}; while
      detailed quantitative data about industrial experiences with a more accepted
      practice such as test-driven development is scarce, the few
      references that indicate quantitative
      measures~\cite{TDD-book,TDD-empirical,MW-ICSE03} report
      test-LOC-to-application-LOC ratios between 0.4 and 1.0 for projects
      of size comparable to EiffelBase. More extensive assessments belong
      to future work beyond the scope of the present paper.}

\runningemphasis{Functional faults}{are the primary target of \autofix, given
      that contracts provide an effective specification of functional
      correctness. This excludes, for example, violation of liveness
      properties (e.g., termination) or low-level I/O runtime errors
      (Section~\ref{sec:subjects}). Nonetheless, the expressiveness of
      contracts is significant, and in fact we could identify various
      categories of contract-violation faults that \autofix can or cannot
      fix (Section~\ref{sec:feasible-faults}).}

 \runningemphasis{Correctness of contracts}{is assumed by \autotest, which uses
      them as oracles, and by \autofix, which fixes implementations
      accordingly. Since contracts have errors too, this may affect the
      behavior of \autofix on certain faults (see
      Section~\ref{sec:feasible-faults}). Anyway, the line for correctness
      must be drawn somewhere: test cases may also include incorrect
      usages or be incorrectly classified.}

\runningemphasis{Types of fixes}{generated by \autofix include only a subset of
      all possible actions (Section~\ref{sec:synthesis-of-fix-actions})
      and are limited to simple schema
      (Section~\ref{sec:generating-candidate-fixes}). This limits the
      range of fixes that \autofix can generate; at the same time, it
      helps reduce the search space of potential fixes, focusing on the
      few schema that cover the majority of
      cases~\cite{dallmeier:extraction:2007,Martinez2013}.}

\paragraph{Threats to validity.} \label{sec:threats}
While we designed the evaluation of \autofix targeting a broad scope and repeatable results, a few threats to generalizability remain.

    \runningemphasis{Automatically generated test cases}{were used in all our
      experiments. This provides complete automation to the debugging
      process, but it also somewhat restricts the kinds of projects and
      the kinds of faults that we can try to those that we can test with
      \autotest. We plan to experiment with manually-written test cases in
      future work.}

    \runningemphasis{Unit tests}{were used in all our experiments, as opposed to
      system tests.  Unit tests are normally smaller, which helps with
      fault localization and, consequently, to reduce the search space of
      possible fixes. The fact that unit tests are produced as part of
      fairly widespread practices such as test-driven
      development~\cite{TDD-book} reflects positively on the likelihood
      that they be available for automated fixing.}

    \runningemphasis{Size}{and other characteristics (type of program, programming
      style, and so on) of the programs used in the evaluation were
      constrained by the fundamental choice of targeting object-oriented
      programs using contracts that can be tested with \autotest. This
      implies that further experiments are needed to determine to what
      extent the algorithms used by \autofix scale to much larger code
      bases---possibly with large-size modules and system-wide
      executions---and which design choices should be reconsidered in that
      context. To partly mitigate this threat to generalizability, we
      selected experimental subjects of non-trivial size exhibiting
      variety in terms of quality, maturity, and available
      contracts---within the constraints imposed by our fundamental design
      choices, as discussed in Section~\ref{sec:subjects}.}

    \runningemphasis{Variability}{of performance relative to different choices for
      the various heuristics used by \autofix has not been exhaustively
      investigated.  While most heuristics rely on well-defined notions,
      and we provided the rationale for the various design choices, there
      are a few parameters (such as $\alpha$, $\beta$, and $\gamma$ in
      Section~\ref{sec:dynamic-analysis}) whose impact we have not
      investigated as thoroughly as other aspects of the \autofix
      algorithm. As also discussed in Section~\ref{sec:dynamic-analysis},
      the overall principles behind the various heuristics are not
      affected by specific choices for these parameters; therefore, the
      impact of this threat to generalizability is arguably limited.}

    \runningemphasis{Limited computational resources}{were used in all our
      experiments; this is in contrast to other evaluations of fixing
      techniques~\cite{GouesDFW12}. Our motivation for this choice is that
      we conceived \autofix as a tool integrated within a personal
      development environment, usable by individual programmers in their
      everyday activity. While using a different approach to automatic
      fixing could take advantage of massive computational resources,
      \autofix was designed to be inexpensive and evaluated against this
      yardstick.}

    \runningemphasis{Classification}{of fixes into proper and improper was done
      manually by the first author. While this may have introduced a
      classification bias, it also ensured that the classification was
      done by someone familiar with the code bases, and hence in a
      good position to understand the global effects of suggested
      fixes. Future work will investigate this issue empirically, as done
      in recent related work~\cite{Kim:ICSE13}.}

    \runningemphasis{Programmer-written contracts}{were used in all our
      experiments. This ensures that \autofix works with the kinds of
      contracts that programmers tend to write. However, as future work,
      it will be interesting to experiment with stronger higher-quality
      contracts to see how \autofix performance is affected. In recent
      work~\cite{PFPWM-ICSE13} we obtained good results with this approach
      applied to testing with \autotest.}

\section{Related work on automated fixing}
\label{sec:related}

We present the related work on automated program fixing in three areas: techniques working on the source code (as \autofix does); applications to specialized domains; and techniques that operate dynamically at runtime.

\subsection{Source-code repairs}
Techniques such as \autofix target the source code to permanently remove the buggy behavior from a program.

\paragraph{Machine-learning techniques.}
Machine-learn\-ing techniques can help search the space of candidate fixes efficiently and support heuristics to scale to large code bases.

Jeffrey et al.~\cite{jeffrey:bugfix::2009} present BugFix, a tool that summarizes existing fixes in the form of \emph{association rules}.
BugFix then tries to apply existing association rules to new bugs.
The user can also provide feedback---in the form of new fixes or validations of fixes provided by the algorithm---thus ameliorating the performance of the algorithm over time.

Other authors applied \emph{genetic algorithms} to generate suitable fixes.
Arcuri and Yao \cite{arcuri:novel:2008,DBLP:journals/asc/Arcuri11} use a co-evolutionary algorithm where an initially faulty program and some test cases compete to evolve the program into one that satisfies its formal specification.

Weimer et al.~\cite{weimer:automatically:2009,weimer:automatic:2010} describe GenProg, a technique that uses genetic programming\footnote{See also Arcuri and Briand's remarks~\cite[Sec.~2]{arcuri:practical:2011} on the role of evolutionary search in Weimer et al.'s experiments~\cite{weimer:automatically:2009}.} to mutate a faulty program into one that passes all given test cases.
GenProg has been extensively evaluated~\cite{GouesNFW12,GouesDFW12} with various open-source programs, showing that it provides a scalable technique, which can produce non-trivial corrections of subtle bugs, and which works without any user annotations (but it requires a regression test suite).

Kim et al.~\cite{Kim:ICSE13} describe Par, a technique that combines GenProg's genetic programming with a rich predefined set of fix patterns (suggested by human-written patches).
Most of the fix patterns supported by Par are covered by \autofix's synthesis strategies (Section~\ref{sec:synthesis-of-fix-actions}); the few differences concern the usage of overloaded methods---a feature not available in the Eiffel language, and hence not covered by \autofix.
Par has also been extensively evaluated, with a focus on \emph{acceptability} of patches: the programmers involved in the study tended to consider the patches generated by Par more acceptable than those generated by GenProg, and often as acceptable as human-written patches for the same bugs.
The notion of acceptability addresses similar concerns to our notion of proper fix, since they both capture quality as perceived by human programmers beyond the objective yet weak notion of validity, although the two are not directly comparable.

Of the several approaches to source-code general-purpose program repair discussed in this section, GenProg and Par are the only ones that have undergone evaluations comparable with \autofix's: the other approaches have only been applied to seeded faults \cite{he:automated:2004,Gopinath:2011:SPR,DBLP:journals/asc/Arcuri11}, to few benchmarks used for fault localization~\cite{jeffrey:bugfix::2009}, or do not aim at complete automation~\cite{weimer:patches:2006}.

GenProg can fix 52\% of 105 bugs with the latest improvements~\cite{GouesDFW12}; Par fixes 23\% of 119 bugs (GenProg fixes 13\% of the same 119 bugs~\cite{Kim:ICSE13}).
In our experiments in Section~\ref{sec:experiment}, we target almost twice as many bugs (204) and \autofix fixes 42\% of them.
Whereas these quantitative results should not directly be compared because they involve different techniques and faults, they demonstrate that all three approaches produce interesting results and have been thoroughly evaluated.
GenProg's and Par's evaluations have demonstrated their scalability to large programs: GenProg worked on 8 C~programs totaling over 5 million lines of code; Par worked on 6 Java~programs totaling nearly 500 thousand lines of code.
\autofix's evaluation targeted a total of 72 thousand lines of Eiffel~code; while lines of code is a coarse-grained measure of effort, more experiments are needed to conclusively evaluate \autofix's scalability on much larger programs.
The test cases used in GenProg's and Par's evaluations (respectively, around 10 thousand and 25 thousand) do not seem to be directly comparable with those used by \autofix: GenProg and Par use manually-written tests, which may include system tests as well as unit tests; \autofix does not require user-written test cases (and uses fewer on average anyway) but uses automatically generated tests that normally exercise only a limited subset of the instructions in the whole program.
The sensitivity of GenProg or Par about the input test suite have not been systematically investigated,\footnote{GenProg's sensitivity to the design choices of its genetic algorithm has been recently investigated~\cite{GouesWF12}.} and therefore we do not know if they could perform well with tests generated automatically.
In contrast, our experiments show that \autofix is robust with respect to the input tests, and in fact it works consistently well with tests randomly generated given the simple contracts available in Eiffel programs.
Another advantage of leveraging contracts is that \autofix can naturally target \emph{functional} errors (such as those shown in Section~\ref{sec:overview}).

Weimer et al.'s evaluation of fix \emph{quality} has been carried out only for a sample of the bugs, and mostly in terms of induced runtime performance~\cite{GouesNFW12}.
It is therefore hard to compare with \autofix's.
Finally, \autofix works with remarkably limited computational resources: using the same pricing scheme used in GenProg's evaluation~\cite{GouesDFW12}\footnote{We consider \emph{on-demand instances} of Amazon's EC2 cloud computing infrastructure, costing \$0.184 per wall-clock hour at the time of GenProg's experiments.}, \autofix would require a mere \$0.01 per valid fix (computed as $0.184 \times \text{total fixing time in hours} \:/\: \text{total number of valid fixes}$) and \$0.03 per proper fix; or \$0.06 per valid and \$0.23 per proper fix including the time to generate tests---two orders of magnitude less than GenProg's \$7.32 per valid fix.

\paragraph{Axiomatic reasoning.}
He and Gupta~\cite{he:automated:2004} present a technique that compares two program states at a faulty location in the program.
The comparison between the two program states illustrates the source of the error; a change to the program that reconciles the two states fixes the bug.
Unlike our work, theirs compares states purely statically with modular weakest precondition reasoning.
A disadvantage of this approach is that modular weakest precondition reasoning may require detailed postconditions (typically, full functional specifications in first-order logic) in the presence of routine calls: the effects of a call to \code{foo} within routine \code{bar} are limited to what \code{foo}'s postcondition specifies, which may be insufficient to reason about \code{bar}'s behavior.
Even if the static analysis were done globally instead of modularly, it would still require detailed annotations to reason about calls to native routines, whose source code is not available.
This may limit the applicability to small or simpler programs; \autofix, in contrast, compares program states mostly dynamically, handling native calls and requiring only simple annotations for postconditions.
Another limitation of He and Gupta's work is that it builds fix actions by \emph{syntactically} comparing the two program states; this restricts the fixes that can be automatically generated to changes in expressions (for example, in off-by-one errors).
\autofix uses instead a combination of heuristics and fix schemas, which makes for a flexible usage of a class's public routines without making the search space of possible solutions intractably large.

\paragraph{Constraint-based techniques.}
Gopinath et al.~\cite{Gopinath:2011:SPR} present a framework that repairs errors due to value misuses in Java programs annotated with pre- and postconditions.
A repairing process with the framework involves encoding programs as relational formulae, where some of the values used in ``suspicious'' statements are replaced by free variables.
The conjunction of the formula representing a program with its pre- and postcondition is fed to a SAT solver, which suggests suitable instantiations for the free variables.
The overall framework assumes an external fault localization scheme to provide a list of suspicious statements; if the localization does not select the proper statements, the repair will fail.
Solutions using dynamic analysis, such as \autofix, have a greater flexibility in this respect, because they can better integrate fault localization techniques---which are also typically based on dynamic analysis.
As part of future work, however, we will investigate including SAT-based techniques within \autofix.

Nguyen et al.~\cite{Nguyen:ICSE13} build on previous work~\cite{Chandra:2011:AD} about detecting suspicious expressions to automatically synthesize possible replacements for such expression; their SemFix technique replaces or adds constants, variables, and operators to faulty expressions until all previously failing tests become passing.
The major differences with respect to \autofix are that SemFix's fault localization is based on statements rather than snapshots, which gives a coarser granularity; and that the fixes produced by SemFix are restricted to changes of right-hand sides of assignments and Boolean conditionals, whereas \autofix supports routine calls, more complex expression substitutions, and conditional schemas.
This implies that \autofix can produce fixes that are cumbersome or impossible to build using SemFix. For example, conditional fixes are very often used by \autofix (Tables~\ref{tab:roll-of-schema-valid} and \ref{tab:roll-of-schema-proper}) but can be generated by SemFix only if a conditional already exists at the repair location; and supporting routine calls in fixes takes advantage of modules with a well-designed API.

\paragraph{Model-driven techniques.}
Some automated fixing methods exploit finite-state abstractions to detect errors or to build patches.
\autofix also uses a form of finite-state abstraction as one way to synthesize suitable fixing actions (Section~\ref{sec:behavioral}).

In previous work, we developed Pachika~\cite{dallmeier:generating:2009}, a tool that automatically builds finite-state behavioral models from a set of passing and failing test cases of a Java program.
Pachika also generates fix candidates by modifying the model of failing runs in a way which makes it compatible with the model of passing runs.
The modifications can insert new transitions or delete existing transitions to change the behavior of the failing model; the changes in the model are then propagated back to the Java implementation.
\autofix exploits some of the techniques used in Pachika---such as finite-state models and state abstraction---in combination with other novel ones---such as snapshots, dynamic analysis for fault localization, fix actions and schema, contracts, and automatic test-case generation.

Weimer \cite{weimer:patches:2006} presents an algorithm to produce patches of Java programs according to finite-state specifications of a class.
The main differences with respect to \autofix are the need for user-provided finite-state machine specifications, and the focus on security policies: patches may harm other functionalities of the program and ``are not intended to be applied automatically'' \cite{weimer:patches:2006}.

\subsection{Domain-specific models} \label{sec:doma-spec-models}

Automated debugging can be more tractable over restricted models of computations.
A number of works deal with fixing finite-state programs, and normally assumes a specification given in some form of temporal logic~\cite{mayer:evaluating:2008,JobstmannSGB12,JobstmannSGB12}.

Gorla et al.~\cite{DBLP:conf/sigsoft/CarzanigaGPP10,DBLP:journals/cai/GorlaPWMP10} show how to patch web applications at runtime by exploiting the redundancy of services offered through their APIs; the patches are generated from a set of rewrite rules that record the relations between services.
In more recent work~\cite{Carzaniga:ICSE2013}, they support workarounds of general-purpose Java applications based on a repertoire of syntactically different library calls that achieve the same semantics.

Janjua and Mycroft~\cite{Mycroft-concurrency} target atomicity violation errors in concurrent programs, which they fix by introducing synchronization statements automatically.
More recently, Jin et al.~\cite{JinSZLL11} developed the tool AFix that targets the same type of concurrency errors.

Abraham and Erwig~\cite{abraham:goal-directed:2005} develop automated correction techniques for spreadsheets, whose users may introduce erroneous formulae.
Their technique is based on annotating cells with simple information about their ``expected value''; whenever the computed value of a cell contradicts its expected value, the system suggests changes to the cell formula that would restore its value to within the expected range.
The method can be combined with automated testing techniques to reduce the need for manual annotations~\cite{abraham:test-driven:2008}.

Samimi et al.~\cite{Samimi:2012:ARH} show an approach to correct errors in print statements that output string literals in PHP applications.
Given a test suite and using an HTML validator as oracle for acceptable output, executing each test and validating its output induces a partial constraint on the string literals.
Whenever the combination of all generated constraints has a solution, it can be used to modify the string literals in the print statements to avoid generating incorrect output.
Constraint satisfaction can be quite effective when applied to restricted domains such as PHP strings; along the same lines, \autofix uses constraint-based techniques when dealing with linear combinations of integer variables (Section~\ref{sec:lin-constraints}).

\subsection{Dynamic patching} \label{sec:dynamic-patching}

Some fixing techniques work \emph{dynamically}, that is at runtime, with the goal of contrasting the adverse effects of some malfunctioning functionality and prolonging the up time of some piece of deployed software.
Demsky et al.~\cite{DBLP:conf/ecoop/DemskyD08,DBLP:journals/tse/DemskyS11} provide generic support for dynamic patching inside the Java language.

\paragraph{Data-structure repair.}
Demsky and Rinard~\cite{demsky:automatic:2003} show how to dynamically repair data structures that violate their consistency constraints.
The programmer specifies the constraints, which are monitored at runtime, in a domain language based on sets and relations.
The system reacts to violations of the constraints by running repair actions that try to restore the data structure in a consistent state.

Elkarablieh and Khurshid~\cite{elkarablieh:juzi::2008} develop the Juzi tool for Java programs.
A user-defined \verb+repOk+ Boolean query checks whether the data structure is in a coherent state.
Juzi monitors \verb+repOk+ at runtime and performs some repair action whenever the state is corrupted.
The repair actions are determined by symbolic execution and by a systematic search through the object space.
In follow-up work~\cite{muhammad:zubair:malik:case:2009,MalikSK11}, the same authors outline how the dynamic fixes generated by Juzi can be abstracted and propagated back to the source code.

Samimi et al.'s work~\cite{SamimiAM10} leverages specifications in the form of contracts to dynamically repair data structures and other applications.
As in our work, an operation whose output violates its postcondition signals a fault.
When this occurs, their Plan~B technique uses constraint solving to generate a different output for the same operation that satisfies the postcondition and is consistent with the rest of the program state; in other words, they \emph{execute the specification} as a replacement for executing a faulty implementation.
Their prototype implementation for Java has been evaluated on a few data-structure faults similar to those targeted by Demsky and Rinard~\cite{demsky:automatic:2003}, as well as on other operations that are naturally expressed as constraint satisfaction problems.

\paragraph{Memory-error repair.}
The ClearView framework \cite{perkins:automatically:2009} dynamically corrects buffer overflows and illegal control flow transfers in binaries.
It exploits a variant of Daikon~\cite{ECGN01} to extract invariants in normal executions.
When the inferred invariants are violated, the system tries to restore them by looking at the differences between the current state and the invariant state.
ClearView can prevent the damaging effects of malicious code injections.

Exterminator~\cite{Berger:2012:SNSAA, Novark:Exterminator:2008} is a framework to detect and correct buffer overflow and dangling pointer errors in C and C++ programs.
The tool executes programs using a probabilistic memory allocator that assigns a memory area of variably larger size to each usage; an array of size $n$, for example, will be stored in an area with strictly more than $n$ cells.
With this padded memory, dereferencing pointers outside the intended frame (as in an off-by-one overflow access) will not crash the program.
Exterminator records all such harmless accesses outside the intended memory frame and abstracts them to produce patches that permanently change the memory layout; the patched layout accommodates the actual behavior of the program in a safe way.

\section{Conclusions}
\label{sec:future}
\label{sec:conclusion}

In the past decade, automated debugging has made spectacular advances:
first, we have seen methods to isolate failure causes automatically;
then, methods that highlight likely failure locations.  Recently, the
slogan ``automated debugging'' has denoted techniques that truly
deserve this name: we can actually generate workable fixes completely
automatically.

The \autofix approach, described in the paper, is an important contribution towards the ideal of automatic debugging.
In experiments with over 200 faults in software of various quality, \autofix generated fixes for 42\% of the faults; inspection reveals 59\% of them are not mere patches but real corrections of quality comparable to those programmers familiar with the faulty programs could write.
\autofix achieves these results with limited computational resources: running on standard hardware, it required an average time per fix under 20 minutes---where the average includes all failed fixing attempts and the automatic generation of test cases that profile the faults.
One of the key ingredients used to achieve these encouraging results is the reliance on \emph{contracts} to boost and automate all debugging steps.
The kinds of contracts required by \autofix are simple and normally available in Eiffel programs; the effort of writing them is, therefore, limited and comparable to other everyday programming activities.

With \autofix, the programmer's debugging effort could be reduced to
almost zero in many cases.  We write ``\emph{almost} zero'', as we still
assume that a human should assess the generated fixes and keep
authority over the code.  One may also think of systems that generate
and apply fixes automatically; the risk of undesired behavior may
still be preferred to no behavior at all, and can be alleviated by
more precise specifications expressed as contracts.  In any case, we
look forward to a future in which much of the debugging is taken over
by automated tools, reducing risks in development and relieving
programmers from a significant burden.

\iftwocolumn \section*{Availability} \else \vskip6pt\textbf{Availability.} \fi
The \autofix source code, and all data and results cited in this article, are available at:

{\centering \website\\}

\iftwocolumn \section*{Acknowledgments} \else \vskip6pt\textbf{Acknowledgments.} \fi
This work was partially funded by the Hasler-Stiftung (Grant no.~2327) and by the Deutsche For\-schungs\-ge\-mein\-schaft (Ze509/4-1) under the title ``AutoFix---Programs that fix themselves''; and by the Swiss National Science Foundation (Project 200021-134976:  ``Automated Support for Invariant Inference'').
We also gratefully acknowledge the support of the Swiss National Supercomputing Centre (CSCS) for the experiments (Project s264).
The concept of generating fixes from differences in passing and failing runs was conceived with Andreas Leitner. Stefan Buchholz and Lucas S.\ Silva contributed to an early implementation of \autofix.

\iftwocolumn
\else
\clearpage
\newpage
\appendix
\appendixpage
\addappheadtotoc

This Appendix contains additional material omitted from the referred main text.
The section titles correspond to those in the main text where the additional material belongs; for greater clarity, the titles also mention, in parentheses, the corresponding section numbers in the main text, when they exist.

\section{\afx in action~\tomainref{sec:overview}}

\subsection{Another error in \code{move_item}~\tomainref{sec:oneerror}}

Another error occurs when \code{index} has value 0 (Figure~\ref{fig:set-remove-at0}), meaning that \code{before} (line~\ref{ln:before}) has value \code{True}; this is a valid position for \code{go_i_th} but not for \code{put_left}, because there is no position ``to the left of 0'' where \code{v} can be re-inserted: the call to \code{put_left} on line~\ref{ln:putleft} violates its precondition (line~\ref{ln:putleft-pre}).

\begin{figure}[h]
  \centering
  \begin{subfigure}{\textwidth}
    \centering
  \begin{tikzpicture}[
  item/.style={rectangle, minimum size=8mm,very thick,rounded corners=2mm,draw=green!50!black!50,font=\scriptsize, top color=white, bottom color=green!50!black!20},
  node distance=4mm,
  pin distance=7mm,
  every pin edge/.style={<-, shorten <=4mm, ultra thick,green!40!black!60},
  ]
  \lstset{basicstyle=\footnotesize}

  \node (i0) [item,dotted,label=above:\code{0},pin=above:\code{index}] {};
  \node (i1) [item,right=of i0,label=\code{1}] {};
  \node (i2) [item,right=of i1,label=above:\code{2}] {};
  \node (i3) [right=of i2,minimum size=8mm] {$\cdots$};
  \node (iv) [item,right=of i3] {\code{v}};
  \node (i4) [right=of iv,minimum size=8mm] {$\cdots$};
  \node (icm1) [item,right=of i4,label={[label distance=-.5pt]above:\code{count - 1}}] {};
  \node (icm) [item,right=of icm1,label={[label distance=.5pt]above:\code{count}}] {};
  \node (iend) [item,right=of icm,dotted,label={[label distance=0pt]above:\code{count + 1}}] {};
  \begin{scope}[-latex,green!40!black!70,thick]
    \foreach \na / \nb in {i0/i1,i1/i2,i2/i3,i3/iv,iv/i4,i4/icm1,icm1/icm,icm/iend}
    {
      \path (node cs:name=\na,angle=20) edge (node cs:name=\nb,angle=160);
      \path (node cs:name=\nb,angle=-160) edge (node cs:name=\na,angle=-20);
    }
  \end{scope}
\end{tikzpicture}
\caption{Calling \code{put_left} in \mbox{\code{move_item}} when \code{index = 0} triggers a failure: there is no valid position to the left of $0$ where $v$ can be moved.}
\label{fig:set-remove-at0}
\end{subfigure}

  \begin{subfigure}{\textwidth}
    \centering
  \begin{tikzpicture}[
  item/.style={rectangle, minimum size=8mm,very thick,rounded corners=2mm,draw=green!50!black!50,font=\scriptsize, top color=white, bottom color=green!50!black!20},
  node distance=4mm,
  pin distance=7mm,
  every pin edge/.style={<-, shorten <=4mm, ultra thick,green!40!black!60},
  ]
  \lstset{basicstyle=\footnotesize}

  \node (i0) [item,dotted,label=above:\code{0}] {};
  \node (i1) [item,right=of i0,label=\code{1}] {};
  \node (i2) [item,right=of i1,label=above:\code{2}] {};
  \node (i3) [right=of i2,minimum size=8mm] {$\cdots$};
  \node (iv) [item,right=of i3] {\code{v}};
  \node (i4) [right=of iv,minimum size=8mm] {$\cdots$};
  \node (icm1) [item,right=of i4,label={[label distance=-.5pt]above:\code{count - 1}}] {};
  \node (icm) [item,right=of icm1,label={[label distance=.5pt]above:\code{count}}] {};
  \node (iend) [item,right=of icm,dotted,label={[label distance=0pt]above:\code{count + 1}},pin=above:\code{index}] {};
  \begin{scope}[-latex,green!40!black!70,thick]
    \foreach \na / \nb in {i0/i1,i1/i2,i2/i3,i3/iv,iv/i4,i4/icm1,icm1/icm,icm/iend}
    {
      \path (node cs:name=\na,angle=20) edge (node cs:name=\nb,angle=160);
      \path (node cs:name=\nb,angle=-160) edge (node cs:name=\na,angle=-20);
    }
  \end{scope}
\end{tikzpicture}
\caption{The cursor \code{index} is on the boundary position \code{count + 1}. Calling \code{remove} in \mbox{\code{move_item}} will remove the element $v$ inside the list without updating \code{index}.}
\label{fig:set-remove-before}
\end{subfigure}

  \begin{subfigure}{\textwidth}
    \centering
    \begin{tikzpicture}[
  item/.style={rectangle, minimum size=8mm,very thick,rounded corners=2mm,draw=green!50!black!50,font=\scriptsize, top color=white, bottom color=green!50!black!20},
  node distance=4mm,
  pin distance=7mm,
  every pin edge/.style={<-, shorten <=4mm, ultra thick,green!40!black!60},
  ]
  \lstset{basicstyle=\footnotesize}

  \node (i0) [item,dotted,label=above:\code{0}] {};
  \node (i1) [item,right=of i0,label=\code{1}] {};
  \node (i2) [item,right=of i1,label=above:\code{2}] {};
  \node (i3) [right=of i2,minimum size=8mm] {$\cdots$};
  \node (iv) [item,right=of i3,draw=none,top color=green!30!black!10,bottom color=green!30!black!10] {};
  \node (i4) [right=of iv,minimum size=8mm] {$\cdots$};
  \node (icm1) [item,right=of i4,label={[label distance=.5pt]above:\code{count}}] {};
  \node (icm) [item,right=of icm1,dotted,label={[label distance=-.1pt]above:\code{count+1}}] {};
  \node (iend) [item,right=of icm,draw=none,top color=green!30!black!10,bottom color=green!30!black!10,label={[label distance=0pt]above:\code{count + 2}},pin=above:\code{index}] {};
  \begin{scope}[-latex,green!40!black!70,thick]
    \foreach \na / \nb in {i0/i1,i1/i2,i2/i3,i4/icm1,icm1/icm,i3/i4}
    {
      \path (node cs:name=\na,angle=20) edge (node cs:name=\nb,angle=160);
      \path (node cs:name=\nb,angle=-160) edge (node cs:name=\na,angle=-20);
    }
  \end{scope}
\end{tikzpicture}
\caption{After removing the item containing \code{v}, \code{count} is decremented. Thus, the cursor \code{index} points to the invalid position \code{count + 2}.}
\label{fig:set-remove-after}
\end{subfigure}
    \caption{Two faults violating contracts in \code{move_item}. In the first fault (\subref{fig:set-remove-at0}), \code{index} is initially ``\code{before}''. In the second fault (\subref{fig:set-remove-before}--\subref{fig:set-remove-after}), \code{index} is initially in position \code{count + 1} (\subref{fig:set-remove-before}), a position which becomes invalid after removing an element inside the list (\subref{fig:set-remove-after}).}
\label{fig:set-bug}
\end{figure}
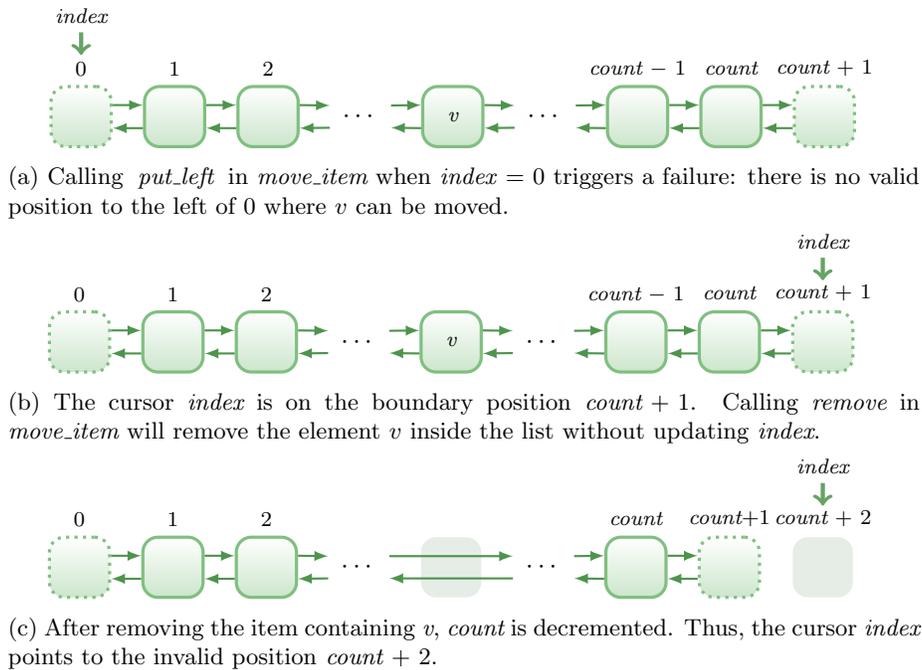

\subsection{Automatic corrections of the errors in \code{move_item} \tomainref{sec:correction-example}}
\afx collects the test cases generated by \autotest that exercise routine \code{move_item}.
Based on them, and on other information gathered by dynamic and static analysis, it produces, after running only a few minutes on commodity hardware without any user input, up to 10 suggestions of fixes for each of the two errors discussed.
The suggestions include only \emph{valid} fixes: fixes that pass all available tests targeting \code{move_item}.
Among them, we find the two ``\emph{proper}'' fixes in Figure~\ref{lst:moveItemCorrections}, which completely correct the errors in a way that makes us confident enough to deploy them in the program.

\lstset{numbers=left}
\begin{figure}[!ht]
\centering
\begin{tabular}{m{.1\textwidth} m{.35\textwidth} m{.35\textwidth}}
&
\begin{lstlisting}
if before then  (*\label{ln:fix1-begin}*)
  forth
end             (*\label{ln:fix1-end}*)
\end{lstlisting}
&
\begin{lstlisting}
if idx > index then     (*\label{app:ln:fix2-begin}*)
  idx := idx - 1
end                     (*\label{app:ln:fix2-end}*)
\end{lstlisting}
\end{tabular}
\caption{Corrections of the two errors in \code{move_item} automatically generated by \afx.}
\label{lst:moveItemCorrections}
\end{figure}

The correction for error occurring when calling \code{put_left} with \code{index = 0}) consists of inserting the lines~\ref{ln:fix1-begin}--\ref{ln:fix1-end} in Figure~\ref{lst:moveItemCorrections} (left) before the call to \code{put_left} on line~\ref{ln:putleft} in Figure~\ref{lst:moveItem}.
When \code{before} holds (i.e., \code{index} is 0), calling \code{forth} moves the cursor to position 1, which is valid for \code{put_left} and is where \code{v} is inserted.
This behavior follows a reasonable interpretation of what it means to insert an element ``to the left of cursor'' when the cursor is already in the leftmost position: insert the element in the leftmost position.

The correction for the other error (occurring when \code{v} is initially in a position to the left of \code{index}) consists of inserting the lines~\ref{ln:fix2-begin}--\ref{ln:fix2-end} in Figure~\ref{lst:moveItemCorrections} (right) before the call to \code{go_i_th} on line~\ref{ln:goith} in Figure~\ref{lst:moveItem}.
The condition \code{idx > index} holds precisely when \code{v} was initially in a position to the left of \code{index}$\,$; in this case, we must decrement \code{idx} by one to accommodate the decreased value of \code{count} after the call to \code{remove}.
This fix completely corrects this error beyond the specific case in Figure~\ref{fig:set-bug} reported by \autotest, even though \code{move_item} has no postcondition that formalizes its intended behavior.

\afx's suggestions for the two errors include other fixes that managed to pass all available tests.
Among the 10 suggestions for the error of calling \code{put_left} with \code{index = 0},\footnote{These data capture an average run of \afx on the two errors. See Section~\ref{sec:success-rate} for an analysis of variability from run to run.} \afx produces two fixes semantically equivalent to the one in Figure~\ref{lst:moveItemCorrections}; for example, it may use \code{index = 0} instead of \code{before}, but the two expressions are equivalent.
The remaining seven suggestions for the same error include fixes that avoid the failure and are consistent with the available contracts, but they arguably fail to completely capture the intended implicit semantics of \code{move_item}.
One of these fixes suggests to replace line~\ref{ln:putleft} in Figure~\ref{lst:moveItem} with \code{if not before then put_left (v) end}: call \code{put_left} only when its precondition is satisfied; otherwise, just discard the element \code{v}.
The original intent of the programmer implementing \code{move_item} probably did not contemplate this behavior as acceptable; therefore, we call fixes such as this one \emph{improper}.
Section~\ref{sec:validating-fixes} gives a more rigorous definition of proper fix, and Section~\ref{sec:fix-quality} discusses how many of the fixes generated by \afx are proper.

\section{Experimental evaluation~\tomainref{sec:experiment}}

\subsection{Experimental subjects~\tomainref{sec:subjects}}

Table~\ref{tab:subject-programs-complexity} presents the same information as Table~\ref{tab:subject-programs-size} averaged over the number of classes.

 \begin{table}[!htbp]
  \tablefont
   \caption{Average size and other metrics of the code bases.}
   \label{tab:subject-programs-complexity}%
   \centering
     \begin{tabular}{l *{7}{r@{.}l}}
     \toprule
     \multicolumn{1}{c}{\textbf{Code base}} & \multicolumn{2}{c}{$\frac{\textbf{\#kLOC}}{\textbf{\#C}}$} & \multicolumn{2}{c}{$\frac{\textbf{\#R}}{\textbf{\#C}}$} & \multicolumn{2}{c}{$\frac{\textbf{\#Q}}{\textbf{\#C}}$}
     & \multicolumn{2}{c}{$\frac{\textbf{\#Pre}}{\textbf{\#C}}$} & \multicolumn{2}{c}{$\frac{\textbf{\#Post}}{\textbf{\#C}}$} & \multicolumn{2}{c}{$\frac{\textbf{\#Inv}}{\textbf{\#C}}$} \\
     \midrule
     \efbase & 2&4 & 136&7 & 15&4 & 104&3 & 115&5 & 19&0 \\
     \txtlib & 1&2 & 78&0 & 4&8 & 9&7 & 13&4 & 1&1  \\
     \cardgm & 0&6 & 46&2 & 2&5 & 4&9 & 18&3 & 1&8  \\
     \elearn & 0&5 & 39&1 & 0&7 & 5&3 & 5&5 & 1&4 \\
     \toprule
     \textbf{Total} & \textbf{0}&\textbf{9} & \textbf{60}&\textbf{2} & \textbf{4}&\textbf{0} & \textbf{19}&\textbf{3} & \textbf{26}&\textbf{7} & \textbf{4}&\textbf{0} \\
     \bottomrule
     \end{tabular}%
 \end{table}%

\subsection{Experimental results~\tomainref{sec:experimental-results}}
Section~\ref{sec:subject-program-quality} is a preliminary discussion of how often \autotest provided tests of good quality suitable for fixing.

%==========================================================
\subsubsection{Testability of the experimental subjects}
\label{sec:subject-program-quality}
%==========================================================

For the evaluation, what matters most is the number and quality of fixed produced by \afx. It is interesting, however, to look into the results of \autotest sessions to get a more precise characterization of the experimental subjects and to see how the four code bases differ in their testability.
The data provides more evidence that the four code bases have different quality and are diverse subjects for our experiments.

\begin{figure}[bp]
	\centering
	\begin{subfigure}{.5\textwidth}
  		\centering
    	\includegraphics*[width=1\linewidth]{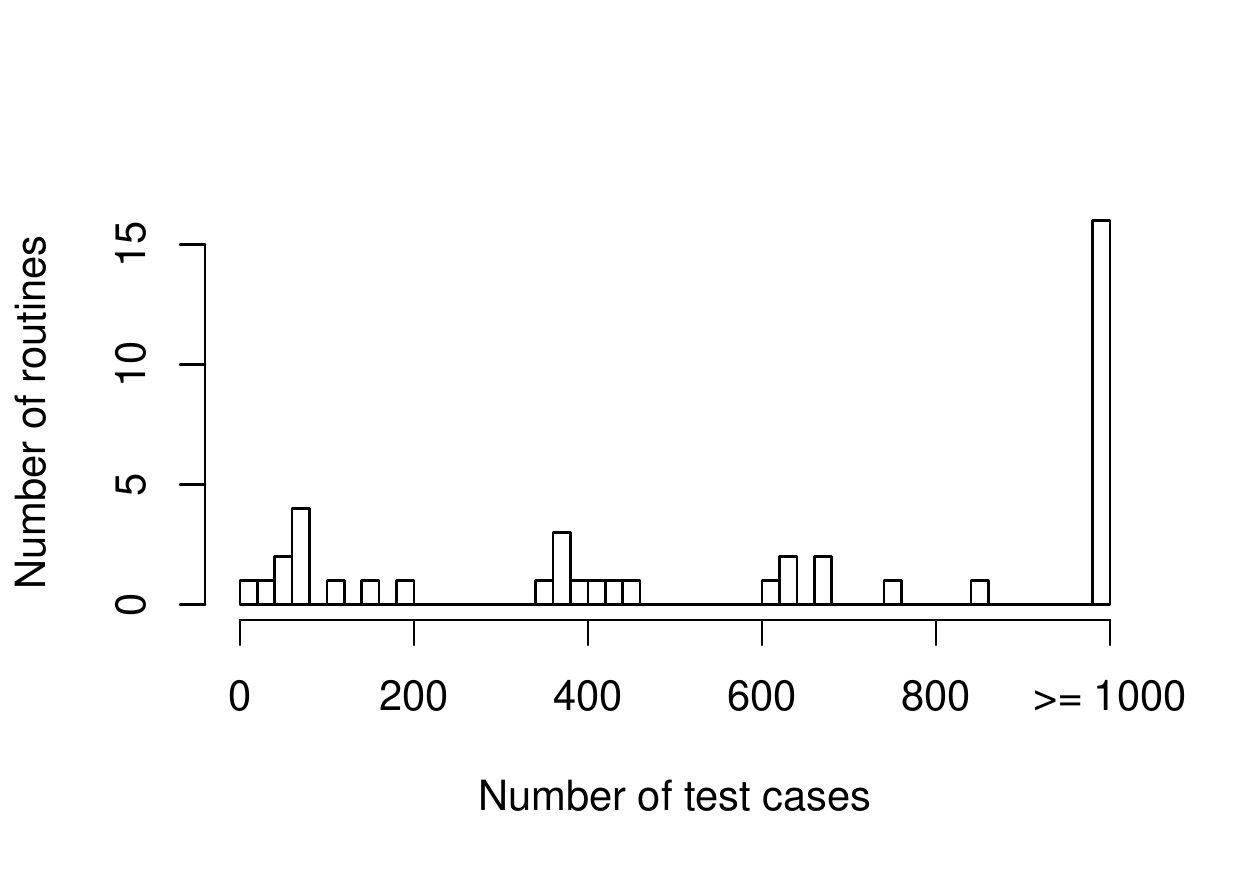}
    	\caption{\efbase}
		\label{fig:tc-efbase}
	\end{subfigure}%
	\begin{subfigure}{.5\textwidth}
  		\centering
    	\includegraphics*[width=1\linewidth]{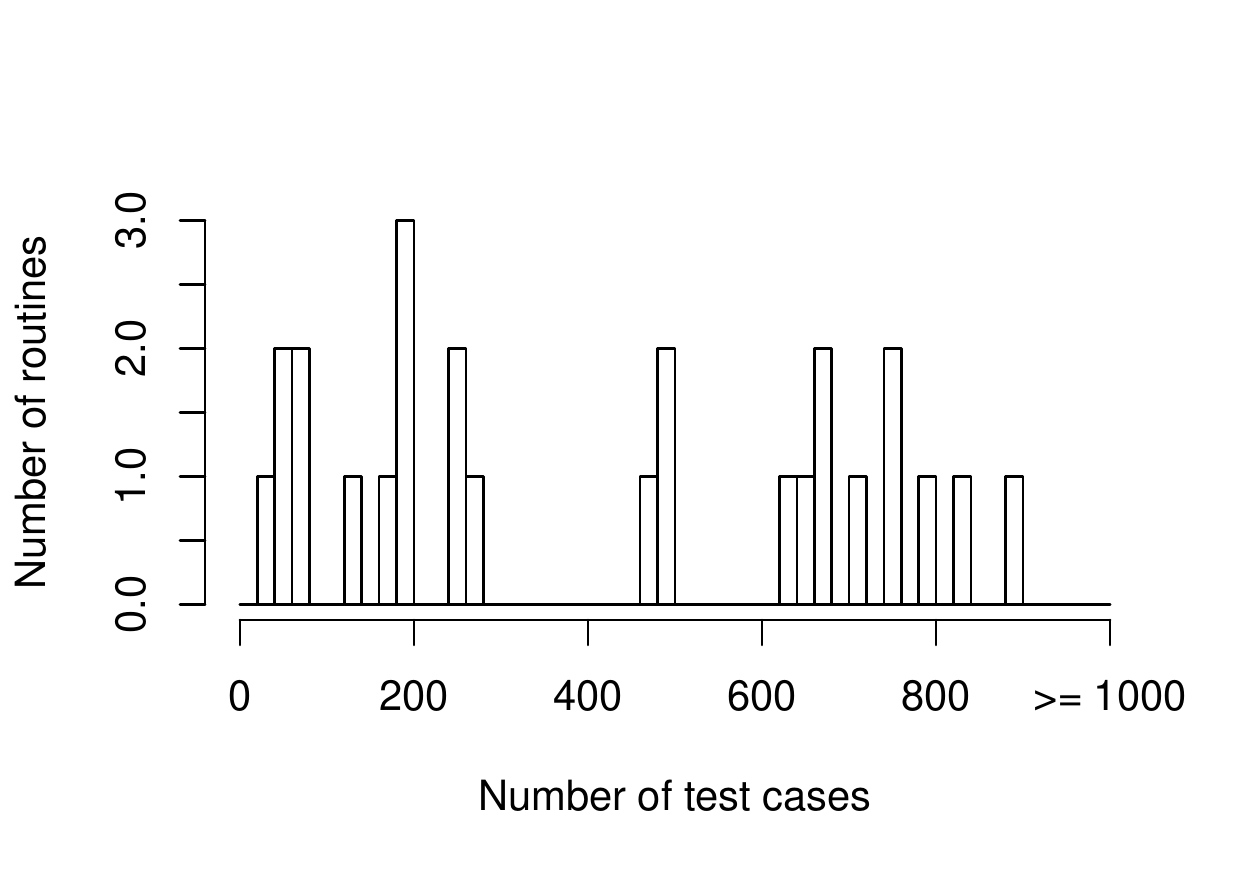}
    	\caption{\txtlib}
		\label{fig:tc-txtlib}
	\end{subfigure}%
	
	\begin{subfigure}{.5\textwidth}
  		\centering
    	\includegraphics*[width=1\linewidth]{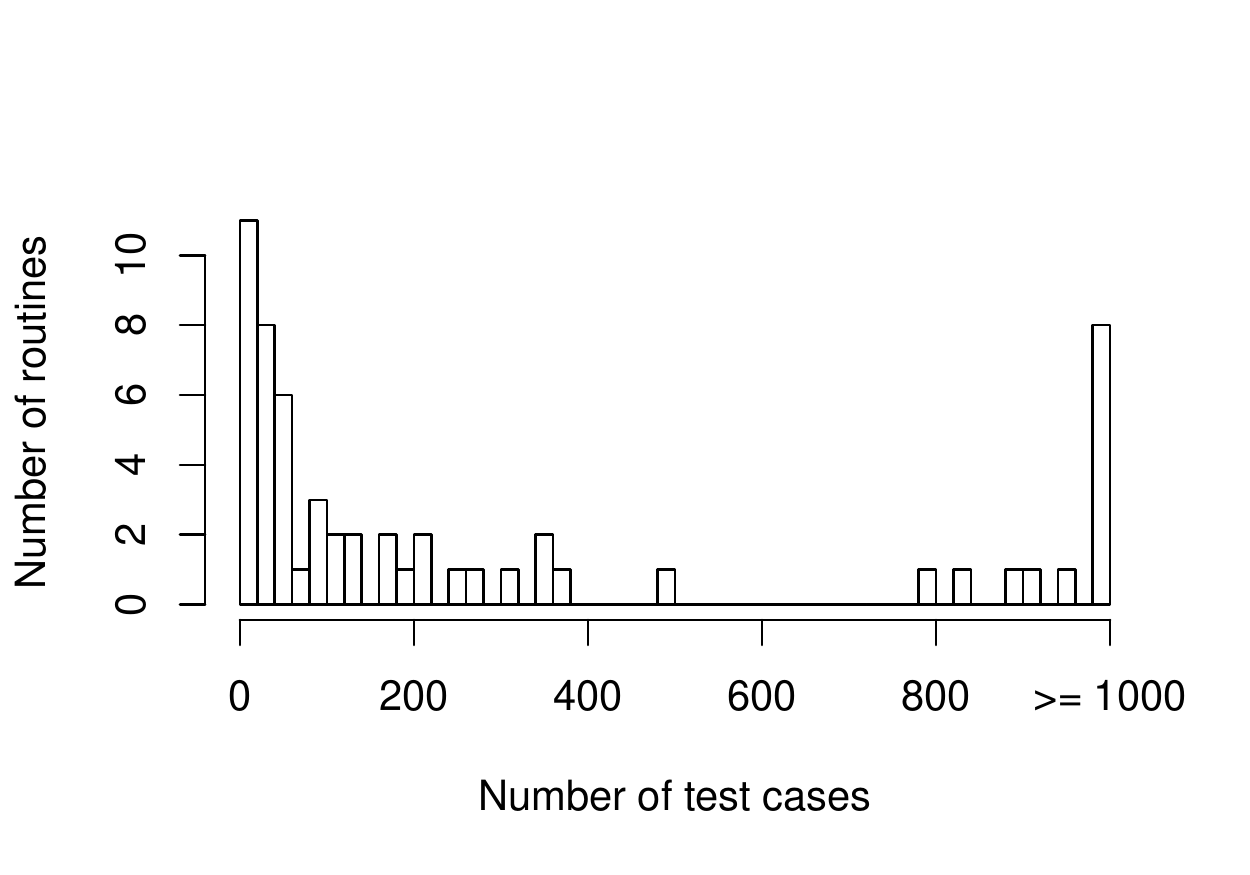}
    	\caption{\cardgm}
		\label{fig:tc-cardgm}
	\end{subfigure}%
	\begin{subfigure}{.5\textwidth}
  		\centering
    	\includegraphics*[width=1\linewidth]{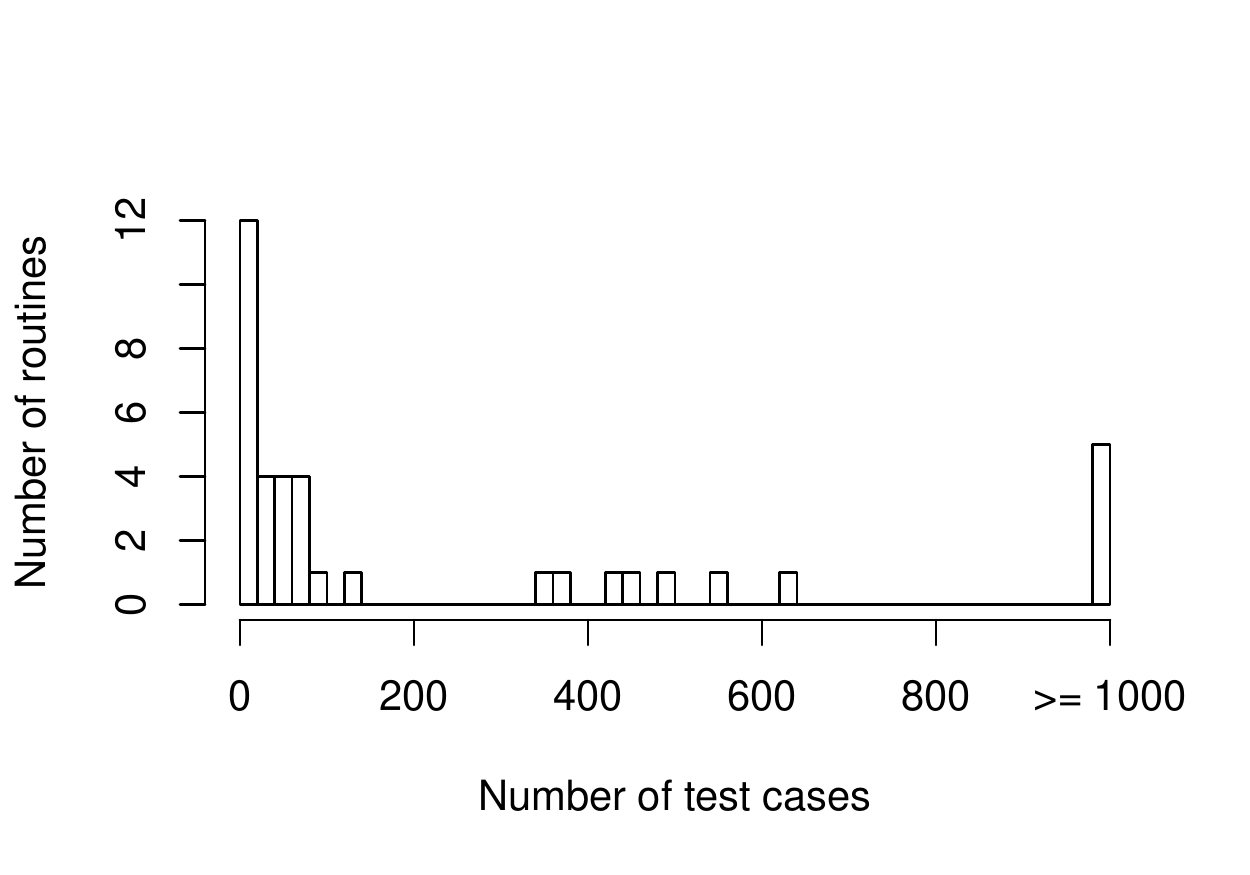}
    	\caption{\elearn}
		\label{fig:tc-elearn}
	\end{subfigure}%
	
	\begin{subfigure}{1\textwidth}
  		\centering
    	\includegraphics*[width=0.5\linewidth]{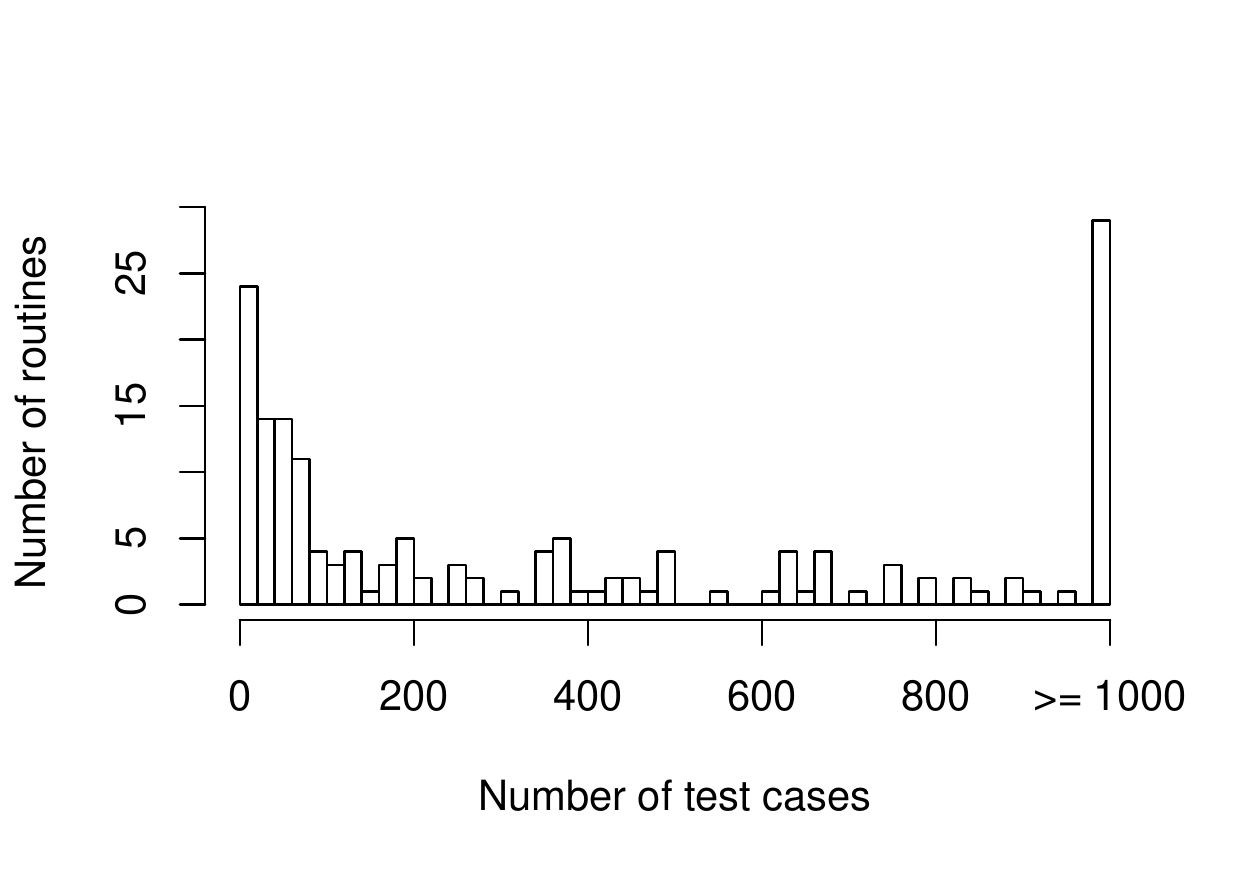}
    	\caption{All code bases}
		\label{fig:tc-all}
	\end{subfigure}%
	\caption{Number of tests generated by \autotest on the experimental subjects.}
	\label{fig:number-of-test-cases}
\end{figure}

\paragraph{Total number of tests.}
Each histogram in Figure~\ref{fig:number-of-test-cases} depicts the distribution of the mean number of test cases generated by \autotest in the 30 repeated 60-minute sessions for each routine.
That is, a bar at position $x$ reaching height $y$ denotes that there exist $y$ routines $r_1, \ldots, r_y$ such that, for each $1 \leq j \leq y$, the mean number $|T|$ of tests $T$ in the 60-minute series on some fault of $r_j$ is $x$.
Figures~\ref{fig:tc-efbase}--\subref{fig:tc-elearn} show the distributions of the individual code bases, while Figure~\ref{fig:tc-all} is the overall distribution.

The figures suggest that \efbase is normally easily testable---probably a consequence of its carefully-designed interface and contracts.
In contrast, \cardgm and \elearn are hard to test on average; and \txtlib is a mixed case.
A Mann-Whitney $U$ test confirms that the differences are statistically significant: if we partition the four code bases into two groups, one comprising \efbase and \txtlib and the other \cardgm and \elearn, intra-group differences are not statistically significant (with $692 \leq U \leq 1272$ and $p > 0.06$) whereas inter-group differences are (with $264 \leq U \leq 1754$ and $p < 0.03$).\footnote{In this section, the sample sizes for the $U$ tests are the number of faults in each code base.}

\begin{figure}[btp]
	\centering
	\begin{subfigure}{.5\textwidth}
  		\centering
    	\includegraphics*[width=1\linewidth]{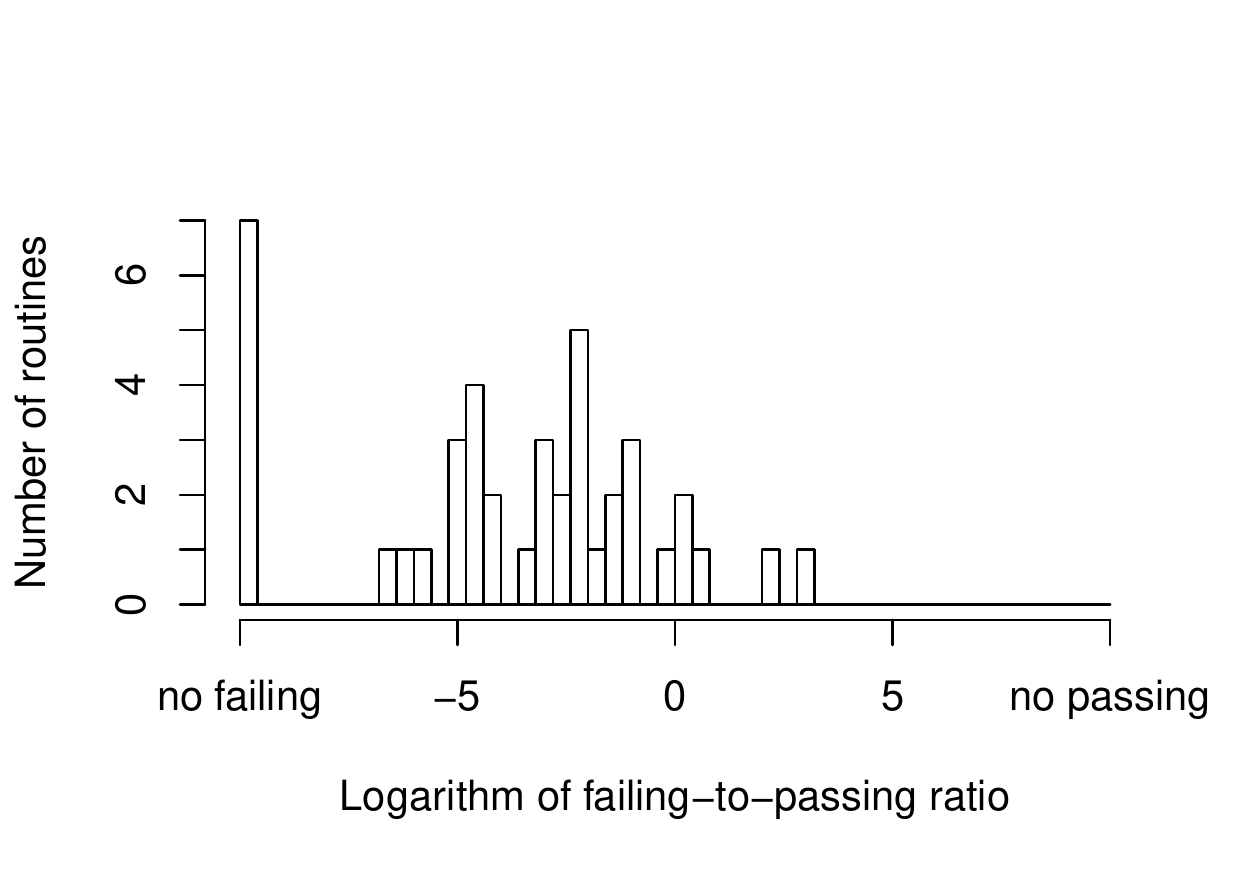}
    	\caption{\efbase}
		\label{fig:rl-efbase}
	\end{subfigure}%
	\begin{subfigure}{.5\textwidth}
  		\centering
    	\includegraphics*[width=1\linewidth]{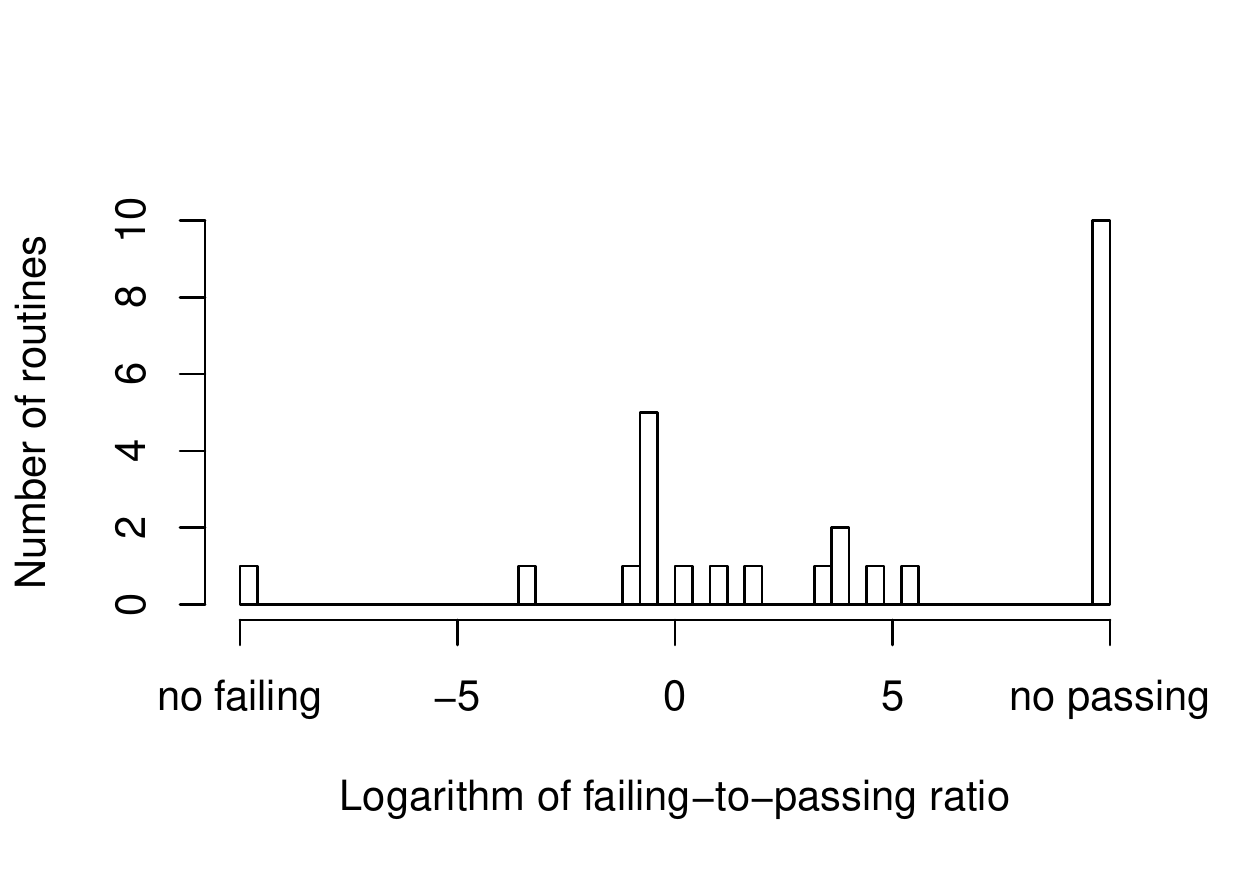}
    	\caption{\txtlib}
		\label{fig:rl-txtlib}
	\end{subfigure}%
	
	\begin{subfigure}{.5\textwidth}
  		\centering
    	\includegraphics*[width=1\linewidth]{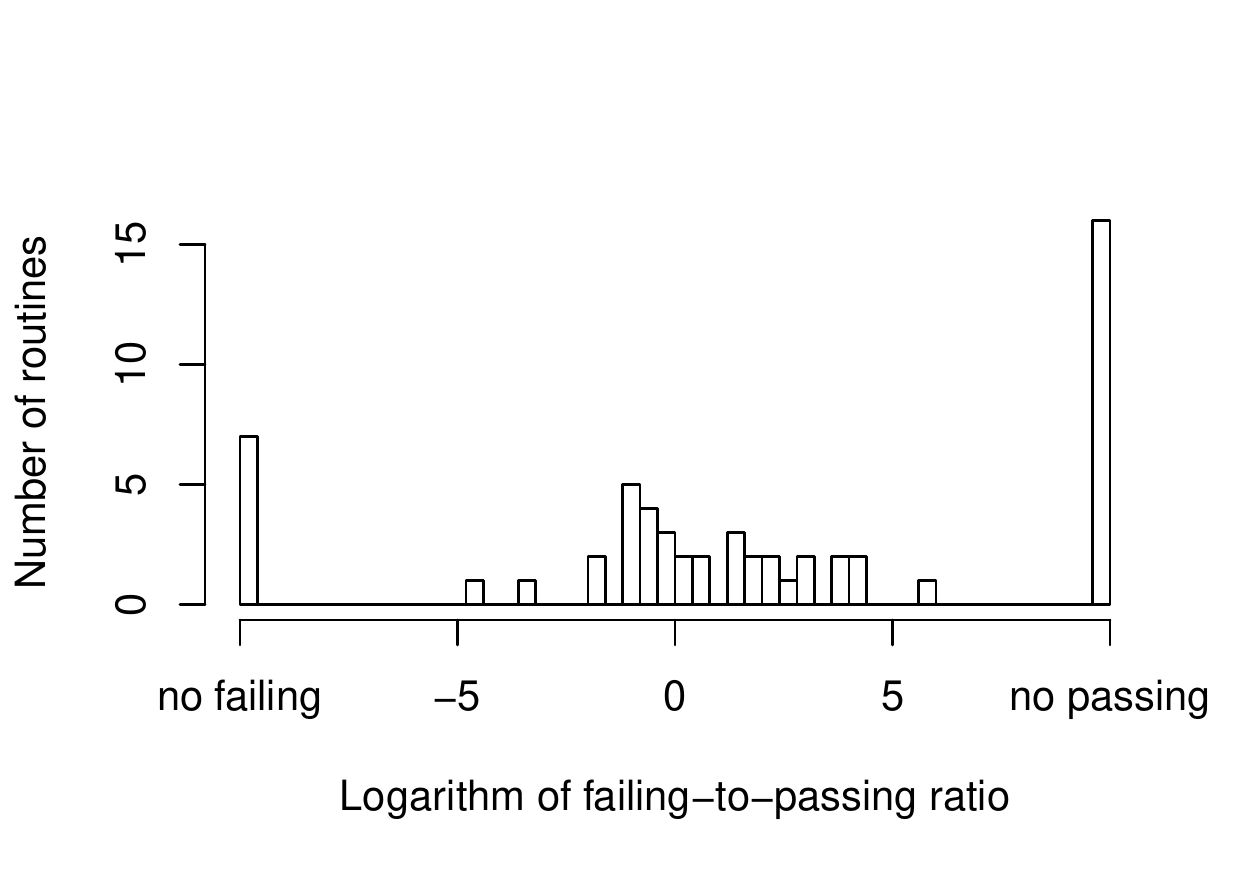}
    	\caption{\cardgm}
		\label{fig:rl-cardgm}
	\end{subfigure}%
	\begin{subfigure}{.5\textwidth}
  		\centering
    	\includegraphics*[width=1\linewidth]{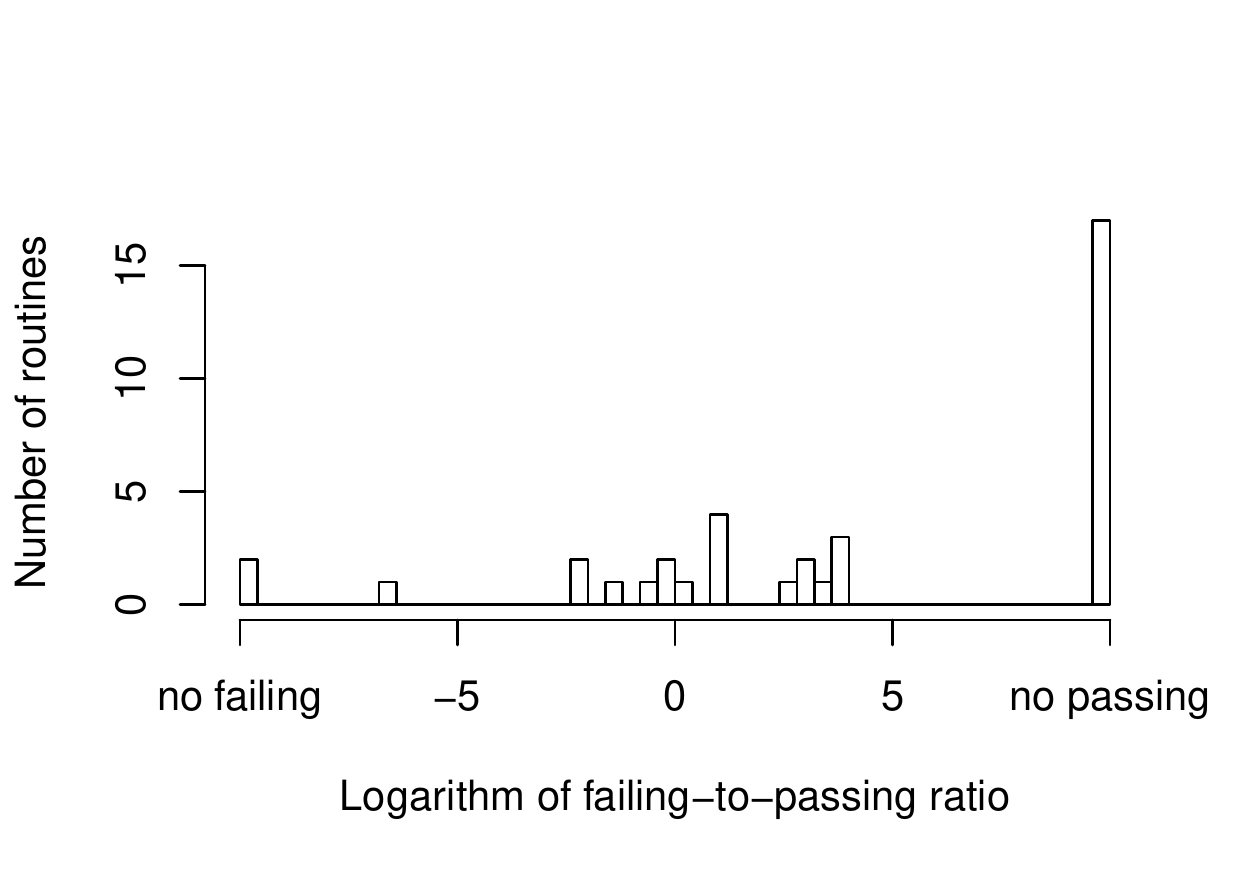}
    	\caption{\elearn}
		\label{fig:rl-elearn}
	\end{subfigure}%
	
	\begin{subfigure}{1\textwidth}
  		\centering
    	\includegraphics*[width=0.5\linewidth]{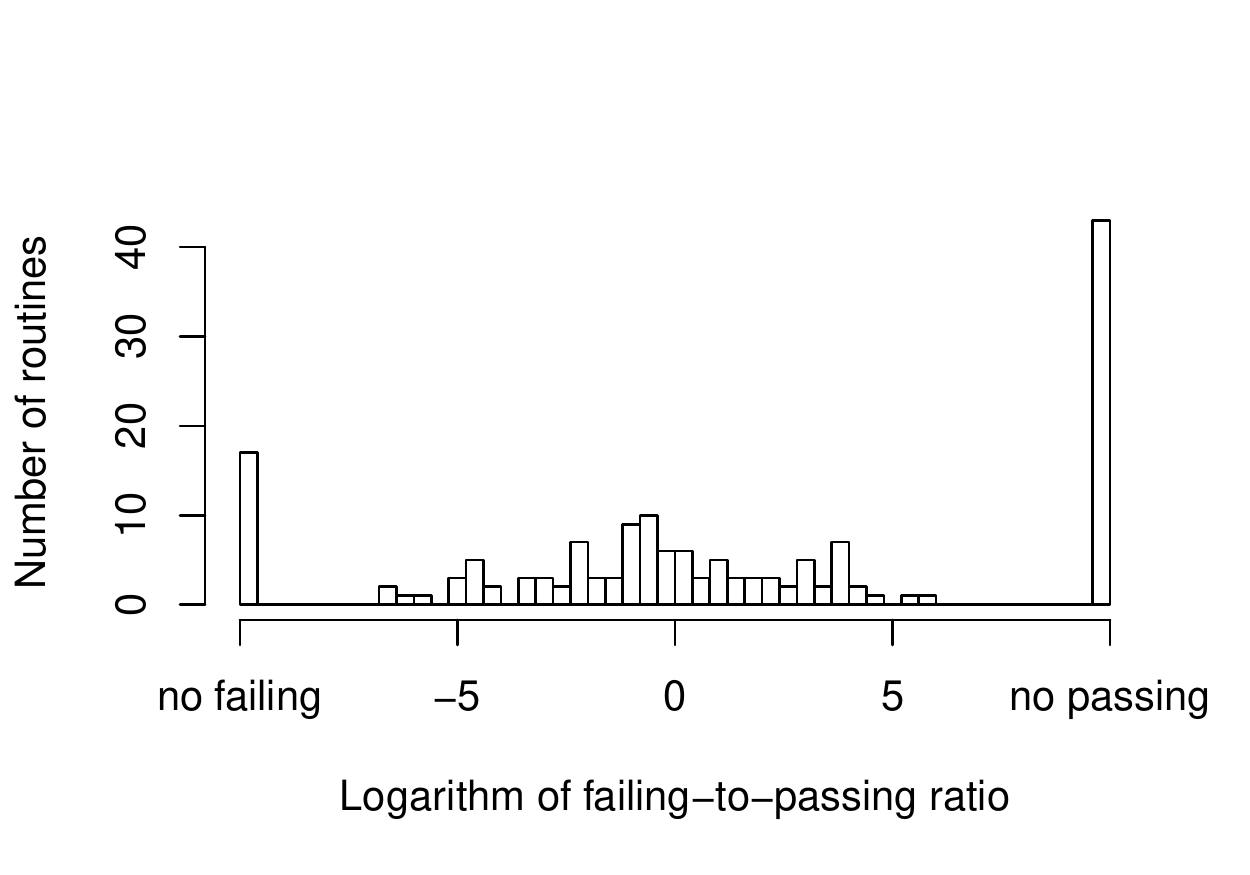}
    	\caption{All code bases}
		\label{fig:rl-all}
	\end{subfigure}%
	\caption{Failing-to-passing ratio of tests generated by \autotest on the experimental subjects; the horizontal scales are logarithmic.}
	\label{fig:failing-passing-ratio}
\end{figure}

\paragraph{Failing to passing tests.}
Another interesting measure is the average ratio of failing to passing tests generated in one session, which gives an idea of how frequent failures are.
Each histogram in Figure~\ref{fig:failing-passing-ratio} depicts the distribution of the mean failing-to-passing ratio for the test cases generated by \autotest in the 30 repeated 60-minute sessions for each routine; notice that the horizontal scale is logarithmic.
That is, a bar at position $x$ reaching height $y$ denotes that there exist $y$ routines $r_1, \ldots, r_y$ such that, for each $1 \leq j \leq y$, the mean ratio $|F|/|P|$ of failing tests $F$ to passing tests $P$ in the 60-minute series on some fault of $r_j$ is $e^x$.

Consistently with Figure~\ref{fig:number-of-test-cases}, Figure~\ref{fig:failing-passing-ratio} suggests that it is harder to produce failing tests for \efbase than for the other code bases.
A Mann-Whitney $U$ test confirms that the difference between \efbase and the other three code bases is statistically significant (with $105\leq U\leq 445$ and $p < 10^{-7}$) whereas the differences among \txtlib, \cardgm, and \elearn are not (with $515\leq U\leq 859$ and $p > 0.06$).

\section{Related work~\tomainref{sec:related}}

\afx integrates numerous techniques to detect, locate, and correct programming errors automatically. This section reviews the related work in these areas, focusing on those most closely related and that include an extensive evaluation.

%----------------------------------------------------------------------------
\subsection{Fault detection: automatic testing} \label{sec:rw-fault-detection}
%----------------------------------------------------------------------------
During the last decade, automatic testing has become an effective technique to detect faults, in programs and systems, completely automatically.
Among the approaches to automatic testing, random testing is one of the simplest, yet it has been successfully applied to a variety of programs including Java libraries and
applications~\cite{Pacheco.EAg.2005,Csallner2004,SharmaGAFM11}; Eiffel libraries~\cite{CPOLM11}; and Haskell programs~\cite{Claessen00quickcheck:a}.
The research on random testing has produced a variety of tools---including our own \autotest~\cite{meyer:programs:2009}, Randoop~\cite{PachecoLEB07}, JCrasher~\cite{Csallner2004}, Eclat~\cite{Pacheco.EAg.2005}, Jtest~\cite{Jtest}, Jartege~\cite{Oriat2004}, Yeti~\cite{DBLP:conf/iceccs/Oriol2010}, and RUTE-J~\cite{Andrews2006a}---as well as rigorous analysis~\cite{ArcuriRT} justifying its practical success on theoretical grounds.

Search-based test-case generation refines random testing with the goal of improving its performance and accuracy.
McMinn~\cite{McMinn04} and Ali et al.~\cite{AliBHP10} survey the state of the art in search-based techniques.
Genetic algorithms are a recurring choice for searching over unstructured spaces in combination with random exploration; Tonella~\cite{Tonella04} first suggested the idea, and Andrews et al.~\cite{AndrewsML11} show how to use genetic algorithms to optimize the performance of standard random testing.
Our previous work~\cite{WeiICST10,WRFPHSNM-ASE11} also extended purely random testing with search-based techniques.
Other approaches to automatic testing introduce white-box techniques such as symbolic execution~\cite{MajumdarS07} and fuzzying~\cite{GodefroidLM12}, or leverage the availability of formal specifications in various forms~\cite{HieronsBBCDDGHKKLSVWZ09,ZimmermanN10}.

%----------------------------------------------------------------------------
\subsection{Fault localization} \label{sec:rw-fault-localization}
%----------------------------------------------------------------------------
Fault localization is the process of locating statements that should be changed in order to correct a given fault.
Many of the approaches to automated fault localization rely on measures of code coverage or program states.

\paragraph{Code coverage.}
Code coverage metrics have been used to rank instructions based on their likelihood of triggering failures.
Jones et al.~\cite{jones:visualization:2002}, for example, introduce the notion of \textit{failure rate}: an instruction has a high failure rate if
it is executed more often in failing test cases than in passing test cases.
A block of code is then ``suspicious'' of being faulty if it includes many instructions with high failure rate; Jones et al.\ also implemented visualization support for their debugging approach in the tool Tarantula.

Renieris and Reiss's fault localization technique~\cite{renieris:fault:2003} is based on the notion of \textit{nearest neighbor}: given a test suite, the nearest neighbor of a faulty test case $t$ is the passing test case that is most similar to $t$.
Removing all the instructions mentioned in the nearest neighbor from the faulty test produces a smaller set of instructions; instructions in the set are the prime candidates to be responsible for the fault under consideration.
Artzi et al.~\cite{artzi:fault:2012}~apply similar techniques to rank statements together with their runtime values to locate execution faults in PHP web applications.
For better fault localization effectiveness, Artzi et al.\ also exploit concolic test-generation techniques to build new test cases that are similar to the failing one, the basic idea being that the differences between similar passing and failing test executions highly correlate with the fault cause.

Many other authors have extended code coverage techniques for fault localization.
For example, Zhang et al.~\cite{zhang:capturing:2009} address the propagation of infected program states;
Liu et al.~\cite{liu:sober::2005} rely on a model-based approach; and Wong et al.~\cite{wong:family:2010} perform an extensive comparison of variants of fault localization techniques and outline general principles behind them (which we follow in Section~\ref{sec:dynamic-analysis}).
Pytlik et al.~\cite{pytlik:automated:2003} discuss the limitations of using only state invariants for fault localization, a limitation that \autofix avoids by combining snapshots based on state invariants with snapshots based on enumeration (Section~\ref{sec:state-assess}).

\paragraph{Program states.}
The application of code coverage techniques produces a set of instructions likely to be responsible for failure; programmers still have to examine each instruction to understand what the problem is.
Fault localization techniques based on program states aim at providing more precise information in such contexts: state-based analyses are finer-grained than those based only on code coverage because they can also report suspicious state values that should be changed.
Huang et al~\cite{huang:automated:2007}, for example, suggest to insert check points in the program to mark ``points of interest''.
Then, a dynamic analysis---applied to program states rather than locations---can identify a set of suspicious states; furthermore, the usage of check points introduces the flexibility to skip uninteresting parts of the computation, for example repeated iterations of a loop.
\emph{Delta debugging}~\cite{zeller:isolating:2002,DBLP:journals/tse/ZellerH02} addresses similar issues: isolating the variables, and their values, relevant to a failure by analyzing the state difference between passing and failing test cases.

Angelina~\cite{Chandra:2011:AD} is a technique that repeatedly runs a program against a group of tests with the intent of discovering a list of expressions from the suspected faulty code such that: changing the value of any such expression at runtime could make the failing tests pass, while still letting the originally passing tests pass.
Such expressions are then reported to the programmer as suggestions: building the actual corrections is still the programmer's job.

Most fault localization techniques target each fault individually, and hence they perform poorly when multiple bugs interact and must be considered together.
To address such scenarios, Liblit et al.~\cite{liblit:scalable:2005} introduce a technique that separates the effects of multiple faults and identifies predictors associated with each fault.

While the research on automated fault localization has made substantial progresses, effectively applying fault localization in practice to help programmers still poses open challenges.
Parnin and Orso~\cite{ParninO11} demonstrate that most automated debugging techniques focus on tasks (mostly, localization) that represent only a small part of the real debugging activity.
Automated fixing techniques can help in this regard by providing an additional layer of automation that includes synthesizing suitable validated corrections.

\paragraph{Fault localization for automatic fixing.}
The program fixing techniques of the present paper include fault localization techniques (Section~\ref{sec:fault-analysis}).
To generate fixes completely automatically fault localization must be sufficiently precise to suggest only a limited number of ``suspicious'' instructions.
In our case, using contracts helps to restrict the search to the boundaries of the routine where a contract-violation fault occurs.
Then, we combine dynamic analysis techniques based on those employed for fault localization (Section~\ref{sec:dynamic-analysis}) with simple static analyses (Section~\ref{sec:static-analysis}) to produce a ranking of state snapshots within routines that is sufficiently accurate for the fixing algorithm to produce good-quality results.

Coker and Hafiz~\cite{Zack:ICSE13} show how to identify, through static analyses based on types, unsafe integer usages in C programs; simple program transformations can automatically patch such unsafe usages.

%%% Local Variables: 
%%% mode: latex
%%% TeX-master: "autofix.tex"
%%% End: 

\fi

% For arXiv
% \iftwocolumn
% \bibliographystyle{IEEEtran}
% \else
% \bibliographystyle{plain}
% \fi
% \bibliography{autofix}

\iftwocolumn
\input{bios/bios.tex}
\fi

\end{document}